**Forming native shortcut networks to simulate protein folding**

**Susan Khor**

## Abstract

Native shortcut networks (SCN0) are sub-graphs of native protein residue networks (PRN0). They are identified by the Euclidean Distance Search (EDS) algorithm we proposed previously. EDS exploits the navigable small-world network character of PRN0s. In this paper, we propose the Network Dynamics (ND) model, which reconstructs a PRN0 by adding back its edges according to some recipe, while its nodes are fixed at native (PDB) locations. A PRN0 reconstruction will eventually reconstruct its SCN0, but only after exploring several non-native shortcut network (SCN) configurations. It is these other SCN configurations that are of interest to us, as they produce the statistics to evaluate the different recipes. The recipes vary from each other slightly to investigate the effect of different edge orderings on protein folding. An edge ordering is deemed more successful if it produces a stronger correlation with experimental folding rates. Over proteins of different chain fold types, this basic requirement is best satisfied when a recipe favours earlier restoration of edges with smaller sequence separation, and earlier restoration of edges incident on nodes with larger remaining degree within the 1-hop neighborhood of a previously selected node. Further, this recipe generated more route-like trajectories over SCN space and a wider range of simulated folding rates, both of which signal cooperative two-state folding behavior. It also produced better correspondence between ND calculated phi-values and experimental phi-values obtained from a set of ten transition state ensembles (TSE). Also introduced is the local centrality measure which uses centrality of initial fold substructures to yield calculated phi-values from PRN0s with even better overall correspondence than ND calculated phi-values. Preliminary investigation with mixed ND recipes by chain fold type suggests that concocting further context-sensitive recipes by chain fold type and combining them could improve ND. This search task may be better undertaken by machine learning.

**Keywords**: protein residue network; small-world navigation; crony node selection; local centrality; simulated folding rate; calculated phi-value

## 1 Introduction

Proteins as networks of interacting nodes representing amino acid residues have been used to study protein structure and dynamics [1-7]. Typically, these studies propose a formula to construct a distance-based contact map from 3D coordinates of native-state protein molecules (PDB coordinates [8]), and techniques to analyze the resultant adjacency matrices. This paper presents the outcomes of our effort to use a network description of native-state proteins to model their folding as a graph *re*-construction process.

The proposed model is called Network Dynamics (ND). ND is based on the protein residue network abstraction we introduced in [9]. Later we found that a special sub-graph of native-state protein residue networks (PRN0) called native-state shortcut networks (SCN0) could be used to identify reasonable folding pathways for a set of canonical proteins [10]. Also observed in [10] was a significant Pearson correlation of -0.77 between the logarithm form of SCN0 Contact-Order (SCN0_lnCO) and experimental folding rate for two-state proteins. These two findings hint at the possibility of using this network abstraction to model protein folding. The Contact-Order of a set of edges is their average sequence distance [11]. Sequence distance of an edge or contact between two amino acids, is one plus the number of intervening amino acids on the protein chain. Typically, an edge is denoted by residue ids from the PDB file ($u$, $v$), and $|u\text{-}v|$ the absolute difference between $u$ and $v$ is its sequence distance. SCN0_lnCO is the average of the natural logarithm of sequence distance over all edges in SCN0: $\frac{1}{m}\sum_{i}^{m}\ln(e_i)$.

Shortcut networks are derived from protein residue networks by the Euclidean Distance Search (EDS) algorithm [9]. EDS operates like the message passing protocol of Milgram's small-world experiment [12], and depends on both a node's 3D coordinates and its connectivity in a network. Since EDS is a greedy local search algorithm, it may need to backtrack to reach a target node. A shortcut is an edge whose presence removes the need for EDS to backtrack on a particular path.

EDS was motivated by the desire to demonstrate the role of clustering in the construction of intra-protein pathways. Here clustering refers to the arrangement of edges in a graph that promotes short cycles [13]. Helical secondary structure elements in folded proteins are a major source of clustering in protein residue networks, and network abstractions of protein structure in general. While the small-world character of network abstractions of native-state protein structure has been widely recognized [1, 3, 5], intra-protein pathways were mainly built with a global search algorithm such as shortest paths, which is not as sensitive to the clustering particulars of protein structures as EDS paths are [14]. One of the main structural effects of protein chain collapse is increase in clustering (accompanied by shortening of average path length) through the formation of highly specific structures such as helices, beta-sheets and higher order combinations thereof. By demonstrating the dependency of intra-protein pathways on clustering, the link between protein folding and intra-protein pathways is fore-grounded.

Different from protein folding models where nodes representing amino acid residues move and contacts are made when node-pairs come within an acceptable Euclidean distance from each other, the ND model fixes nodes at their native-state locations (PDB coordinates [8]), and the set of all possible contacts are predetermined by the edges of a native-state protein residue network. The idea is for ND to order the formation of these contacts to simulate protein folding behaviors, by way of two parameters that govern node selection probability and edge creation probability. ND results are better when these two parameters are set in accordance with the major factors driving protein folding, namely slow chain entropy loss, and prioritization of hydrophobic and nucleation effects [15, 16]. We believe ND breaks new ground in that it attempts to extend the utility of a complex network description of native protein structures to the folding process.

The current design of ND is accessed at three levels: (i) macroscopic: correlation between experimentally observed folding rate and peak ND energy (a network-wide statistic); (ii) mesoscopic: agreement between first fold substructure identified previously from SCN0 [10] and ND suggested first fold substructure based on clustering coefficients of sub-networks covering pairs of adjacent secondary structure elements; and (iii) microscopic: correspondence between experimental phi-values and ND calculated phi-values using node level statistics such as degree and centrality. Results point to a single ND variant ('a') as the best performer at all three levels.

## 2 Materials and Method

### 2.1 Network construction

*PRN0* refers to a native-state Protein Residue Network. The nodes of a PRN0 represent amino acid residues of a protein chain. The edges of a PRN0 represent (putative) native contacts between pairs of amino acid residues. Two amino acid residues are in contact if they are not covalently bonded with each other (their sequence separation is > 1), and their (native-state) interaction strength $I_{uv}$ is ≥ 5.0. $I_{uv} = \frac{n_{uv} \times 100}{\sqrt{R_u \times R_v}}$ where $n_{uv}$ is the number of distinct atom-pairs, one from each of the residues $u$ and $v$, that are within 7.5 Å; the denominator corresponds to the geometric mean of the sizes of the amino acids, with $R_u$ and $R_v$ obtained from [17]. The cutoff Euclidean distance of 7.5 Å, and the $I_{uv} \geq 5.0$ threshold, were set through trial and error with a larger set of PDB structures previously [9], with the goal of creating networks that are singly connected without being unnecessarily dense.

A *PRN* is a PRN0 in progress; it contains all the nodes of a PRN0, but may be missing some of the edges. Networks generated by ND are PRNs.

*SCN0* refers to a native-state ShortCut Network; it is a sub-graph of a PRN0. A SCN0 is identified when the Euclidean Distance Search (EDS) algorithm maps out all EDS paths in a PRN0 (Figs. A1 & A2 Supp. Info.). Edges that help EDS avoid backtracking are labeled as shortcuts. Native shortcuts make the edges of a SCN0.

*SCN* refers to a shortcut network; it is a sub-graph of a PRN. Since a PRN is an edge-incomplete PRN0, EDS may identify shortcuts for a PRN that are not native shortcuts. Such shortcut edges are labeled *non-native*. Strictly speaking, non-native shortcuts are native contacts (in the context of a PRN0) since they are PRN0 edges. The edges of a SCN comprise native and non-native shortcuts. SCN is the object on which ND energy of a PRN is calculated (section 2.3).

**2.2 The ND model**

The basic steps a ND model makes are outlined in Figure 1. ND is a native-centric model; it assumes the existence of a PRN0 with nodes fixed at PDB locations, and is inspired by two major factors in productive protein folding: (i) slow loss in chain entropy, and (ii) prioritization of hydrophobic and nucleation effects.

The first factor is implemented by associating the probability of creating an edge with its sequence distance, so that sequentially local contacts can be made more likely to form than sequentially non-local contacts.

The second factor is implemented by associating the probability of selecting a node (to create edges) with its degree (connectivity). This implementation was prompted by the positive association between PRN0 node degree with both burial extent of residues (Fig. B1 Supp. Info.), and key folding residues (Figs. B2-B12 Supp. Info.). Key folding residues are residues involved in frequently formed contacts in transition-state structures. They are sourced from literature, and include reported folding nucleation sites, hydrophobic core sites, conserved rigid residues, and hydrogen-deuterium exchange sites that are slowest to exchange out or first to gain protection.

**Initialization:**

1. Build the native-state protein residue network (PRN0) and extract its native shortcut edges (SCN0).
2. For each node in the PRN0, record its edges (adjacency list) and number of edges (node degree).
3. Remove all edges from the PRN0. The PRN0 edges are added back in the following loop, and shortcut edges are identified by running EDS on the current PRN.

**Repeat until all native shortcuts are identified**:

4. Select a node $v$ with probability $p_n$ (according to node selection policy).
5. Create $v$'s edges with probability $p_e$, as follows:
   Inspect $v$'s adjacency list in random order. For a node $u$ in $v$'s adjacency list, if edge ($u$, $v$) is not yet created, create ($u$, $v$) with probability $p_e$.
6. If new edges are created, run EDS over the set of edges created so far to identify native and non-native shortcut edges (see note below Fig. A2 Supp. Info.). A shortcut edge is native if it belongs to SCN0, and non-native otherwise.
7. If ND variant is 'c' or 'a', creation of an edge ($u$, $v$) affects $p_n$. The node degrees of both $u$ and $v$ are decremented by one each. If ND variant is 'a', create the crony set.
8. Go to step 4.

**Figure 1** Main steps in a ND model.

**Table 1** Parameters of the five ND variants.

| Description | ND variant | (1) $p_n$ | (2) $p_e$ |
|---|---|---|---|
| unbiased | 'r' | 1/*N* | 0.5 |
| biased | 'x' | 1/*N* | scaled(1.0 / *ln*(\|x-y\|)) |
| | 'c' | *Simple* | scaled(1.0 / *ln*(\|x-y\|)) |
| | 'a' | *Crony* | scaled(1.0 / *ln*(\|x-y\|)) |
| improbable | 'y' | 1/*N* | 1.0 - scaled(1.0 / *ln*(\|x-y\|)) |

(1) Node selection probability. *N* denotes the number of nodes in a PRN0. Both *Simple* and *Crony* policies select a node according to a probability mass function (*p.m.f.*) that is proportional to a node's remaining degree. The difference is the *p.m.f.* is over all nodes with *Simple*, but over nodes in a crony set only with *Crony* (nodes not in the crony set has 0.0 probability). A node's 1-hop neighborhood (including itself) in the PRN0 comprises its crony set.

(2) Edge creation probability. *ln* denotes the natural logarithm function; $|a|$ denotes absolute value of $a$. It turns out that scaling has a major impact on the results as it dramatically reduces $p_e$ for longer range contacts (Fig. A3 Supp. Info.). Scaling is done as follows: ($p$ - min_$p$ + 0.0001) / (max_$p$ - min_$p$) where $p$ = 1.0/*ln*(|x-y|), and min_$p$ and max_$p$ are respectively the minimum and the maximum $p$ over all PRN0 edges. The constant 0.0001 is added to the numerator to ensure *scaled* $p$ > 0.0 for all edges.

ND allows a node to created zero or more contacts each time it is selected. This is a design choice that we believe is more physically realistic.

Only PRN edges are formed directly by ND; ND exerts no direct control over the formation of native or non-native shortcuts. Shortcut edges "emerge" when EDS finds them in a PRN. Unlike PRN edges which persists throughout a ND run once formed, the existence of shortcut edges depend on the current state of a PRN. A consequence of this is contact probability of an edge at a given point on the reaction coordinate, may differ considerably, depending on whether the edge is viewed as a shortcut or a PRN edge. Contact probability is the frequency with which the contact occurs in a collection of networks.

Five ND variants ('r', 'x', 'c', 'a' and 'y') are investigated in this paper. They differ in how the node selection probability ($p_n$) and edge creation probability ($p_e$) parameters are set (Table 1). 'r' is an *unbiased* ND model: nodes are selected uniformly at random, and edges are created with probability 0.5.

'x', 'c' and 'a' are ND models *biased* in different ways. 'x' exploits difference in sequence distance ($p_e$ is the inverse of the natural logarithm of the sequence distance of an edge scaled to be within (0.0, 1.0]), but pays no heed to node selection (nodes are selected uniformly at random). 'c' not only exploits difference in sequence distance like 'x', but also difference in *remaining* node degree (number of edges incident on a node in a PRN0 that is not yet created in a PRN). By preferentially selecting nodes with larger (remaining) degree, the 'c' ND variant aims to give key folding residues more chances to create their contacts (at least in the beginning of simulations), and thus establish nucleation networks or hydrophobic cores essential for subsequent folding activity.

The 'a' ND variant is like 'c', but aims to increase selection pressure on larger degree nodes and nodes in their direct neighborhood via a *crony node selection policy*. This policy chooses the next node (for edge creation) from a crony set, which comprises the currently selected node and its direct neighbors in the PRN0. Node selection is still proportional to remaining node degree within a crony set. Nodes outside a crony set are considered (selected proportional to remaining degree) only when no new contacts can be established by the current crony set (the sum of their remaining node degrees is 0). The currently selected node is included in a crony set to be consistent with the other node selection policies, which allows a node to be selected successively.

'y' is an *improbable* ND model: nodes are selected uniformly at random, and edges are purposely created in opposition to protein folding theory (long-range contacts are preferred over short-range contacts).

**2.3 ND energy calculation**

At minimum, a model should produce a random variable that correlates significantly with some important independent measure of protein folding; experimentally observed folding rates in our case. For the ND model, such a random variable is ND energy denoted ***E***, which is calculated using Miyazawa-Jernigan (MJ) inter-residue potentials [18].

For a ND network with $f$ native shortcuts and $g$ non-native shortcuts, ***E*** is the sum of the inter-residue potential of the $f$ native shortcuts minus the sum of the inter-residue potential of the $g$ non-native shortcuts.

The formulation of ***E*** is native-centric [19, 20] and minimally frustrated [21]. Folding of small single-domain proteins is a process of gaining native contacts while keeping entropy loss minimal [20, 22]. Transition-states (free energy barriers) arise in Go-models because loss in chain entropy occurs earlier than gain in stability [20]. These two factors are expressed in the formulation of ***E***: the presence of native shortcuts is stabilizing and lowers ***E***; while the presence of non-native shortcuts raises ***E***, as a penalty for premature (chain) entropy loss. On average, native shortcuts have smaller sequence separation than (potential) non-native shortcuts (Figure 2). We take this to imply that the presence of native shortcuts involve smaller entropy loss than the presence of non-native shortcuts. ND energy eventually follows a downward trend as the number of non-native shortcuts decreases, and the number of native shortcuts increases.

The negation of MJ scores for non-native shortcuts in the formulation of ***E*** is merely a convenient way to make some contacts repulsive to inject frustration into the system. When interpreted in terms of MJ potentials (which scores contacts according to their favourability of forming), the formulation of ***E*** may seem counter-intuitive with respect to the non-native shortcuts (which are nonetheless native in PRN0). It turns out that the use of MJ potentials may not be necessary; comparable results at the macroscopic level is achievable with other ways of contact scoring (Fig. A4 Supp. Info.).

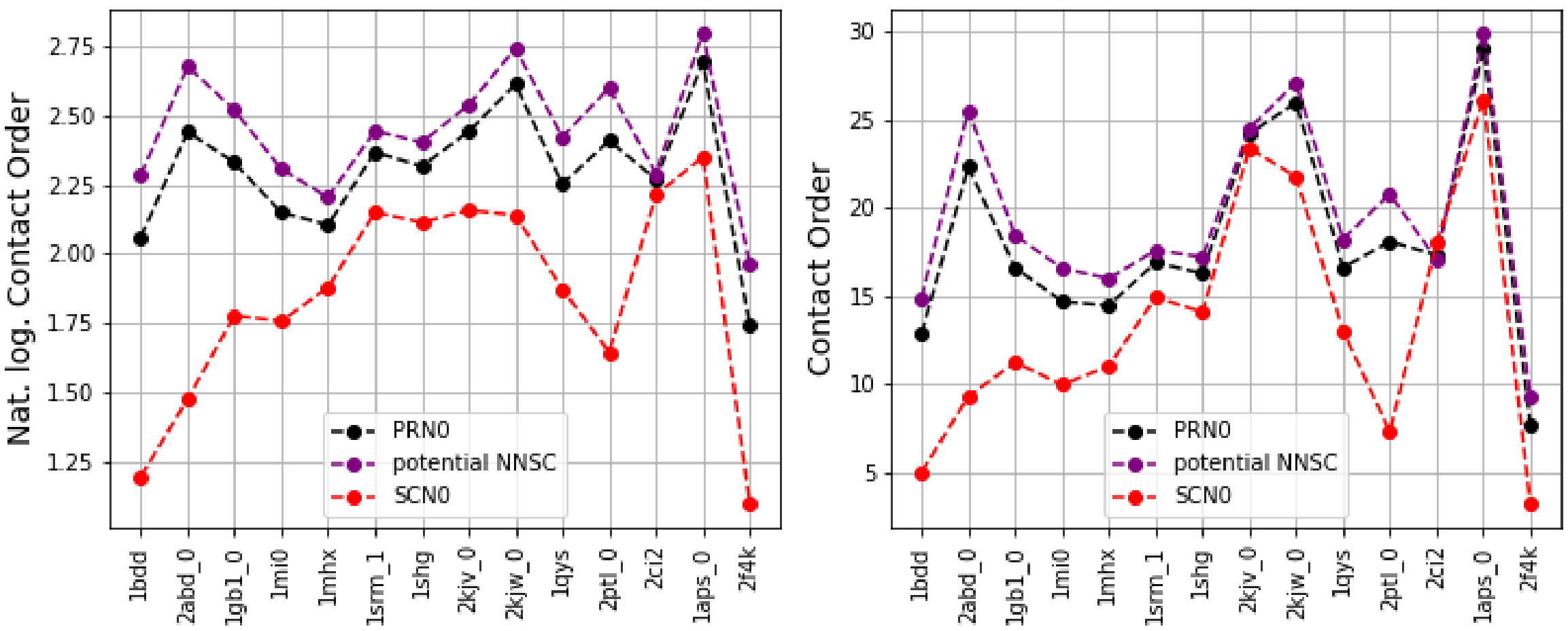

**Figure 2** Average natural logarithm of sequence separation [23] **(left)**, and average sequence separation [11] **(right)** for PRN0 edges, native shortcuts (SCN0) and potential non-native shortcuts (NNSC). Potential NNSC is PRN0 edges without SCN0 edges. The set of all non-native shortcuts (pooled from the 100 independent ND PRNs at each $Q$ over all $Q$s) cover almost all potential non-native shortcuts, and is highly similar for the ND variants except 'y' (Fig. G2 Supp. Info.).

### 2.4 ND energy profile for a protein

100 independent ND runs are made for each experiment to take advantage of the effects of the Weak Law of Large Numbers. 10 snapshots are taken per ND run. Each snapshot captures the ND network when its proportion of native shortcuts ($Q$) first equals or exceeds a multiple of 0.1. For each ND run, there is a snapshot at $Q$ = 0.1, 0.2, 0.3, ... 0.9, 1.0.

The $\boldsymbol{E}$ of ND networks are calculated, and averaged at each $Q$. A ND energy profile captures such sample mean $\boldsymbol{E}$ values as a function of $Q$. Peak $\boldsymbol{E}$ is the largest mean $\boldsymbol{E}$, and peak $Q$ is the argument where the maxima of an energy profile (peak $\boldsymbol{E}$) is found (Figure 3).

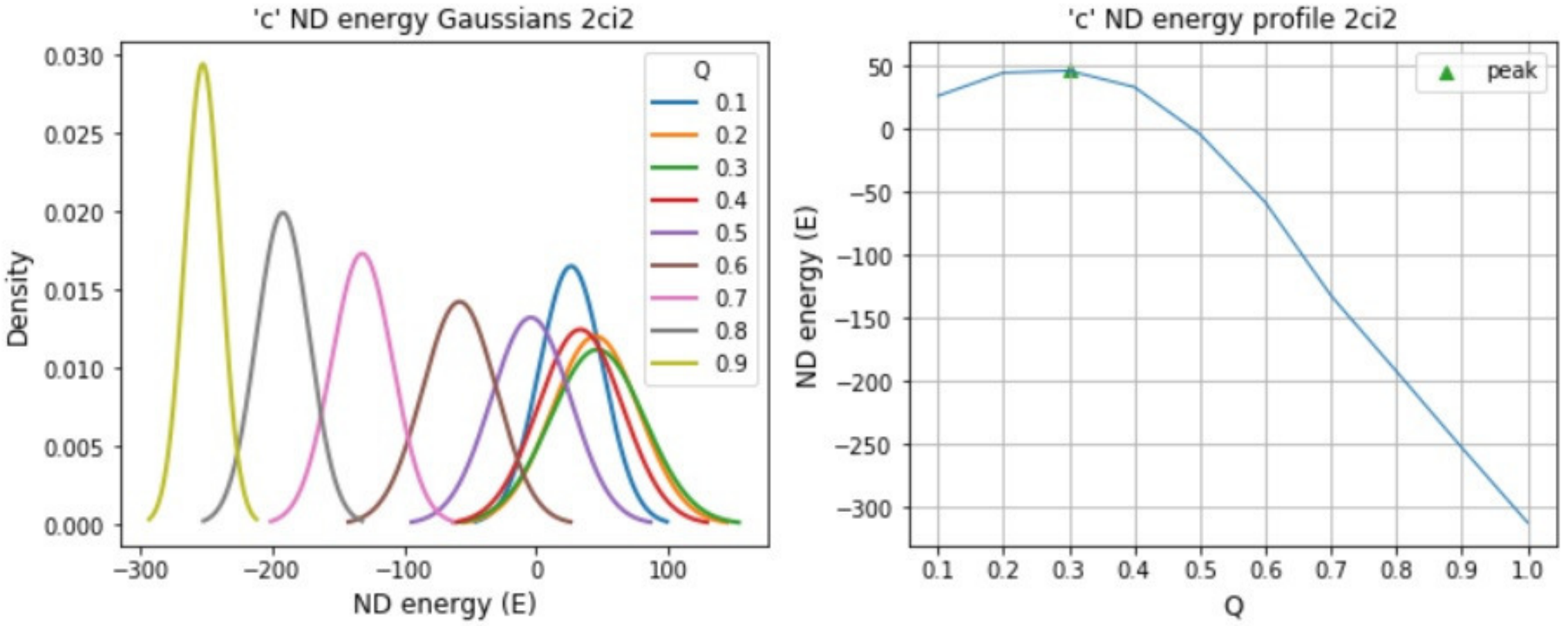

**Figure 3** ND energy ($\boldsymbol{E}$) Gaussians and corresponding energy profile for 2CI2, produced with 100 independent 'c' ND networks. Peak $Q$ is at 0.3, and peak $\boldsymbol{E}$ is almost 50.

In general, except for the 'y' ND variant, the ***E*** profiles of ND networks at $Q < 1.0$ are normally distributed per the Shapiro-Wilk test for normality at $\alpha=0.005$. The ***E*** values at $Q = 1.0$ are skewed right, with most of the density focused around the ND energies of SCN0s.

### 2.5 Model evaluation

Four datasets are used to evaluate ND:

(i) a dataset comprising 52 two-state folders, referred to as the Uzunoglu dataset [24];

(ii) a dataset comprising 20 multi-state folders, referred to as the Kamagata dataset [25];

(iii) a dataset of 14 small proteins whose folding pathways were studied previously [10], referred to as the Khor dataset, and

(iv) a dataset of 10 two-state folders with transition-state ensemble (TSE) structures and experimental phi-values used to generate them, referred to as the Paci dataset [26]. The TSE structures are first converted into PRNs (Figs. A5 & A6 Supp. Info.). This conversion preserves statistical associations reported in [27], namely Contact-Order of TSEs correlate with both experimental folding rate, and Contact-Order of native structures (Table A1 Supp. Info.).

The first two datasets evaluate the ND model against experimental folding rate. The last two smaller datasets are used to compare ND trajectories against known folding pathways, to corroborate the ND transition-state region, and to study the structural properties of ND networks against experimental phi-values.

## 3 Results and Discussion

### 3.1 Macroscopic

ND variant 'a' yields the best macroscopic level results: it produces the strongest Pearson correlation between peak ***E*** and experimental folding rate for the two larger datasets (Uzunoglu and Kamagata). Nonetheless, the r2 score (coefficient of determination) remained around 0.3 for the Kamagata dataset. This suggests that the present design of ND is better suited for modeling two-state folding processes. The significant correlation obtained with ND for the Kamagata dataset are possible only with the exclusion of one "outlier" data-point. Efficacy of ND to simulate proteins that fold in a non-two-state manner (multi-state proteins are distinguished by having detectable stable intermediates) is revisited in section 5.

*3.1.1 Effect of sequence distance on simulated folding rate (peak **E**) and folding cooperativity*

When $p_e$ is calibrated to prefer formation of shorter range contacts, the Pearson correlation between peak ***E*** and experimental folding rate switches from insignificant to significant (p-value < 0.013), and strengthens to at least -0.80 for the Paci dataset, -0.70 for the Uzunoglu dataset, and -0.50 for the Kamagata dataset (compare 'r' with 'x', 'c', 'a' rows in Tables 2-4). In contrast, when $p_e$ prefers longer range contacts, the said Pearson correlation is insignificant ('y' rows in Tables 2-4).

The reported correlations use peak ***E*** values for the simple reason that the peak of an ND energy profile is an unique landmark. It turns out that the peak ***E*** values are close to the best results possible over all *Q*s (Fig. C2 Supp. Info.).

**Table 2** Results with peak values for the Paci dataset, averaged over 100 independent ND runs.

| ND | (1) Fig. C1<br>Pearson r, p-value | (2)<br>peak *Q* | (3)<br>peak ***E*** range | (4)<br>r2 score | (5)<br>Predicted range |
|---|---|---|---|---|---|
| 'r' | -0.056, 0.879 | 0.190, 0.03 | 1.417 | 0.003 | 0.223 |
| 'x' | -0.896, 0.000 | 0.250, 0.05 | 2.249 | 0.802 | 3.453 |
| 'c' | -0.833, 0.003 | 0.260, 0.05 | 1.871 | 0.693 | 3.433 |
| 'a' | -0.848, 0.002 | 0.250, 0.05 | 1.848 | 0.719 | 3.482 |
| 'y' | 0.257, 0.474 | 0.150, 0.05 | 1.622 | 0.066 | 1.003 |

Column descriptions for Tables 2-5:
(1) peak ***E*** correlation with experimental fold rate.
(2) peak *Q* mean and standard deviation.
(3) Range of log10 peak ***E*** values which have been shifted (so they are all positive) and scaled (by squaring).
(4) r2 score (coefficient of determination) with a simple linear regression model. A r2 score closer to 1.0 indicates that a larger proportion of the variability in fold rates has been explained by the regression with peak *Q*.
(5) Range of folding rates predicted with the simple linear regression model in (4).

**Table 3** Results with peak values for the Uzunoglu dataset, averaged over 100 independent ND runs.

| ND | (1) Fig. C2<br>Pearson r, p-value | (2)<br>peak *Q* | (3)<br>peak ***E*** range | (4)<br>r2 score | (5)<br>Predicted range |
|---|---|---|---|---|---|
| 'r' | -0.214, 0.127 | 0.175, 0.04 | 1.928 | 0.046 | 1.507 |
| 'x' | -0.700, 0.000 | 0.250, 0.06 | 4.307 | 0.489 | 4.773 |
| 'c' | -0.711, 0.000 | 0.238, 0.06 | 4.331 | 0.505 | 4.745 |
| 'a' | -0.727, 0.000 | 0.227, 0.06 | 4.241 | 0.528 | 4.655 |
| 'y' | 0.066, 0.640 | 0.142, 0.05 | 1.866 | 0.004 | 0.388 |

**Table 4** Results with peak values for the Kamagata dataset, averaged over 100 independent ND runs.

| ND | (1) Fig. C3<br>Pearson r, p-value | (2)<br>peak *Q* | (3)<br>peak ***E*** range | (4)<br>r2 score | (5)<br>Predicted range |
|---|---|---|---|---|---|
| 'r' | -0.278, 0.235 | 0.185, 0.04 | 1.689 | 0.077 | 1.301 |
| 'x' | -0.554, 0.011 | 0.270, 0.06 | 4.198 | 0.307 | 2.574 |
| 'c' | -0.552, 0.012 | 0.260, 0.07 | 4.228 | 0.305 | 2.516 |
| 'a' | -0.562, 0.010 | 0.220, 0.06 | 4.081 | 0.316 | 2.263 |
| 'y' | -0.091, 0.703 | 0.160, 0.05 | 1.727 | 0.008 | 0.382 |

**Table 5** Results with peak values for the Khor dataset, averaged over 100 independent ND runs.

| ND | (2)<br>peak $Q$ | (3)<br>peak $\boldsymbol{E}$ range |
|---|---|---|
| 'r' | 0.157, 0.05 | 1.367 |
| 'x' | 0.271, 0.07 | 3.982 |
| 'c' | 0.236, 0.06 | 4.096 |
| 'a' | 0.214, 0.05 | 3.924 |
| 'y' | 0.143, 0.05 | 1.448 |

The presence of $p_e$ that favours shorter range contacts widens the range of peak $\boldsymbol{E}$. Range increase in simulated folding rate signals increased folding cooperativity in a model [28]. Cooperativity describes how conformational choices local to segments of a protein chain result in a globally optimal conformation and so solve Levinthal's paradox [29]. Results from the 'r', 'x' and 'y' ND variants confirm that sequence distance is a major factor in ND's ability to capture this salient aspect of protein folding. For the Uzunoglu dataset, the SRIM model [24] achieved a simulated folding rate range of 4.74. With this same dataset, ND achieves simulated ranges of about 4.3 with the biased ND variants (rows 'x', 'c' and 'a' in Table 3).

*3.1.2 Effect of sequence distance on peak Q*

The presence of $p_e$ that favours shorter range contacts delays peak $Q$, while $p_e$ that favours longer range contacts hastens peak $Q$. ND variants that permit long-range (>10 sequence distance) contacts to appear earlier tend to register smaller peak $Q$. The ND variants in decreasing order of permissiveness are: 'y', 'r', 'a', 'c' and 'x' (Fig. C4 Supp. Info.); peak $Q$ generally increases in this order (column (2) in Tables 2-5). The threshold of 10 is used so that contacts within secondary structure elements would generally be considered short-range. On average, an α-helix is 11 residues in length, and a β-strand, 6 residues [2].

*3.1.3 Effect of node degree and crony selection on peak Q*

The main effect of selecting nodes by remaining node degree is a tendency for peak $Q$ to decrease (rows 'x', 'c', 'a' in Tables 2-5). This tendency is related to the previous discussion on the timing of the appearance of long-range contacts in the ND variants. The creation of long-range contacts is made easier when nodes are preferentially selected by remaining degree (*Simple*), and easier still when coupled with a node selection strategy that favours the local neighborhood of a selected node (*Crony*).

This finding confirms an expected behavior of the ND model. A premise of ND is larger node degree relates positively with key folding residues, which include hydrophobic residues (Figs. B1-B12 Supp. Info.). A basic role of hydrophobic residues in protein folding is to induce the establishment of sequentially non-local but spatially local contacts [29].

Identifying factors that influence the selective formation of long-range contacts in early protein folding is crucial since specific long-range contacts are detected in transition-state structures, and play a role in stabilizing the folding nucleus [30, 31]. Long-range contacts are present in TSE PRNs (Fig. C5 Supp. Info.).

*3.1.4 ND-TS region and significance of peak Q as transition-state placement*

For two-state folders, it is tempting to associate peak $Q$ (where $\boldsymbol{E}$ is maximum in an ND energy profile) with the location where transitions occur between unfolded and folded states. An inspection of the percentage of native shortcuts present in each of the TSEs hints at the plausibility of this proposal; they are all below 53% (column SC0 in Table A1 Supp. Info.). The peak $Q$ values for two-state folders also all fall below 0.5 (Tables 2, 3, and 5). This agreement prompts us to name the region where $Q < 0.5$, the *ND-TS region*.

Further support for this classification is presented by locating the 10 TSEs from the Paci dataset, in terms of their $\boldsymbol{E}$ values, within the ND-TS region. This test also serves as a check on the reasonableness of our formulation of $\boldsymbol{E}$. Two approaches are used to locate a TSE within an ND energy profile (Fig. D1 Supp. Info.): (i) Kolmogorov-Smirnoff goodness-of-fit (KS), and (ii) maximum likelihood (MAP). With KS, we search from the nine ND energy Gaussians at each $Q$ (the ND energy Gaussian at $Q$=1.0 is excluded), for the one that best fits the $\boldsymbol{E}$ values of a TSE (produces the smallest test statistic). With MAP, we search from the nine *a priori* equally likely ND energy Gaussians at each $Q$ (the ND energy Gaussian at $Q$=1.0 is excluded), for the one that is most likely to generate the $\boldsymbol{E}$ values of a TSE, which we assume are independent and identically distributed, and can be reasonably described by a single Gaussian. Except for the 1LMB4 TSE, the TSE $\boldsymbol{E}$ values are normally distributed by the Shapiro-Wilk test at $\alpha = 0.01$.

The two approaches yield similar results; the KS and MAP estimated $Q$ values for the 10 TSEs are summarized in Table 6 and visualized in Fig. D2 (Supp. Info.). For all five ND variants, the bulk of the estimated $Q$ values fall within the ND-TS region, with averages ranging between 0.34 and 0.41. Of the biased ND variants, 'a' yields the smallest estimated $Q$ (with the MAP approach); the TSEs are "better tucked-in" within its ND-TS region.

**Table 6** KS and MAP estimates of the location (mean $Q$) of TSEs from the Paci dataset within ND energy profiles, and their respective Mean Absolute Difference (MAD) from ND peak $Q$s.

| ND | ND peak $Q$ (mean, std. dev.) | KS estimated $Q$ | MAD | MAP estimated $Q$ | MAD |
|---|---|---|---|---|---|
| 'r' | 0.19, 0.03 | 0.39, 0.11 | 0.20 | 0.38, 0.13 | 0.21 |
| 'x' | 0.25, 0.05 | 0.41, 0.16 | 0.20 | 0.39, 0.20 | 0.22 |
| 'c' | 0.26, 0.05 | 0.38, 0.18 | 0.20 | 0.35, 0.21 | 0.21 |
| 'a' | 0.25, 0.05 | 0.39, 0.18 | 0.20 | 0.34, 0.22 | 0.23 |
| 'y' | 0.15, 0.05 | 0.34, 0.09 | 0.21 | 0.35, 0.10 | 0.22 |

### 3.2 Mesoscopic

ND variant 'a' yields the best mesoscopic level results: the initial fold step of nine out of the 14 short protein chains examined, is made in the ND-TS region with at least 50% probability, and the $Q$ at which this event first occurs is closer on average to its peak $Q$ for these proteins.

#### *3.2.1 Initial fold substructures*

Folding pathways based on the native shortcut networks of 14 small proteins were proposed previously [10]. These folding pathways are built by pairing up adjacent secondary structure units guided by their clustering coefficients in native shortcut networks, $C_{\mathrm{SCN0}}$ (Fig. E1 Supp. Info.). In this section, we assess how closely ND trajectories track such folding pathways derived from native structures. Of particular interest is the initial fold step, since the presence of these initial substructures signal nascent productive folding activity for two-state folders, and may even be the rate-limiting structure. For example, the initial fold substructures selected by our folding pathway method for the *Ig* domain of proteins G and L are the C-terminus and N-terminus β-hairpins, respectively. Supporting evidence for the selected initial fold step of each protein examined here were discussed in ref. [10].

To perform this assessment, folding pathways were extracted from all ND native shortcut networks with the method outlined in Fig. E1 (Supp. Info.), and $P_{\mathrm{init}}$ which is the proportion of pathways where the initial fold step matches the one proposed by the native shortcut network, is calculated for each $Q$. $P_{\mathrm{init}}$ is inspired by the $P_{fold}$ concept which classifies conformations by their probability of folding [32]. At its transition-state, a protein with two-state kinetics has equal chance of folding or unfolding ($P_{fold}$ = 50%) [33]. Our query is whether the $Q$ values where $P_{\mathrm{init}}$ = 50% fall within the ND-TS region ($Q < 0.5$). The results of this assessment are summarized in Table 7, and visualized in Figs. E2 and E3 (Supp. Info.).

**Table 7** $P_{init}$ results for 14 small proteins from the Khor dataset.

| ND | ND Peak | $P_{init}$ first approaches 50% | | Number of structures (3) | | |
|---|---|---|---|---|---|---|
| | $Q$ (mean, std. dev.) | $Q$ (1) | MAD (2) | $Q < 0.4$ | $Q < 0.5$ | $Q < 0.6$ |
| 'r' | 0.157, 0.05 | 0.478, 0.21 | 0.321 | 4 | 8 | 10 |
| 'x' | 0.271, 0.07 | 0.407, 0.22 | 0.178 | 7 | 8 | 11 |
| 'c' | 0.236, 0.06 | 0.443, 0.24 | 0.221 | 6 | 7 | 8 |
| 'a' | 0.214, 0.05 | 0.378, 0.23 | 0.193 | 8 | 9 | 11 |
| 'y' | 0.143, 0.05 | 0.600, 0.15 | 0.457 | 0 | 3 | 6 |

(1) $Q$ values are interpolated at $P_{init} = 50\%$, and truncated since $Q = q$ refers to $[q, q+0.1)$ for $q < 1.0$.
(2) Mean Absolute Difference between estimated $Q$ and peak $Q$.
(3) Number of proteins structures whose estimated $Q$ lies below the stated threshold.

The results give a positive answer to our query. Averaged over the 14 small proteins, $Q$ where $P_{init}$ first approaches 50% fall within the ND-TS region for all ND variants except the improbable ND variant 'y'. With this finding, it could be argued that our designation of the range of the reaction coordinate covered by $Q < 0.5$. as the ND-TS region is justified, for the four ND variants. The problematic structures (upper quadrant in Fig. E3 Supp. Info. is not crossed at all, or crossed only by either 'r' or 'y' trajectories) are: 2PTL and the synthetic 1MI0 and 1QYS structures. 2CI2, the exemplar of the nucleation mechanism [30], has a $Q$ where $P_{init}$ first approaches 50% at 0.5, with all three biased ND variants.

ND variant 'a' yields the best results for this mesoscopic investigation: it has the smallest estimated $Q$ (0.378) and the bulk of its estimated $Q$ values stay within the ND-TS region (Fig. E2 Supp. Info.); its MAD is one of the smallest (0.193) which means that its peak $Q$s may be more indicative of transition-state placement; and it reports the largest number of structures (9) whose estimated $Q$ lies within the ND-TS region, and at two other thresholds of $Q$. Impressively, at $Q < 0.4$, 'a' has twice as many structures whose $P_{init}$ first approaches 50% than 'r'. But at slightly larger $Q$s, ND variant 'r' catches up with the biased ND variants, on this metric.

*3.2.2 Substructure centrality*

An observation that stands out from the $P_{init}$ results is that a process which is agnostic to both sequence distance and node degree differences such as 'r', is also capable of forming the selected initial fold substructures within the ND-TS region for at least half of the dataset. An extreme interpretation of this observation is that unbiased search in the early stages could initiate productive folding [34, 35]. Protein chains may be primed (via biological evolution [36]) for their native folds [37]; traces of native-like initial fold substructures have been observed in denatured proteins [38, 39].

Nonetheless, the ease with which the selected initial fold substructures can form is a desirable property in light of their centrality to EDS paths, and is in harmony with the functional view that the purpose of protein folding is to create productive intra-protein pathways. We define *substructure centrality* as the number of non-edge EDS paths (paths of length > 1) that traverse more than one of the secondary structure elements of the substructure, divided by the number of residues in the substructure.

The substructure centrality of all candidate initial fold substructures for the 14 small proteins are reported in Table E1 (Supp. Info.). Candidate initial fold substructures with the highest (largest) substructure centrality coincide with the first fold steps, except for 2ABD, 2PTL and 2CIC. Further, candidate initial fold substructures with the two highest substructure centralities are involved in the first two fold steps, except 2ABD, 1APS and 2KJW. These observations indicate that fold substructures that are conduits for many EDS paths, tend to form in the earlier stages of folding, and that such fold substructures can be identified from native shortcut networks.

### 3.3 Microscopic

ND variant 'a', together with 'c', yield the best microscopic level results: normalized node centrality from its ND PRNs produced phi-values that are competitive with phi-values calculated with normalized node degree from TSE PRNs; and its ND PRNs that produce the best calculated phi-values are those that lie within the ND-TS region. However, our investigation reveal that phi-values can be calculated directly from native-state (NS) PRNs quite effectively as well with local node centrality.

#### *3.3.1 Calculated phi-values with PRN node statistics and local node centrality*

Experimental phi-values are produced by the protein engineering method [40] which analyzes how single-site mutations affect folding rate and stability of mutants relative to their wild-types [33]. Residues with large positive phi-values are presumed to be more important to the formation and stabilization of transition-state structures, while residues with phi-values close to 0.0 are as unstructured in the transition-state as in the native-state.

In this section, we use the degree and centrality of PRN nodes to compare the relative importance of residues to folding, as assessed by phi-values from experiments. The connection between node centrality and key residues (the two or three residues critical for forming a protein's folding nucleus) was first made in ref. [1]. They found that node between-ness (normalized by the total number of shortest paths) derived from transition-state ensembles were more predictive of regions that contain key residues, than those derived from native-state structures. Unlike the study in [1], our evaluation is done for all residues with given experimental phi-values in the Paci dataset, and our node centrality measure is based on the set of all EDS paths for a network.

Further, we introduce the notion of *local* node centrality. In contrast to the usual notion of node centrality, which is global since it is computed with a network's set of all paths, local node centrality is calculated with a *subset* of all paths. Inspired by the proclivity of *selected* initial fold substructures to have highest substructure centrality (section 3.2, Table E1 Supp. Info.), we choose this subset to be the set of EDS paths that contribute to substructure centrality of the *candidate* initial fold substructure with highest substructure centrality. This choice of EDS path subset turns out to be satisficing: the selection criterion is clear, and the results are close to optimal (Fig. F1 Supp. Info.).

For comparability with the experimental phi-values, which range between 0.0 and 1.1, PRN node degree, (global) node centrality, and local node centrality are normalized (divided) by the highest value in a sequence. The correspondence between these normalized node statistics (calculated phi-values) and phi-values from experiments is quantified by taking their Mean Absolute Difference (MAD); the interpretation being smaller MAD scores indicate better correspondence between experimental and calculated phi-values. Admittedly, MAD is only one way to reduce this evaluation to a single number per protein. phi-values are calculated with node statistics obtained from three types of PRNs: native-state (NS), transition-state ensemble (TSE), and ND generated. Node statistics from TSE PRNs and ND PRNs, are averages.

Our way of normalizing node statistics permits the use of native-state node degree, and bounds node centrality statistics to within [0.0, 1.0] for this analysis. Conventionally, calculated phi-value for a residue is the ratio of its node degree in an ensemble to its native-state node degree [33, 35]. This formulation fits with the interpretation of experimental phi-values as quantifying the extent to which a residue's native contacts are formed in the transition-state ensemble [27, 31, 39]. But this interpretation applies to node degree statistics from TSE or ND PRNs only. Further, since backtracking is possible with EDS paths, a node may be traversed more than once by an EDS path; as such a node's centrality may exceed the total number of EDS paths. There are also subtle implications for dividing by a constant (per node) when dealing with ND node statistics. Our normalization procedure emphasizes relative importance of nodes within a context, rather than relative to the native-state. We evaluated both ways of normalizing node degree from TSE PRNs, and found no significant difference between the two in terms of average MAD for the Paci dataset (TSE_deg and TSE_deg2 in Figure 4), although the difference can be large for specific protein chains, e.g. 1IMQ (Fig. F2 top-right Supp. Info.).

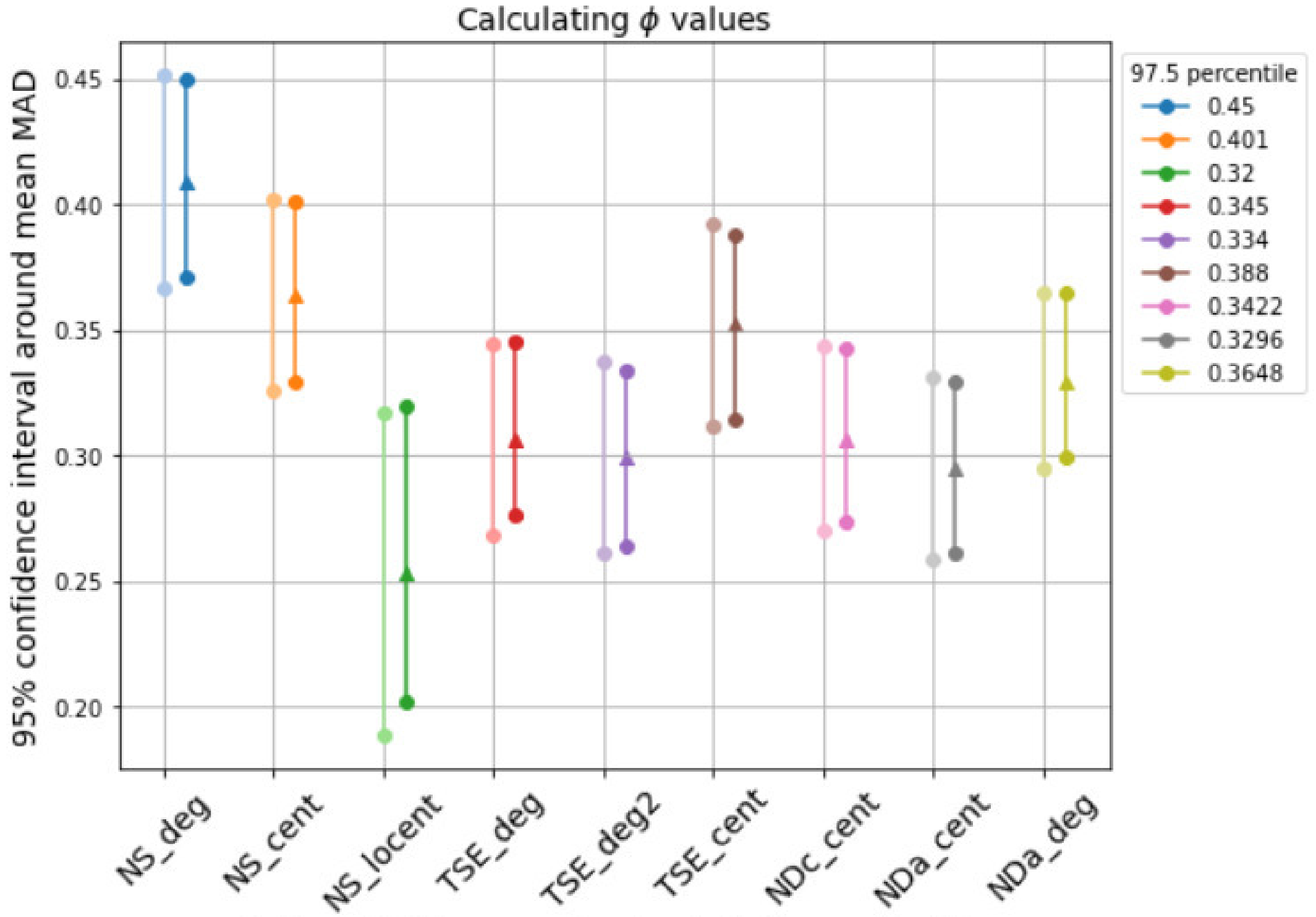

**Figure 4** Range estimates (95% CI) for the average mean absolute difference (MAD) between phi-values from experiments and phi-values calculated with normalized node statistics from native-state (NS) PRNs, transition-state ensemble (TSE) PRNs, and PRNs generated by the best performing ND variants 'a' and 'c', for the 10 proteins in the Paci dataset. The numbers in the legend give the 97.5 percentile estimate for average MAD; it is desirable that this number be small for a method. For TSE_deg2, normalization is by native-state node degree. The two CIs per method are obtained with CLT and bootstrap methods respectively; a triangle marks the average MAD on the bootstrap CI.

### *3.3.2 Results*

For ease of exposition, all mention of node statistics assume aforementioned normalization. For the TSE PRNs, generating calculated phi-values with node degree produces better correspondence (significantly smaller average MAD, p-value = 0.0235) with experimental phi-values than generating calculated phi-values with node centrality. We regard the TSE_deg2 results (the purple coloured 95% confidence intervals in Figure 4) as the benchmark for the other methods.

For NS PRNs, generating calculated phi-values is better with local node centrality (NS_locent) than with (global) node centrality (NS_cent). NS_locent produces a significantly smaller average MAD (p-value = 0.0054) than NS_cent.

NS_locent produces a significantly smaller average MAD than TSE_deg at $\alpha = 0.1$ (p-value 0.0924). In the worst case, NS_locent outperforms TSE_deg2. Worst case is defined as the largest average MAD score within the 95% confidence interval (the 97.5 percentile estimate). However, NS_locent has the largest margin-or-error of the methods; although it can produce some of the smallest MAD scores, it can also produce the worst, e.g. 1LMB4 (Fig. F2 top-left Supp. Info.).

For ND PRNs, MAD scores are calculated at each *Q*, and the minimum MAD score attained, together with min*Q* (the *Q* at which the MAD score is minimum) per protein are identified (Figs. F4 and F5 Supp. Info.). ND variants are evaluated by their average (over the proteins in a dataset) minimum MAD score, and their average min*Q* (Fig. F3 Supp. Info.). Since the experimental phi-values is a property of transition-state structures, it is preferable for the *Q* at which correspondence between experimental and calculated phi-values is strongest (min *Q*), to stay within the ND-TS region.

Generating calculated phi-values from ND PRNs is better with node centrality than with node degree. The phi-values calculated with node centrality from ND variant 'a' PRNs (NDa_cent) is competitive with those calculated with node degree from TSE PRNs. There is no significant difference between the average minimum MAD produced by NDa_cent and the average MAD produced by TSE_deg2.

On average, the biased ND variants ('c', 'x', 'a') perform equally well on generating calculated phi-values with node centrality. There is no significant difference between their average minimum MADs, and their average min*Q*s stay within the ND-TS region even in the worst case (Fig. F3 left Supp. Info.). The 'c' ND variant has a slight advantage over the other two in the sense that its min*Q*s are closer to its peak *Q*s (smallest MAD at 0.14 in Table 8). It is desirable for these two *Q*s to be closer to each other as they both indicate transition-state placement from two levels of observation: microscopic and macroscopic.

**Table 8** *Q* where MAD scores between experimental and calculated phi-values are minimum, and MAD between these *Q* values and peak *Q*s, averaged over the 10 proteins in the Paci dataset.

| ND | Node centrality | | Node degree | |
|---|---|---|---|---|
| | Min *Q* (mean, std. dev.) | MAD from peak *Q* | Min *Q* | MAD from peak *Q* |
| 'r' | 0.290, 0.375 | 0.22 | 0.370, 0.350 | 0.26 |
| 'x' | 0.290, 0.284 | 0.20 | 0.510, 0.363 | 0.40 |
| 'c' | 0.260, 0.184 | 0.14 | 0.420, 0.326 | 0.30 |
| 'a' | 0.290, 0.296 | 0.20 | 0.220, 0.257 | 0.21 |
| 'y' | 0.430, 0.435 | 0.34 | 0.270, 0.306 | 0.20 |

When generating calculated phi-values with node degree, ND variant 'a' yields the smallest worst case MAD (0.3653), and it is the only biased ND variant worst case min*Q* (0.39) lies within the ND_TS region (Fig. F3 right Supp. Info.). Its min*Q*s are also closest to its peak *Q*s (MAD of 0.21 in Table 8).

To summarize, both ND variants 'c' and 'a' can produce node centrality statistics suited for calculating phi-values. This result justifies the use of a node selection policy in ND. However, ND variant 'a' is more flexible in that it can generate calculated phi-values which meet our two above stated criteria with either node centrality or node degree. This finding supports the use of *Crony* node selection in ND. The three best worst case results (97.5 percentile estimate of average MAD scores) are found with NS_locent (0.32), NDa_cent (0.33) and TSE_deg2 (0.334).

**3.4 Contact-Order and Route-Order of ND contact probability maps**

Network clustering of native shortcuts is used as proxy for initial substructure formation in section 3.2. A candidate initial fold substructure is selected as the initial fold step if it belongs to a combination with the largest $C_{SCN0}$ (Fig. E1 Supp. Info.). A closer look at the ND networks is warranted to grasp secondary structure element content, and the overall pattern of contact formation made by the ND variants.

Fig. G1 (Supp. Info) shows the contact probability maps of four model proteins at *Q*=0.5, and their respective native-state (*Q*=1.0) contact maps. A *contact probability map* for *Q*=*q* is a weighted adjacency matrix where each element (*i*, *j*) is the probability of contact (*i*, *j*) occurring at *Q*=*q*. Probability of a contact at *Q*=*q* is the frequency with which the contact occurs in the collection of ND networks at *Q*=*q*.

A visual inspection of Fig. G1 (Supp. Info.) reveals the following:

(i) The 'y' and 'r' contact probability maps are more similar to native-state contact maps, than the contact probability maps of the biased ND variants ('x', 'c' or 'a'), in terms of the contacts made (number of non-white cells). This closer resemblance explains why the Contact-Order of both 'y' and 'r' networks are more strongly correlated with experimental folding rate within the ND-TS region than the Contact-Order of the networks generated by the biased ND variants (Figure 5). A similar observation is made with the Uzunoglu dataset (Fig. G3 Supp. Info.). Of the biased ND variants, 'a' records stronger correlation between Contact-Order and experimental folding rate at most $Q$ in both datasets.

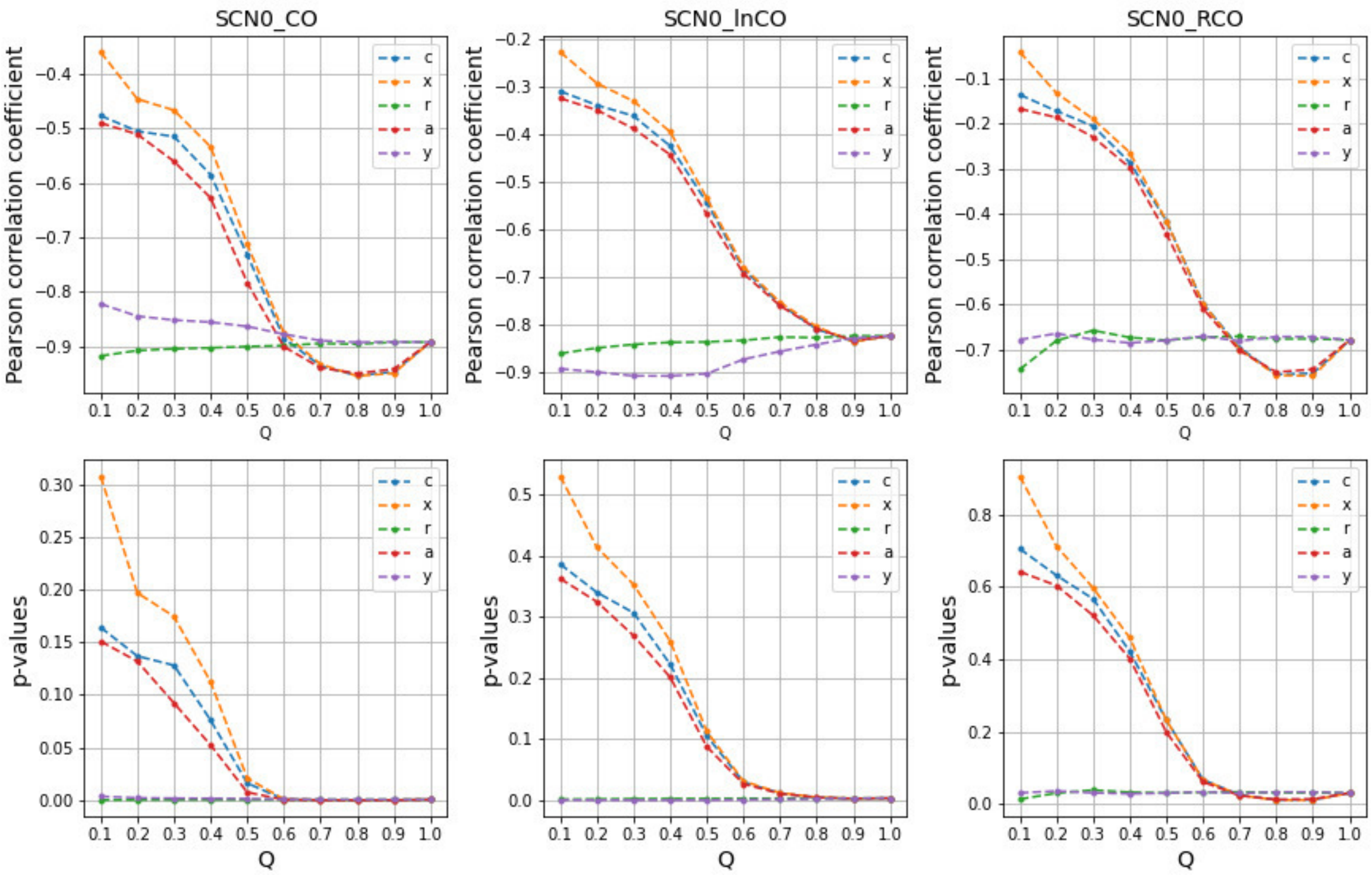

**Figure 5** Pearson correlation (top: coefficient, bottom: p-value) between experimental folding rate and Contact-Order of native shortcuts, as a function of $Q$, for the Paci dataset. For a set of contacts, CO is the average of their sequence distances, lnCO is the average of the natural logarithm of their sequence distances, and RCO (Relative Contact-Order) is CO divided by the number of unique nodes in the set of contacts. Note difference in scale on the y-axes.

(ii) Contacts closer to the main diagonal (shorter-range contacts) have higher probability; this is more evident in the 'x', 'c' and 'a' contact probability maps, and applies to both PRN edges and native shortcuts. The 'r' contact probability maps show evidence of equally developed helical and strand secondary structure elements. In contrast, helical secondary structure elements are better developed than the strand secondary structure elements in the 'x', 'c' and 'a' contact probability maps.

(iii) The biased ND variants are more selective in contact formation than the unbiased ND variant. Put in another way, the contact probabilities produced by the 'y' and the biased ND variants are more heterogeneous (have larger variance), while the contact probabilities produced with 'r' are more homogeneous (have smaller variance) (Fig. G4 Supp. Info.).

Point (iii) suggests that because some contacts have high probability, networks generated by the biased ND variants at $Q=q$ are more similar to each other, and therefore their trajectories are more route-like, i.e. confined to a smaller set of possibilities. Such a qualitative analysis of contact probability maps has been summarized quantitatively as Route-Order (RO) [41, 42].

RO at $Q=q$ is the variance of contact probabilities at $q$ for the *full* set of contacts, divided by the maximum variance of contact probabilities at $q$ (Figure 6). The variance is maximum when $Q=q$ of the contacts have probability 1.0 and $(1 - q)$ of the contacts have probability 0.0 (variance of a Bernoulli random variable with probability $Q=q$). Larger (smaller) RO values indicate fewer (more) routes to the native-state, and is associated with a narrower (wider) opening of the funnel leading to the native structure [41].

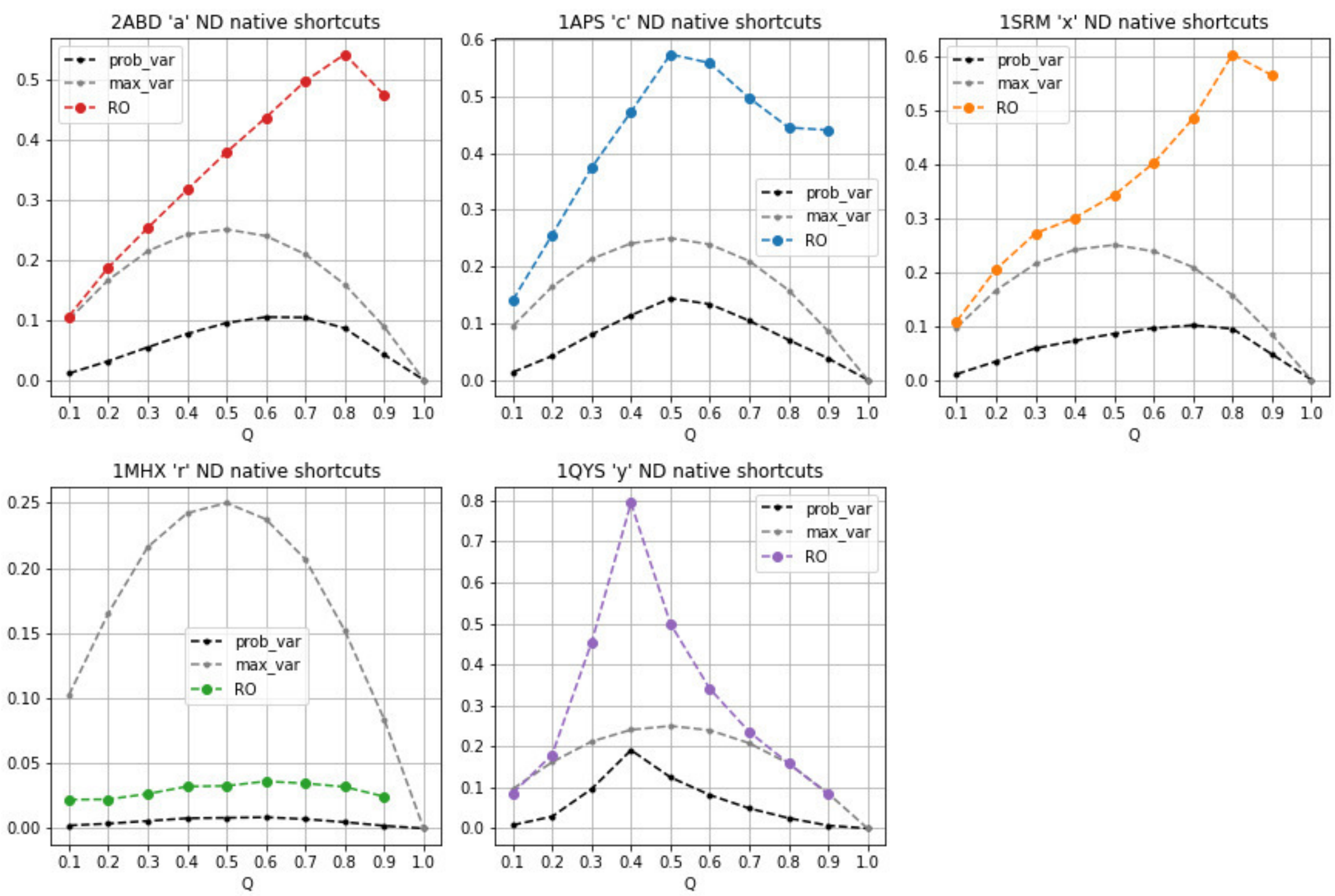

**Figure 6** Route-Order (RO) of native shortcuts generated by the ND variants for five proteins, as a function of $Q$. RO = prob_var/max_var (see text for details), is a real in [0.0, 1.0], and is undefined (0/0) at $Q$ = 1.0.

The *width* of the funnel opening is an important consideration in folding landscape theory [28]: too narrow and search for the native structure becomes needle-in-a-haystack-like; too wide and many intermediate conformations exist to act as stepping stones to the native structure. For folding to be two-state, the width of the funnel opening needs to be of the right size such that the conformation distribution is highly bimodal, and cooperative folding becomes a key narrative for passing between the unfolded and folded states.

Fig. G5 (Supp. Info.) plots RO of native shortcuts as a function of $Q$ for 14 small single-domain proteins. The biased ND variants have similarly larger RO values than the unbiased ND variant. The 'r' RO values hover close to 0.0, while RO values for 'x', 'c' and 'a' are clearly above 0.0 but well below 1.0 in the ND-TS region; they range between 0.2 and 0.6 at $Q = 0.5$.

This result confirms that the biased ND variants produce more route-like trajectories than the unbiased ND variant, and corroborates an earlier finding on diversity of simulated folding rates (section 3.1). More route-like trajectories imply folding funnels with narrower openings, which suggests that folding (ordering of PRN0 edges in ND) with the biased ND variants is more two-state like. Since cooperativity is a key to solve Levinthal's paradox for two-state folders, it is fitting that 'x', 'c' and 'a' report higher diversity in simulated (wider peak $\boldsymbol{E}$ range) folding rates.

However, this line of reasoning can break down when RO becomes "too high", as the 'y' ND variant demonstrates. The 'y' ND variant yields RO values which peak between 0.6 and 0.8 within $0.4 \leq Q \leq 0.6$, for the 14 single-domain proteins. However, the 'y' ND variant produces a much narrower peak $\boldsymbol{E}$ range, and peak $\boldsymbol{E}$ values that do not correlate significantly with experimental folding rates (Tables 2 and 3).

## 4 Conclusion

A network-based protein folding model called Network Dynamics (ND) is proposed and its behavior is examined at three levels of observation: macro, meso and micro. ND views protein folding as a problem of ordering the contacts in native-state protein residue networks (PRN0). Different orderings give rise to the formation of different shortcut network (SCN) configurations. SCNs are discovered from ND generated PRNs by the EDS algorithm.

We found that ND yields the most meaningful protein folding results when it prefers the creation of shorter over longer range contacts, and when nodes are preferentially selected by remaining node degree from a local neighborhood (*Crony* set) to form contacts. These two network formation policies exploit knowledge about the two major factors of protein folding: slow loss in chain entropy, and priority of hydrophobic and other key folding interactions.

Summary of results with the best ND variant ('a'):

(i) A significant correlation of -0.72 between simulated folding rate (peak ND energy) and experimental folding rate from 52 two-state proteins.

(ii) Greater diversity in simulated folding rate, which signals cooperative folding essential for passing between unfolded and folded states in two-state folding.

(iii) More heterogeneous contact probabilities and more route-like trajectories, which imply narrower funnel openings (but not too narrow).

(iv) Better correspondence between phi-values from protein engineering experiments and phi-values calculated from ND node statistics. Correspondence refers to agreement in relative importance. This result suggests the potential utility of ND networks in the hunt for critical contacts that define TSE structures. However, phi-values can also be calculated from PRN0s using local node centrality, a measure that follows from observing the high substructure centrality of selected initial fold substructures.

## 5 Future explorations

The work reported so far has ignored fold type of protein chains, i.e. primary secondary structure content be it alpha-helices (α), beta-sheets (β) or a substantial mix of both (αβ). When the macro-level results for the Uzunoglu dataset are re-examined by chain fold type, we find that the strength of the correlation between peak ***E*** and experimental folding rate not only depends on ND variant, but also on chain fold type. ND variant 'a' produced the strongest correlation for αβ chains, while ND variant 'x' is best suited for β chains (Fig. H1 top Supp. Info.). Surprisingly, the strongest correlation for α chains is produced with ND variant 'y'. This last observation prompted us to experiment with other edge ordering strategies that vary the probability of edge formation ($p_e$) by secondary structure type. We call such ND variants *mixed*, since they do not implement a global $p_e$, but a context-sensitive one.

By exploring several mixed ND variants, and combining the best results by chain fold type, we were able to improve overall peak ***E*** correlation with experimental folding rate for the Uzunoglu dataset to -0.78, and have all correlations by chain fold type be significant (p-value < 0.005) (Figure 7 top). The combination of ND variants that produced this result is m4-a-m4: peak ***E***s produced by the 'm4' mixed ND variant for α chains and for β chains, are combined with peak ***E***s produced by the 'a' ND variant for αβ chains. In the 'm4' mixed ND variant, the decision to apply scaling to $p_e$ depends on the 1-hop secondary structure neighborhood (SSneighborhood) of a node. If the SSneighborhood of a node does not include residues of strand type, all edges incident on the node is formed with no scaling to $p_{e,}$ which means long-range edges become easier to form.

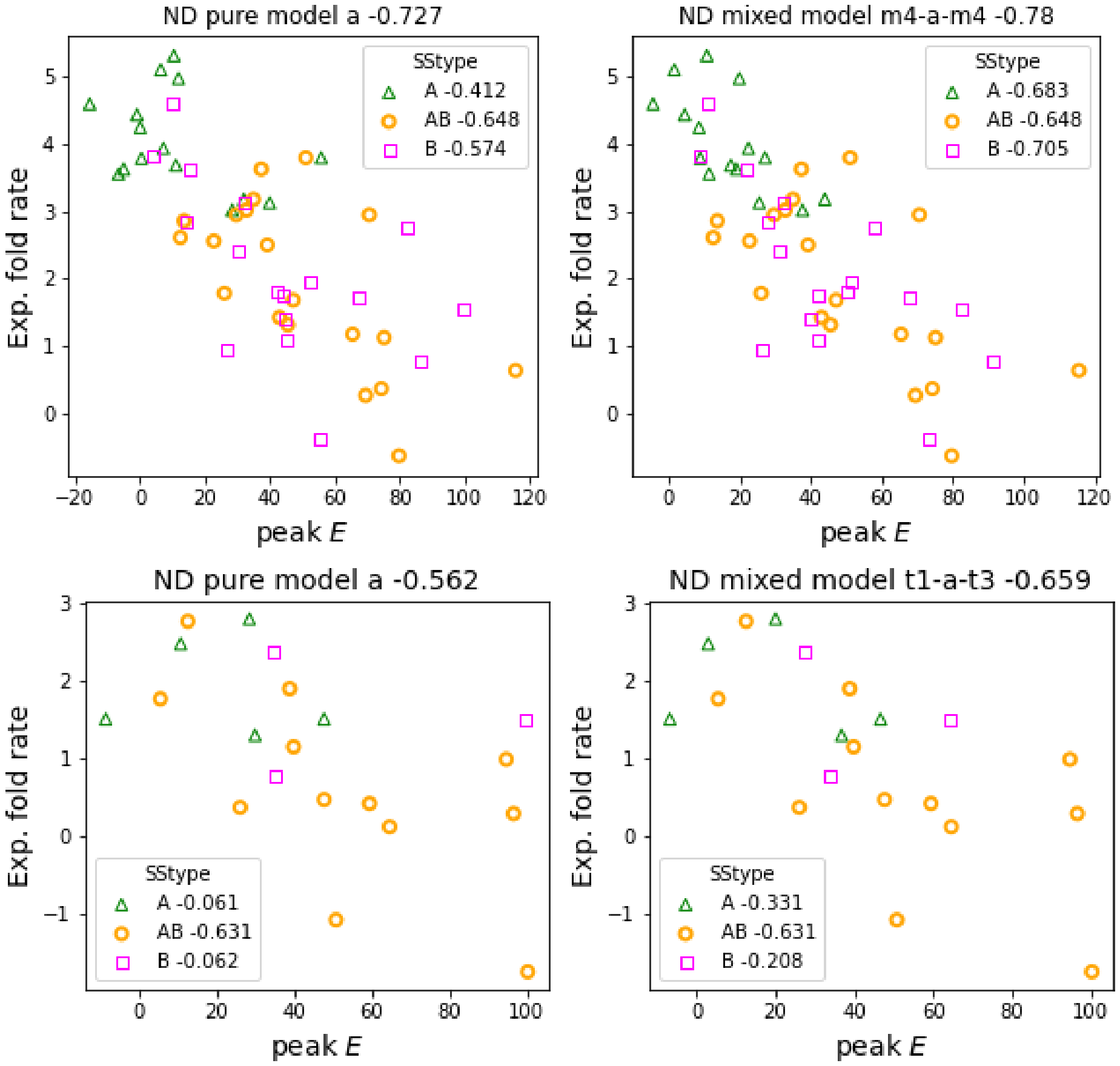

**Figure 7** Experimental folding rate vs. simulated folding rate (peak *E*) scatter plots by chain fold type (SStype) for the Uzunolgo dataset (top) and the Kamagata dataset (bottom). Results from a best pure ND model found so far (left) are compared with results from a best mixed ND model found so far (right). Overall correlations are given in the subplot titles; correlations specific to chain fold types are displayed in the legends.

We applied the above approach to the Kamagata dataset. Again, ND variant 'a' produced the strongest correlation for αβ chains, and ND variant 'y' produced the strongest correlation for α chains (Fig. H1 bottom Supp. Info.). The correlations are poor for both α and β chains, in part due to the composition of the Kamagata dataset which is dominated by αβ chains. The overall correlation for the Kamagata dataset is strengthened to -0.659 (p-value=0.0016) with the t1-a-t3 mixed ND variant, which combines 't1' peak ***E***s for α chains, 'a' peak ***E***s for αβ chains, and 't3' peak $\boldsymbol{E}$s for β chains. Both 't1' and 't3' are mixed ND variants; they add a small value (0.05 and 0.15 respectively) to the scaled $p_e$ when forming PRN0 edges that link residues not of type helix or strand. The improved correlation is better than that obtained with scaled protein chain length $N^{0.5}$ (-0.572, p-value=0.0084); this gives us hope that the ND model could be applicable also to multi-state folders.

The results reported in this section hint at strategies for ordering PRN0 edges to optimize peak $\boldsymbol{E}$ correlation with experimental folding rate, and gain mechanistic insight into the folding process. Applying different contact formation strategies could improve results at other levels too, for example the correspondence between calculated and experimental phi-values. One possible research direction is to use machine learning techniques such as graph neural networks that can learn from simulation data and take on the challenge of fine-tuning ND. A complex-network based model like ND could provide a toehold into this research frontier.

## Supplementary Information

**A. The EDS algorithm** [1].

To identify the shortcut edges in a PRN, the EDS algorithm constructs EDS paths between all pairs of distinct nodes ($i$, $j$) in the PRN. At each step of a path construction, EDS surveys the Cα-Cα Euclidean distance between the target node and each of the current node's direct neighbors in a PRN, and moves to a node $x$ not yet on the path, which is closest amongst all nodes surveyed so far to the target node.

Distance information gathered by nodes along a path are cumulative, and passed to nodes further along a path. So it is possible for $x$ to not be a direct neighbor of the current node (it is a direct neighbor of some earlier node on the path). In this case, EDS retraces the path (*backtrack*) until $x$ becomes reachable. In certain situations, this backtracking is shortened by edges identified as shortcuts.

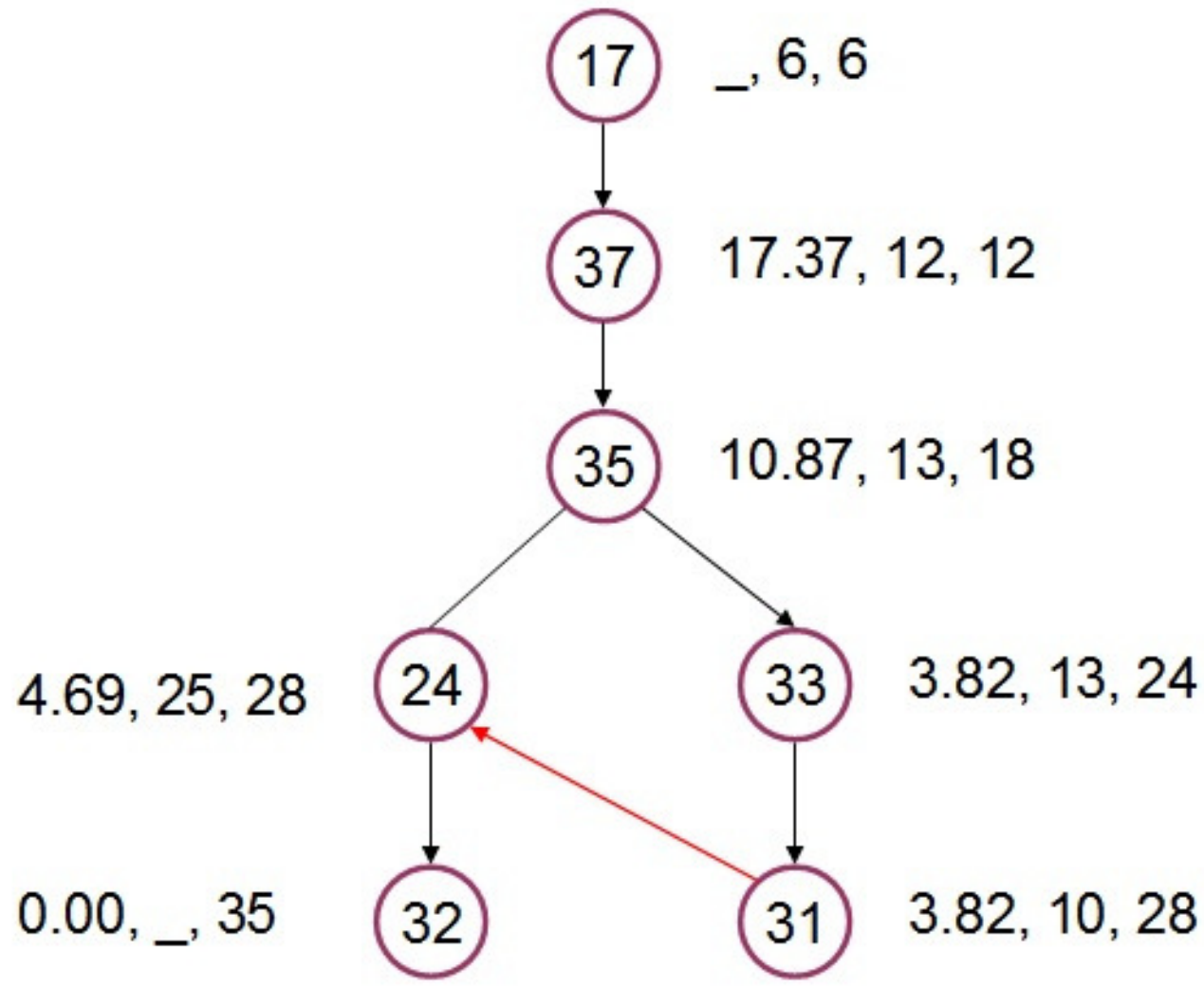

**Fig. A1 Top:** Illustration of the EDS path <17, 37, 35, 33, 31, 24, 32> for 2PTL between residues 17 and 32. This EDS path identifies (31, 24) as a shortcut, since its presence avoids the backtrack from 31 to 24.
The three numbers beside each node denotes the node's Euclidean distance to the target node (32), the node's PRN0 degree, and the number of nodes surveyed so far during the construction of the EDS path, respectively. The last two numbers need not agree, since distance information is passed on to nodes appearing later in an EDS path. But no information is shared between EDS paths. In total, 35 nodes were surveyed to construct this EDS path.
As this EDS path involves more than one SSEs and more than two nodes, it is an example of an EDS path that can contribute to substructure centrality.

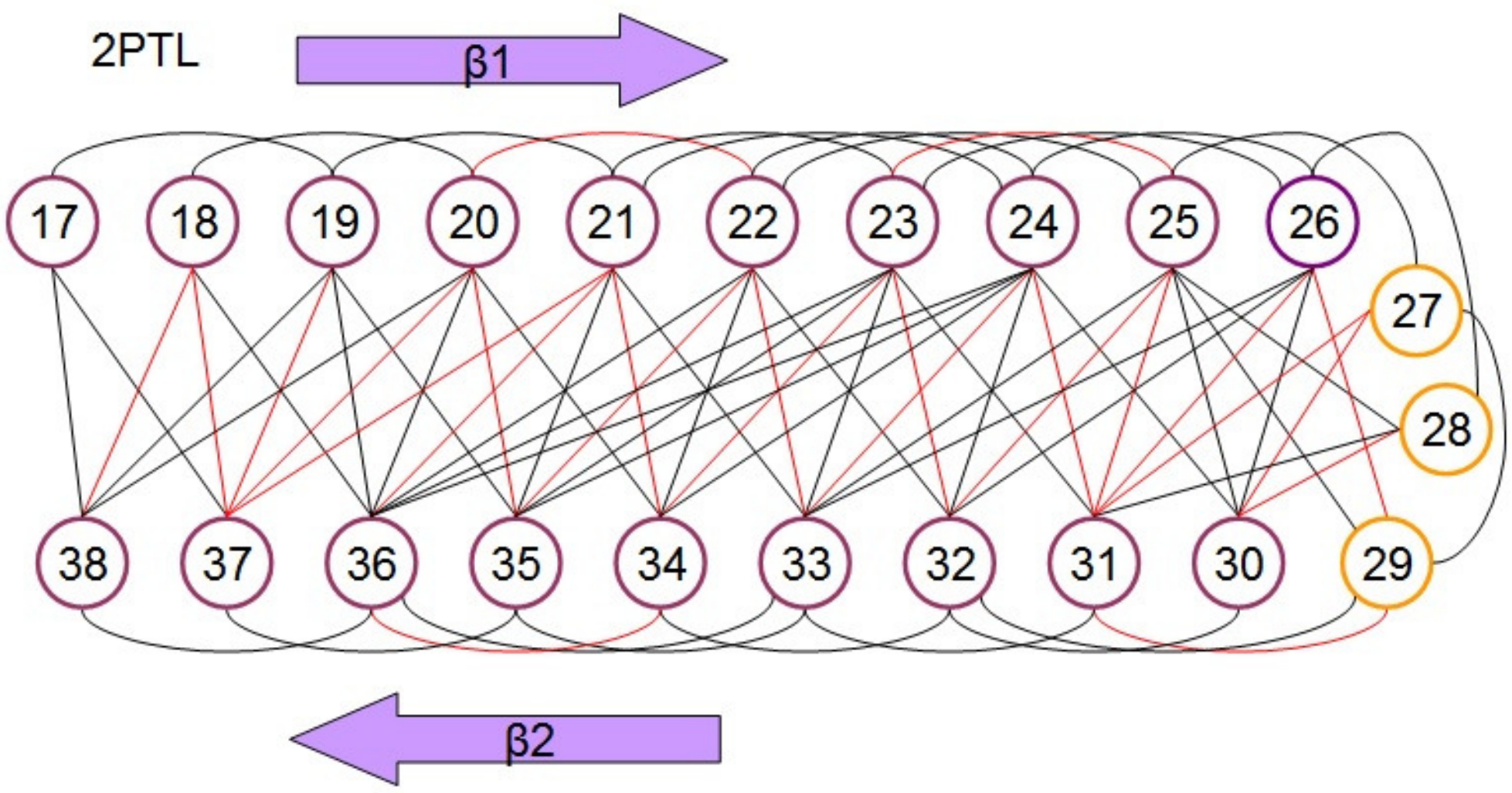

**Fig. A1 Bottom:** Residues of 2PTL's N-terminal β-hairpin and their PRN0 edges (black), and shortcut edges (red). Turn SSE residues are drawn in gold.

```python
def eds(source: int, target: int, d):
    '''d is a numpy array with distances to target node'''
    kpath = [source]; scuts = set()
    ts = np.zeros(N, dtype="int8") # node first-visit timestamps
    ws = np.zeros(N) # accumulated weights (distances to target node of visited nodes so far)
    t = 0; n = source;
    while True:
        p = P[n].copy() # P is the adjacency matrix of a PRN0
        if 1 == p[target]: # check if target is a direct neighbor of current node, edge path
            kpath.append(target)
            break
        t += 1
        np.maximum(ts, p*t, out=ts, where = ts==0)
        w = d * p
        np.maximum(ws, w, out=ws)
        # hide nodes already on the path, ignores -1
        m = np.zeros(N, dtype='bool'); m[kpath] = True
        b = np.ma.MaskedArray(ws, np.logical_or(ws == 0.0, m))
        nn = np.ma.argmin(b) # next node
        # if nn is not a direct neighbor of n, need to backtrack
        if 0 == P[n][nn]:
            kpath.append(-1) # skipping backtrack subpath
        kpath.append(nn)
        # if nn is a direct neighbor of n first visited at an earlier timestep,
        if ts[nn] < t and 1 == P[n][nn]: # (n, nn) is a shortcut
            scuts.add((n, nn))
        n = nn;
    return kpath, scuts
```

**Fig. A2** Example of Python3 code that finds shortcuts (if any) and the EDS path (with backtrack portion represented by -1) between a node-pair, given the PRN0 adjacency matrix $P$, and Euclidean distances $d$ of all nodes to the target node.

It is not necessary to use EDS to check for shortcuts after each edge creation by ND. EDS over all distinct node-pairs is at least an $(N^2 - N)$ operation, and average length of EDS paths scales as $ln\ N$ [1]. EDS needs to be run only when ND creates an edge that connects previously unconnected node-pairs, or an edge that could change an existing path. To discern these possibilities, the set of nodes surveyed when creating each EDS path is stored and updated (the average size of this set per node-pair is some constant of $ln\ N$ [1]). For as yet unconnected node-pairs, this set is empty. For already connected node-pairs, creation of new edges that involve nodes in this set indicate possible change of EDS path for a node-pair. New and changed EDS paths may modify the set of native and non-native shortcut edges.

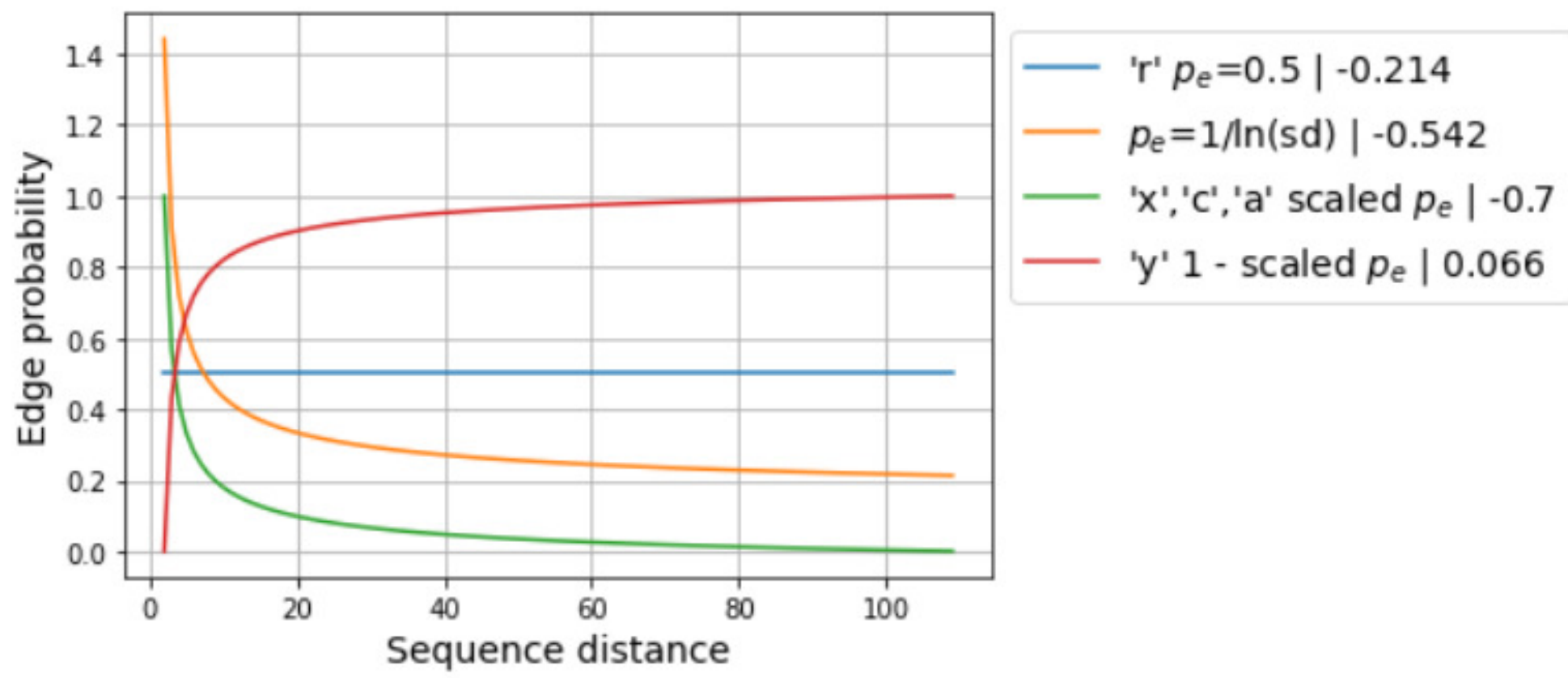

**Fig. A3** Edge formation probability ($p_e$) functions; each takes as input (sequence distance) an integer ≥ 2. The numbers after the vertical line in the legend is the Pearson correlation between experimental folding rate and peak ND energy produced by the 'x' ND model for the Uzunoglu dataset. The correlation is strongest with scaled $p_e$, which assigns the lowest probability (of the four functions) at large sequence distances.

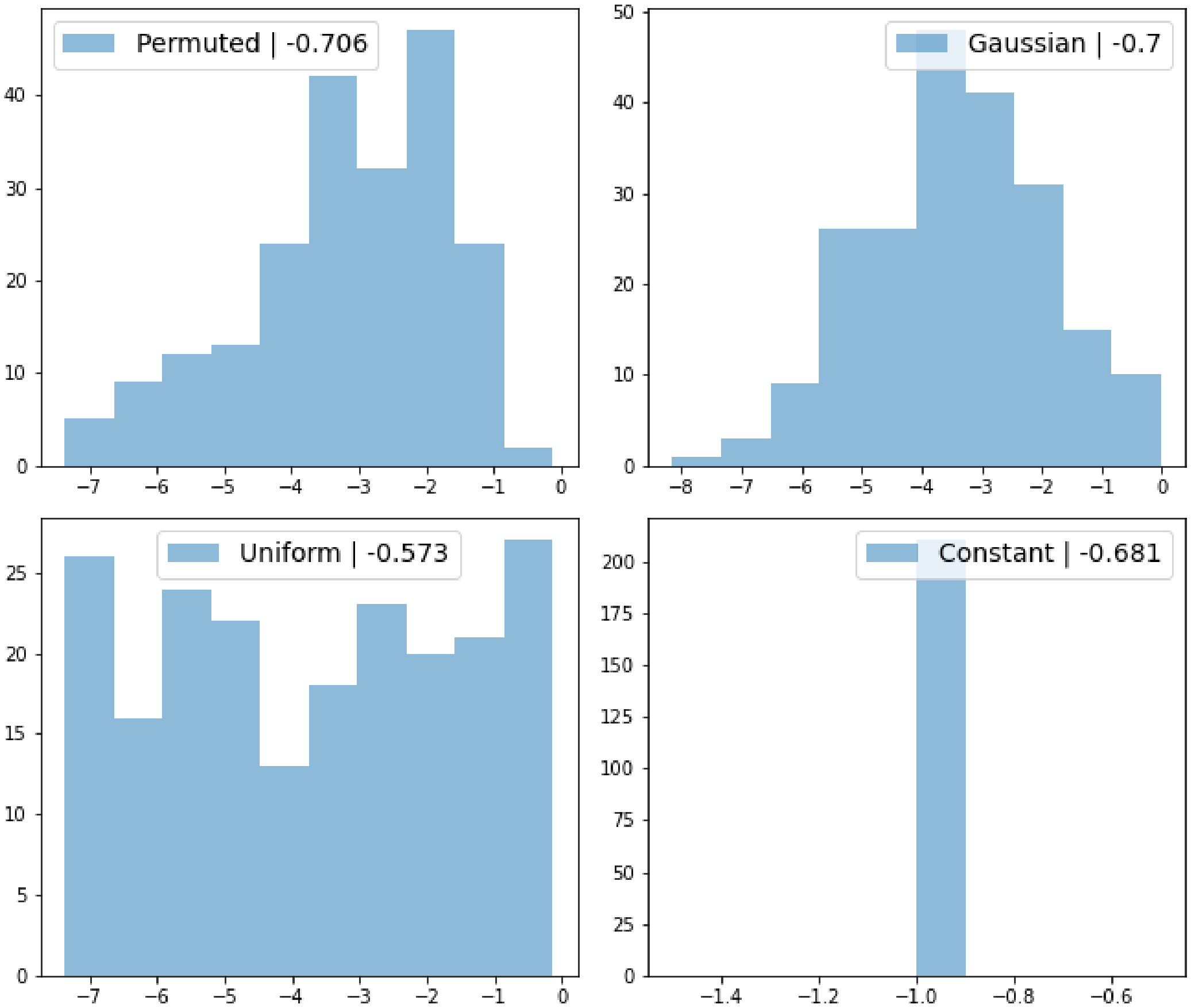

**Fig. A4** In the main text, ND energy is calculated with Miyazawa-Jernigan inter-residue potentials (MJE). It turns out this is not strictly necessary to achieve comparable correlation (-0.7) between peak ND energy and experimental folding rate. The numbers after the vertical line is the Pearson correlation obtained with the 'x' ND model for the Uzunoglu dataset, using four different inter-residue potentials:
(i) Permuted is a permutation of MJE values (maintains the original MJE distribution, which is not normal);
(ii) Gaussian is random values generated from a normal distribution with the mean and standard deviation of the original MJE distribution;
(iii) Uniform is random values generated from a uniform distribution within the range of the original MJE distribution;
and (iv) Constant assigns all amino-acid pairs the value of negative one.
The correlation strength deteriorates slightly with Constant, and is much weaker with Uniform. However, both Permuted and Gaussian produce correlations as strong as when the original MJE is used to calculate ND energy. Gaussian distributed energy interaction values has been successfully employed in theoretical protein folding models [22, 23].

1. Build the native protein residue network (PRN0) from PDB coordinates. Get the native shortcut network (SCN0) of PRN0 with EDS.

2. Build the PRN for the TS structure from its 3D coordinates, but only keep edges that are in PRN0. A TS structure has the same amino acid residues as the native structure, but at different 3D coordinates. This may result in a TS PRN having edges not present in the PRN0; such (non-native) edges are excluded as it is not possible for ND to generate them. In short, a TS PRN is a sub-graph of PRN0.

3. Use EDS on TS PRN to identify shortcut edges. These shortcuts are either native (if it is in SCN0), or non-native.

**Fig. A5** Steps to convert a transition-state (TS) structure into a PRN, and find its native and non-native shortcuts.

**Table A1** Attributes of transition-state ensembles from the Paci dataset.

| PDB id | # TS | # SC0 | % SC0 | NS SC0_RCO | TSE SC0_RCO | NS SC0_C | TSE SC0_C | NS SC0_LE | TSE SC0_LE |
|---|---|---|---|---|---|---|---|---|---|
| 1imq α | 16 | 16 | 36.5 | 0.139 | 0.105 | 0.424 | 0.100 | 26 | 3.250 |
| 1lmb-4 α | 127 | 127 | 48.4 | 0.054 | 0.044 | 0.480 | 0.174 | 6 | 0.354 |
| 1bf4 αβ | 74 | 74 | 53.1 | 0.106 | 0.152 | 0.326 | 0.095 | 22 | 12.230 |
| 2ptl αβ | 126 | 126 | 35.2 | 0.115 | 0.150 | 0.345 | 0.109 | 28 | 8.405 |
| 2ci2 αβ | 184 | 184 | 27.6 | 0.278 | 0.187 | 0.266 | 0.023 | 46 | 4.804 |
| 1aps αβ | 29 | 29 | 40.1 | 0.267 | 0.254 | 0.254 | 0.073 | 79 | 26.034 |
| 1fmk β | 147 | 147 | 35.2 | 0.274 | 0.250 | 0.179 | 0.039 | 48 | 10.401 |
| 1bk2 β | 31 | 31 | 35.6 | 0.253 | 0.261 | 0.248 | 0.070 | 49 | 10.645 |
| 1shf β | 33 | 33 | 42.2 | 0.252 | 0.241 | 0.191 | 0.074 | 47 | 15.909 |
| 1ten β | 90 | 90 | 37.8 | 0.249 | 0.265 | 0.183 | 0.055 | 105 | 34.133 |
| Correlation with folding rate | | | | -0.716 | -0.777 | 0.679 | 0.576 | -0.878 | -0.805 |
| p-value | | | | 0.020 | 0.008 | 0.031 | 0.082 | 0.001 | 0.005 |
| Correlation with native-state | | | | 1.0 | 0.891 | 1.0 | 0.831 | 1.0 | 0.912 |
| p-value | | | | | 0.001 | | 0.003 | | 0 |

# TS: number of structures in a transition-state ensemble (TSE).
# SC0: number of unique sets of native only shortcuts.
NS = Native-state.
TSE = average for TSE structures.
% SC0 = percentage of native shortcuts found in a TS structure.
SC0_RCO = Relative Contact Order computed on native shortcuts only.
SC0_C = Network Clustering coefficient computed on native shortcuts only.
SC0_LE = Number of long-range (sequence separation > 10) native shortcuts.

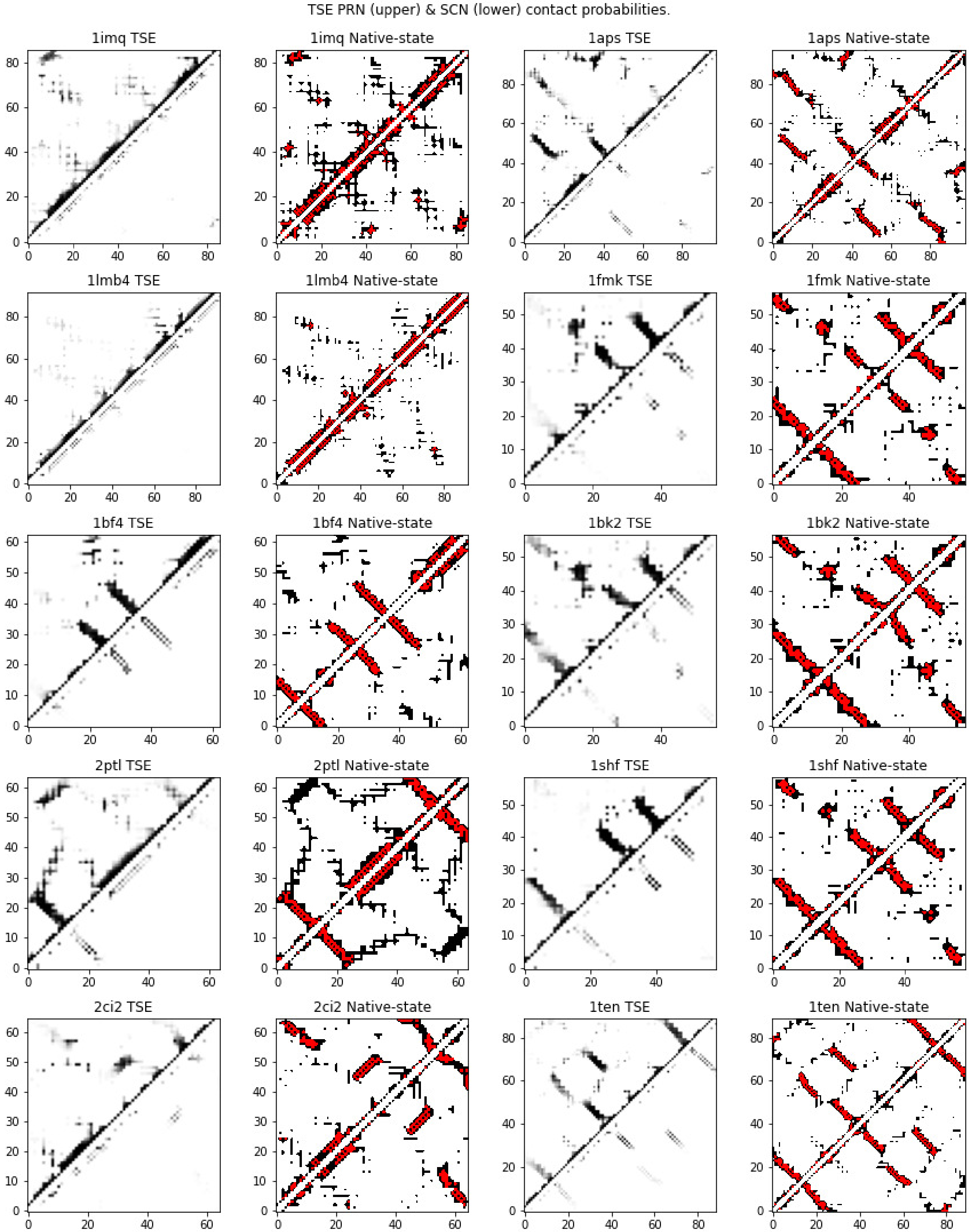

**Fig. A6** Contact probability maps of TSE PRN (upper) and TSE native shortcuts (lower); darker shades of gray denote higher probability. In the native-state contact maps, black and red cells denote PRN0 and SCN0 edges respectively.

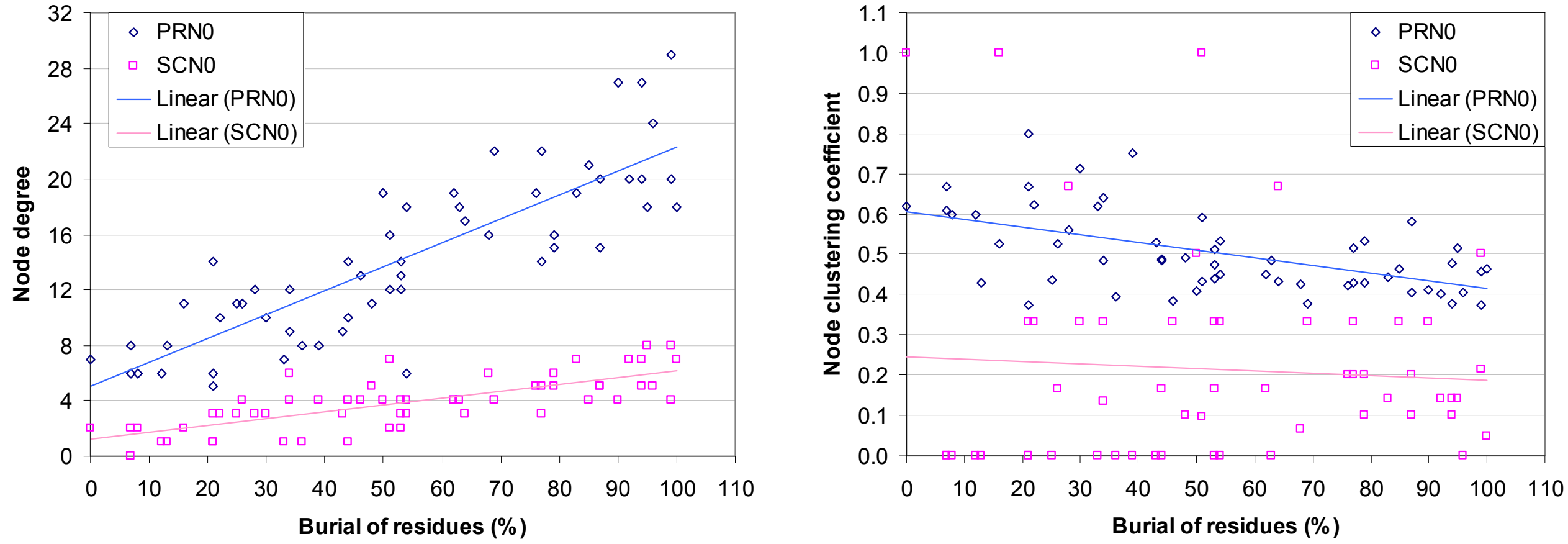

**Fig. B1** Burial of src SH3 domain residues in the native-state (Burial data from Table 1 of ref. [2]) against 1SRM's PRN0 and SCN0 node statistics. **Left**: Pearson's correlation coefficient for PRN0 node degree and Burial of residues is 0.8448 (p-val=2.22E-16), and for SCN0 node degree and Burial of residues is 0.7442 (p-val=4.92E-11). **Right**: Pearson's correlation coefficient for PRN0 node clustering and Burial of residues is -0.5500 (p-val=1.13E-05), and for SCN0 node clustering and Burial of residues is -0.0667 (p-val=0.6249).

The key folding residues are sourced from the references listed below. The conserved rigid residues from Suppl. Mat. of ref. [3] are referred to as mechanically rigid (M-R) sites. For 2IGD and 1SRM, we use the provided data for 1IGD and 1SRL, respectively. The M-R sites confer structural stability in a protein's native-state, but many of these sites also coincide with folding nuclei identified in the literature. The hydrogen-deuterium exchange (H-X) sites (Table 1 in ref. [4]) include amide protons (NHs) which are slowest to exchange out, or first to gain protection. The H-X probes do not identify nucleation sites per se, but rather the neighborhood where they might be found. Key folding sites on turns (which are crucial to trigger β-hairpin formation) are not detected. In Figs. B2 to B12, residues closer to the lower right of a plot are deemed more rigid than those closer to the upper left of a plot.

*1BDD* (Fig. B2): Residues involved in frequently formed contacts in the transition-state structures between the first and second helices (1H-3H) are: F14, L18, F31, I32 and L35; and between the second and third helices (3H-5H) are: L45, F31 and L35 [5]. The H-X sites are Y15...L18, R28, L35, K36, A49, K50 and K51. Except for L18 which is a turn residue immediately after 1H, the key folding residues congregate on the three α-helices. The 3H-5H nucleus residues are more rigid than the 1H-3H nucleus residues, and accordingly folding begins with 3H-5H pairing on the $C_{SCN0}$ folding pathway for 1BDD.

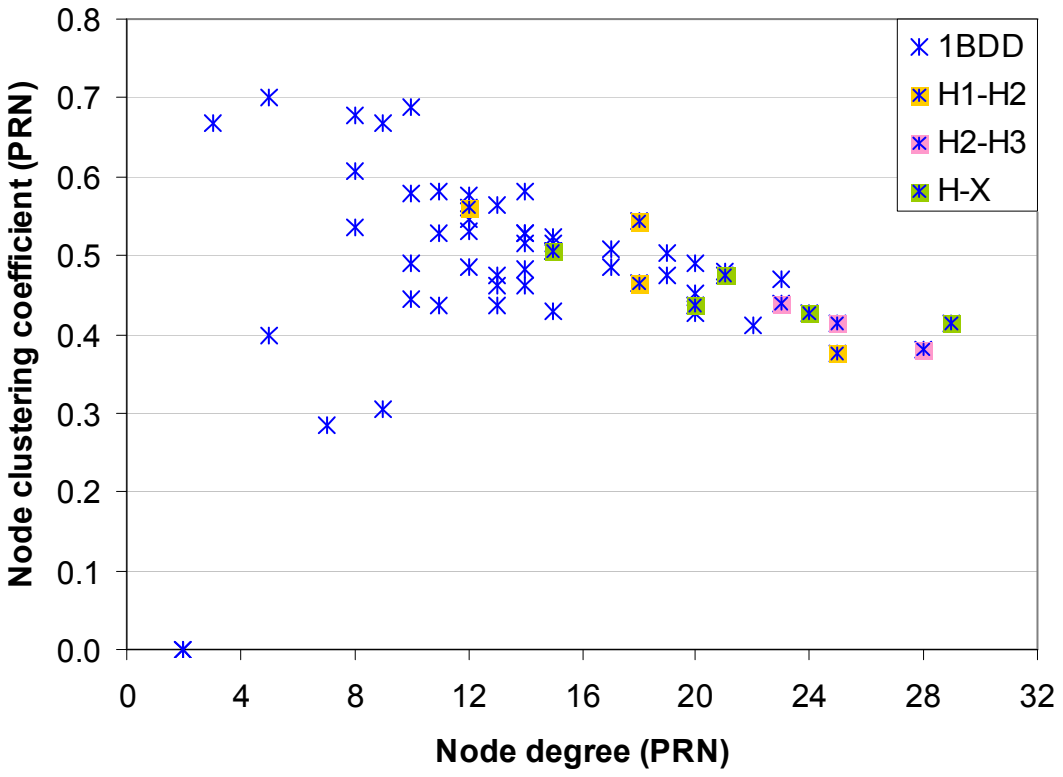

**Fig. B2** Key folding residues for 1BDD are highlighted in pink, orange and green. H-X sites which are not also H1-H2 or H2-H3 nucleation sites are highlighted.

*2IGD* (Fig. B3): The M-R sites are Y8, L10, A31, F35, F57 and V59. These sites are located on the first and forth β-strands (1S and 9S), and the α-helix (5H), and they overlap with 1GB1's key folding residues in a structural alignment (PDBe Fold v2.59). The remaining four most rigid residues belong to 1S, 5H, 6T and 7S. The rigid residue on the third β-strand (7S) is W48, which corresponds to 1GB1's W43 nucleus residue in a structural alignment (PDBe Fold v2.59).

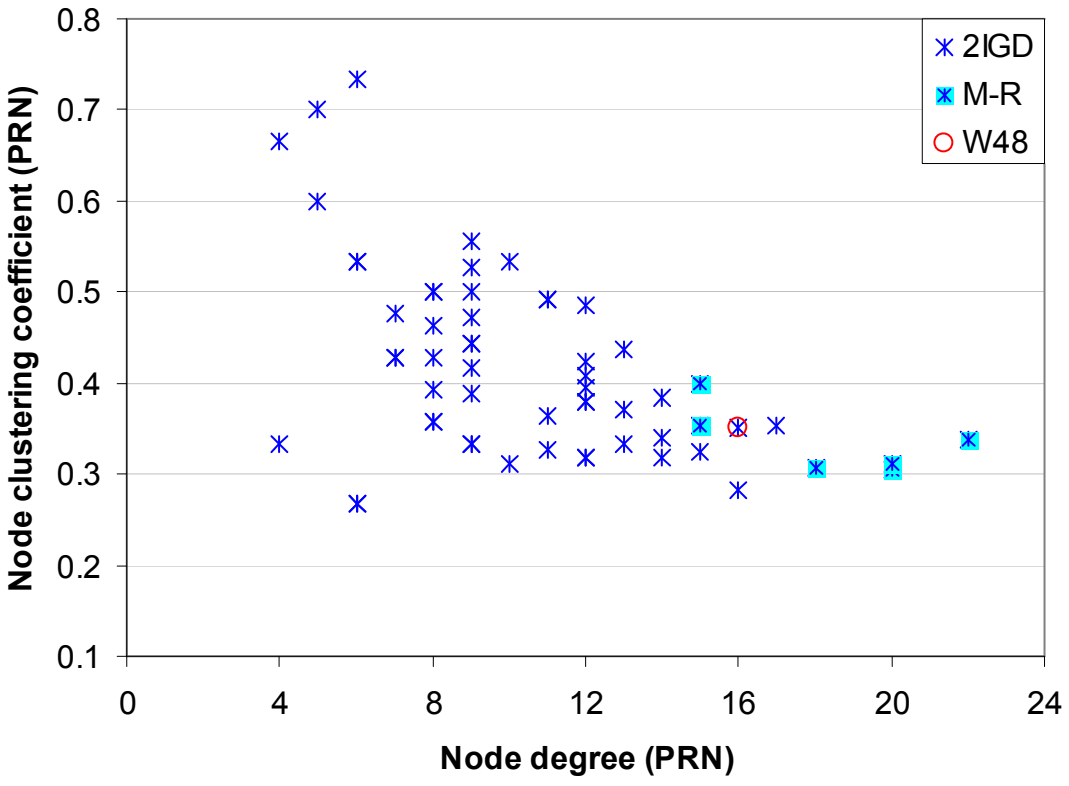

**Fig. B3** Key folding residues for 2IGD are highlighted in blue.

*1GB1* (Fig. B4): The folding nucleation sites are Y3, L5, F30, W43, Y45 and F52 [6, 7]. These six residues are evolutionarily conserved in protein G-like folds, and make frequent long-range native interactions with K4, I6, L7, A26, T51, T53 and V54 as part of the nucleation growth process in simulations (Table 1 in ref. [6]). The H-X sites are L5, I6, T25, A26, E27, F30, T44, K50, and T51...V54. Except for K50 which is a turn residue immediately before 8S, the key folding residues congregate on the first, third and forth β-strands (0S, 6S and 8S), and the α-helix (4H).

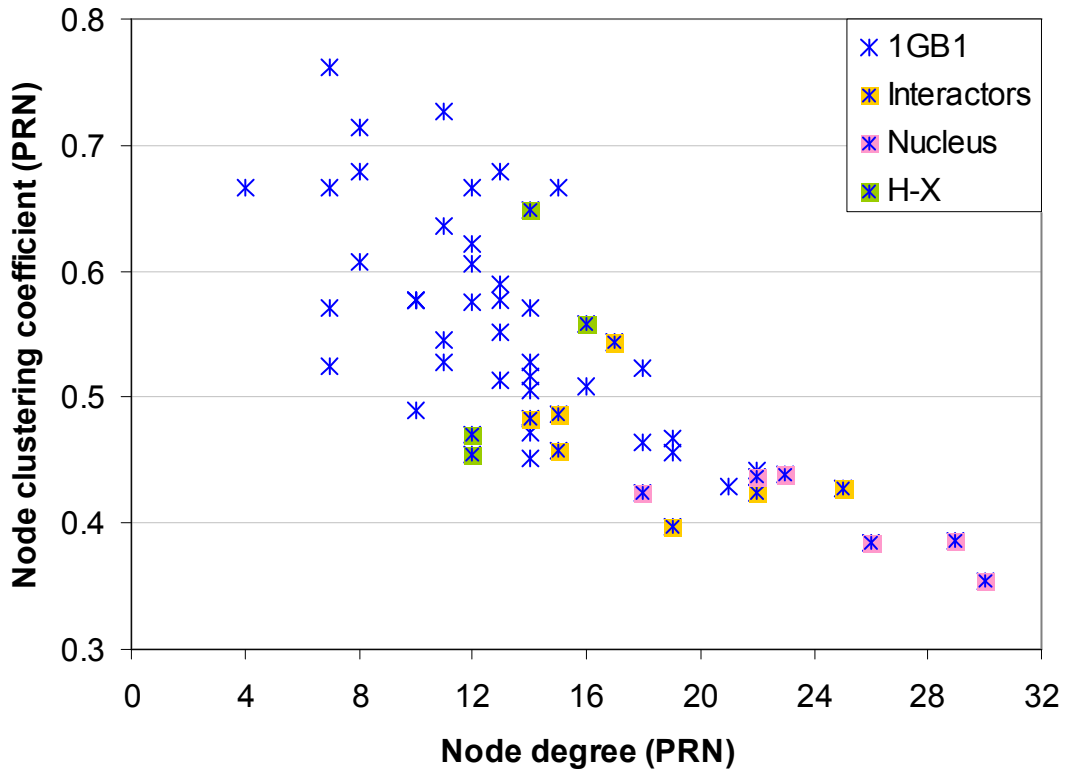

**Fig. B4** Key folding residues for 1GB1 are highlighted in pink and green. Interactor residues (yellow) are those that partner with a Nucleus residue in Table 1 of ref [6]. H-X sites which do not also fall in the other two categories are highlighted.

*2PTL* (Fig. B5): The H-X sites are I20, A22, L24, I25, F36, S45, A47...D52, L72, I74, and K75. The key folding residues congregate on the first, second and forth β-strands (0S, 2S and 8S), and the α-helix (4H). For 2IGD, 1GB1 and 2PTL, key folding residues are absent on one of the β-strands of the β-hairpin that forms later, i.e. 2S for 2IGD and 1GB1, and 6S for 2PTL. This observation is in agreement with the initial folding step for these proteins.

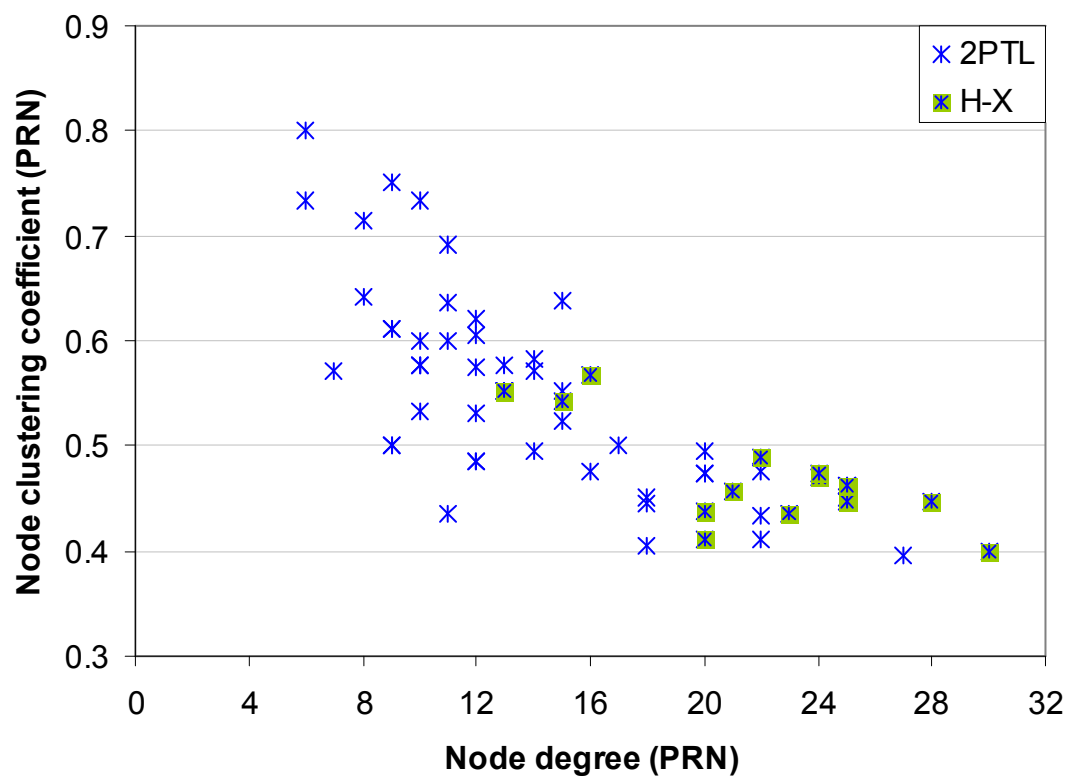

**Fig. B5** Key folding residues for 2PTL are highlighted in green.

*2CI2* (Fig. B6): The folding nucleation sites for 2CI2 are A35, I39, V66, L68 and I76 [8, 9]. Four of these sites (A35, I39, V66, L68), together with R67 and P80 are M-R sites. The H-X sites are K30, I39, L40, I49, L51, V66, L68, F69 and V70. Except for K30, I76, and P80, the key residues congregate on the α-helix (3H), and the second and third β-strands (5S and 7S). K30 is a turn residue immediately before the helix (3H), I76 is a residue in the turn (8T) between the third and forth β-strands (7S and 9S), and P80 is a residue in the forth β-strand (9S). Key folding residues are absent from the first β-strand (1S), which only gets involved at the later stages according to our $C_{\mathrm{SCN0}}$ folding pathway for 2CI2.

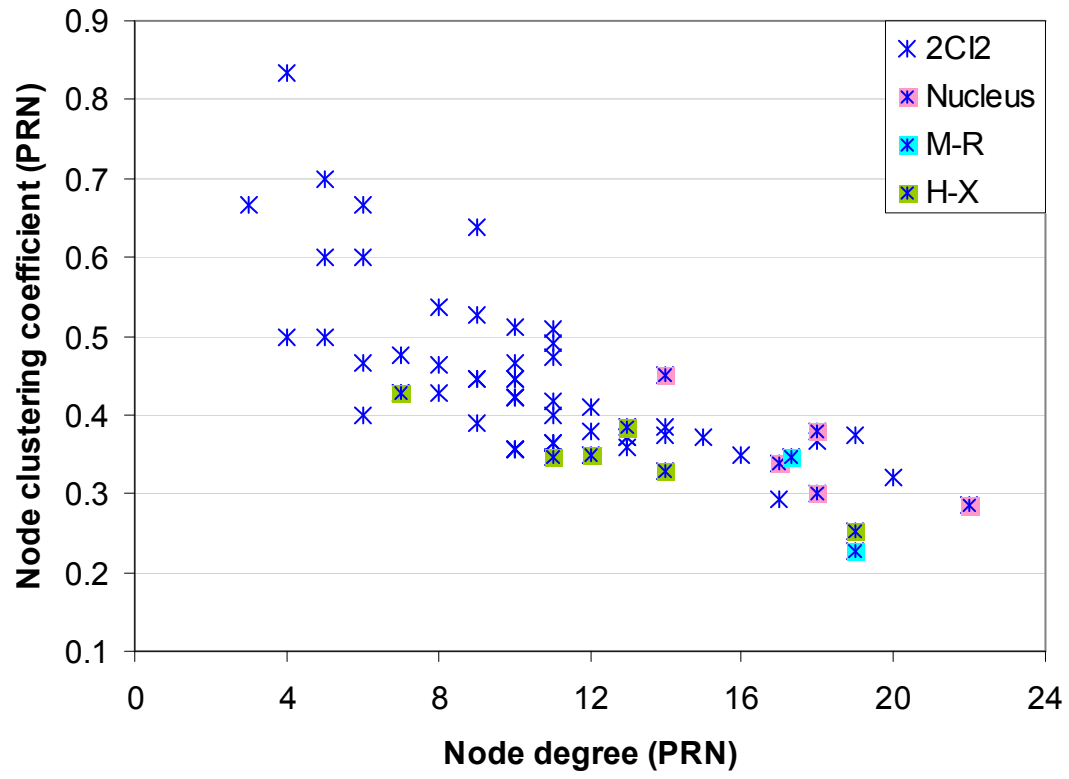

**Fig. B6** Key folding residues for 2CI2 are highlighted in pink, blue and green. M-R sites which are not also Nucleus residues are highlighted. H-X sites which are not also Nucleus or M-R residues are highlighted.

*1SHG* (Fig. B7): The M-R sites are M25, V44 and V53. These three M-R sites, together with V9, A11, V23, L31, L33 and V58, form the hydrophobic core [10]. The M-R sites are found on the third and forth β-strands (5S, 7S) and towards the end of the RT-loop. 5S and 7S are the earliest folding elements in our $C_{\mathrm{SCN0}}$ folding pathway for 1SHG (Table E1). The two other most rigid residues in Fig. B7 are W42 which belongs to 5S, and L31 which belongs to 3S. L31 occupies a protein sequence position that is rigid in all the other proteins studied in the SH-3 domain family [3]. The hydrophobic core sites touch all the β-strands.

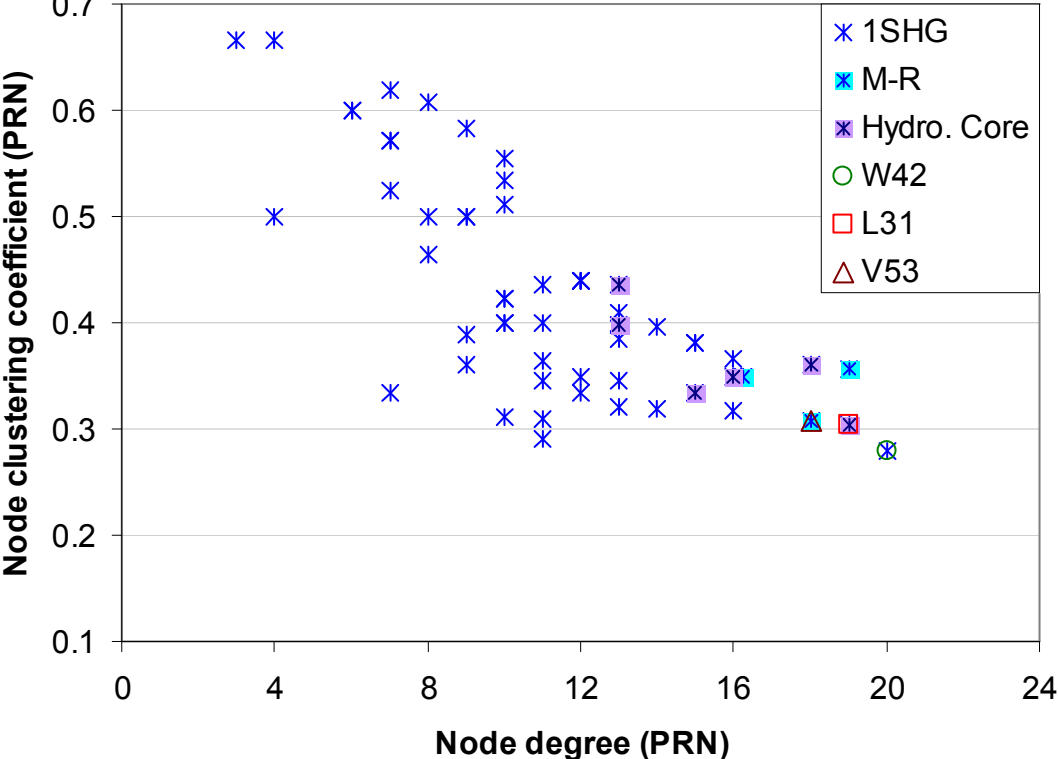

**Fig. B7** Key folding residues for 1SHG are highlighted in blue. Hydrophobic core residues which are not also M-R sites are highlighted in purple.

*1SRM* (Fig. B8): The M-R sites are F26, L32, A45 and I56. These four M-R sites, together with F10, A12, L24, I34, W43 and V61 form the hydrophobic core [2, 11]. The hydrophobic core sites touch all the β-strands, and the diverging turn (26...32). I56 is the most rigid residue and it plays a central role in hydrophobic collapse of src SH3 domain circular permutants (the structurally aligned site in α-spectrin SH3 domain is V53) [12].

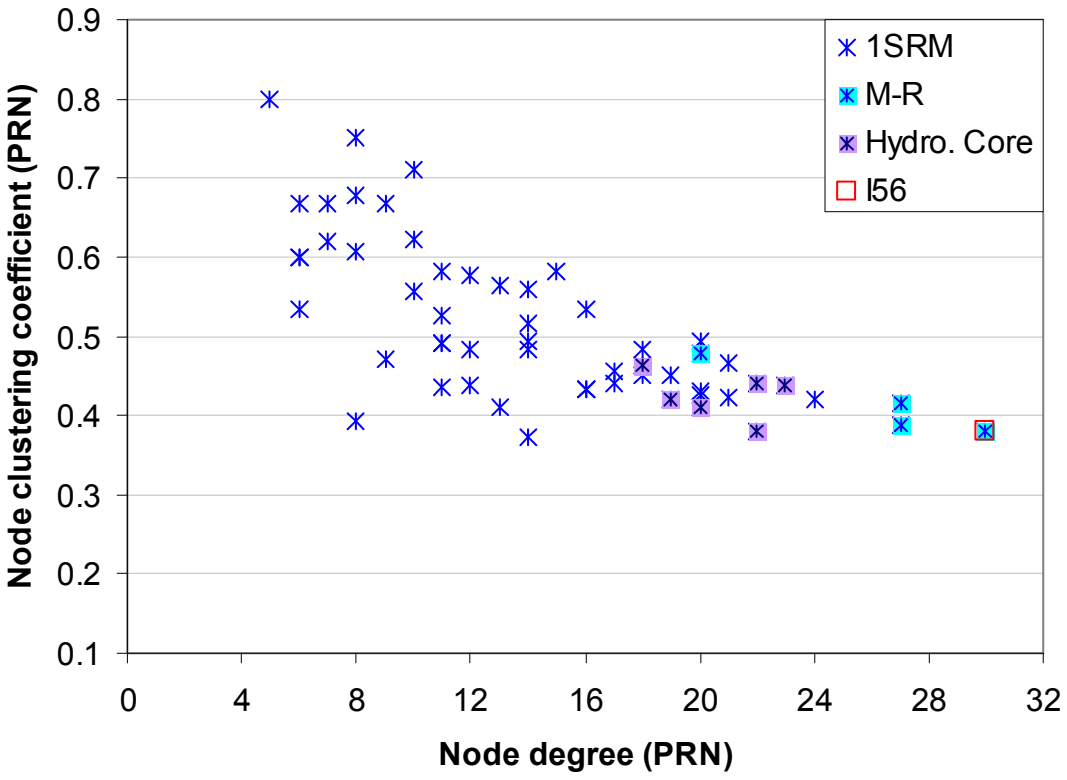

**Fig. B8** Key folding residues for 1SRM are highlighted in blue. Hydrophobic core residues which are not also M-R sites are highlighted in purple.

*2KJV* (Fig. B9) and its p54-55 circular permutant *2KJW* (Fig. B10): The M-R sites for 2KJV are E5, V6, N7, I8, I26, F60, L61, W62, Y63, V65 and L79. Three of these sites (V6, I8 and I26), together with L30 form the hydrophobic core for 2KJV [13]. The M-R sites for 2KJW are L8, Y10, L22, E25, R29, V48, I50, L52, L61, E64, I68 and L72. M-R site L61, together with Y63 and V65 form the hydrophobic core for 2KJW [14]. For both 2KJV and 2KJW, the M-R sites congregate on the first and third β-strands (1S and 7S, or β1 and β3 for both 2KJV and 2KJW) and the two α-helices (3H and 9H). β1, α1 and β3 make up one of S6's two folding cores; the other comprises β1, α2 and β4 [15].

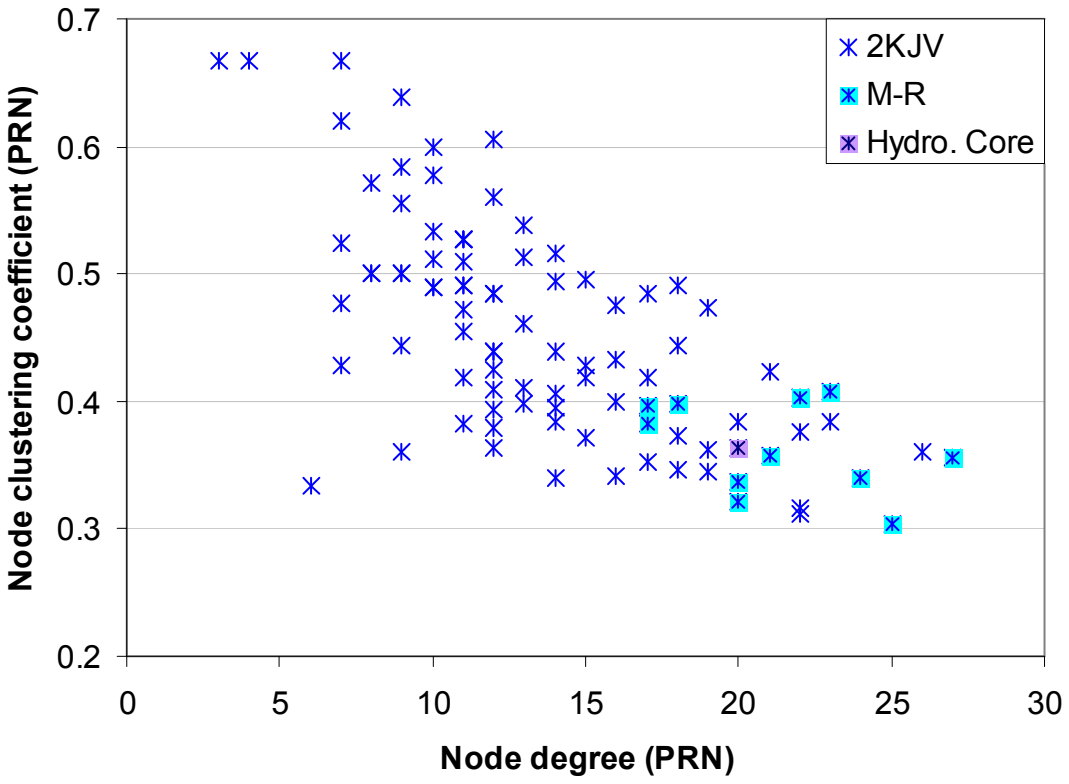

**Fig. B9** Key folding residues for 2KJV are highlighted in blue. Hydrophobic core residues which are not also M-R sites are highlighted in purple.

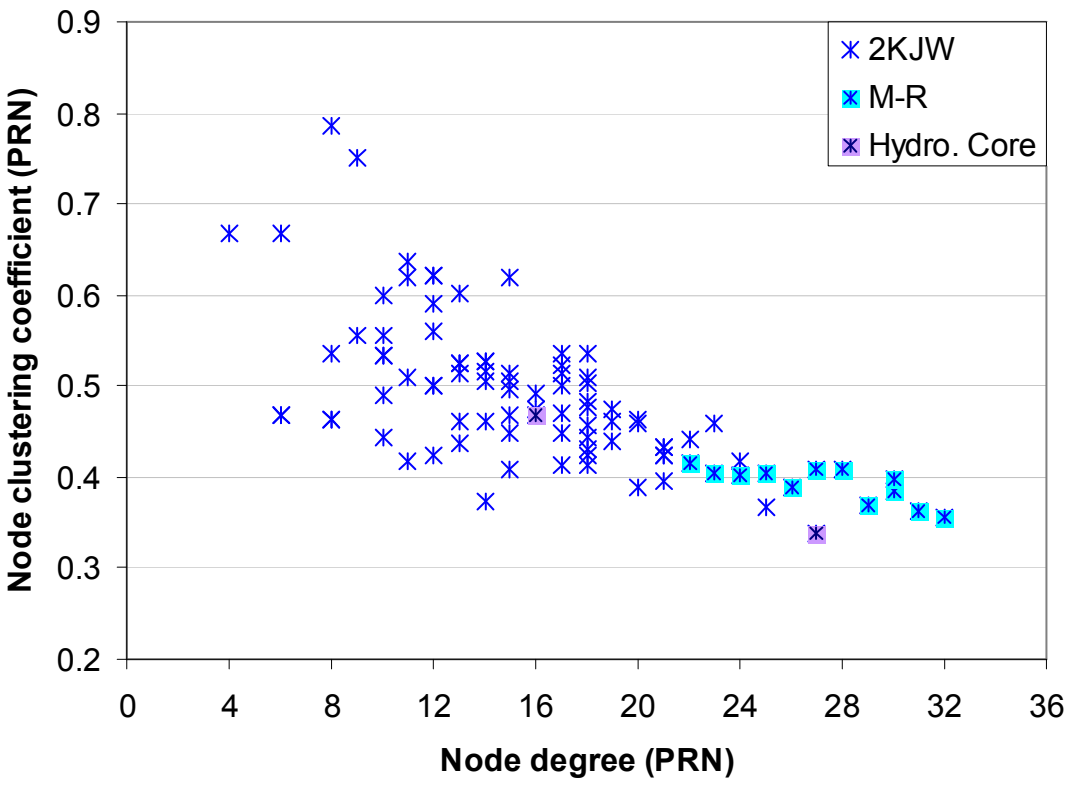

**Fig. B10** Key folding residues for 2KJW are highlighted in blue. Hydrophobic core residues which are not also M-R sites are highlighted in purple.

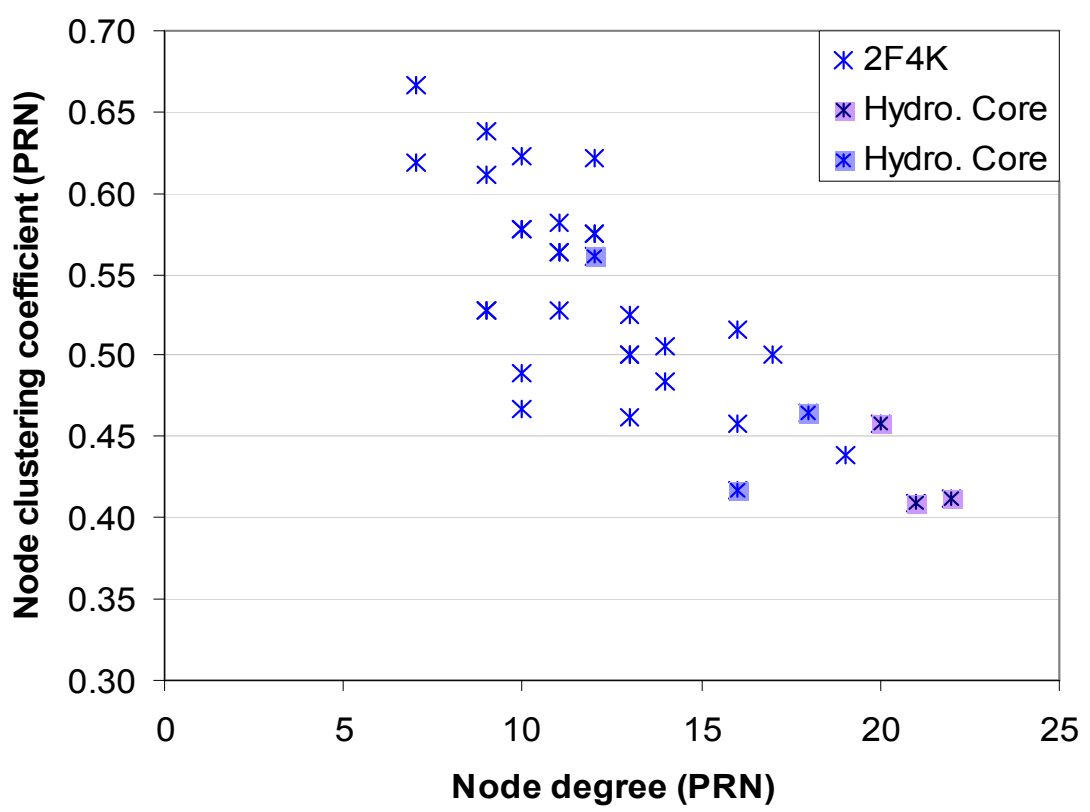

**Fig. B11** Phenylalanine (F) residues 47, 51 and 58 in the hydrophobic core of 2F4K are highlighted in purple. The other hydro core residues L42, V50 and L69 mentioned in [16] are highlighted in periwinkle.

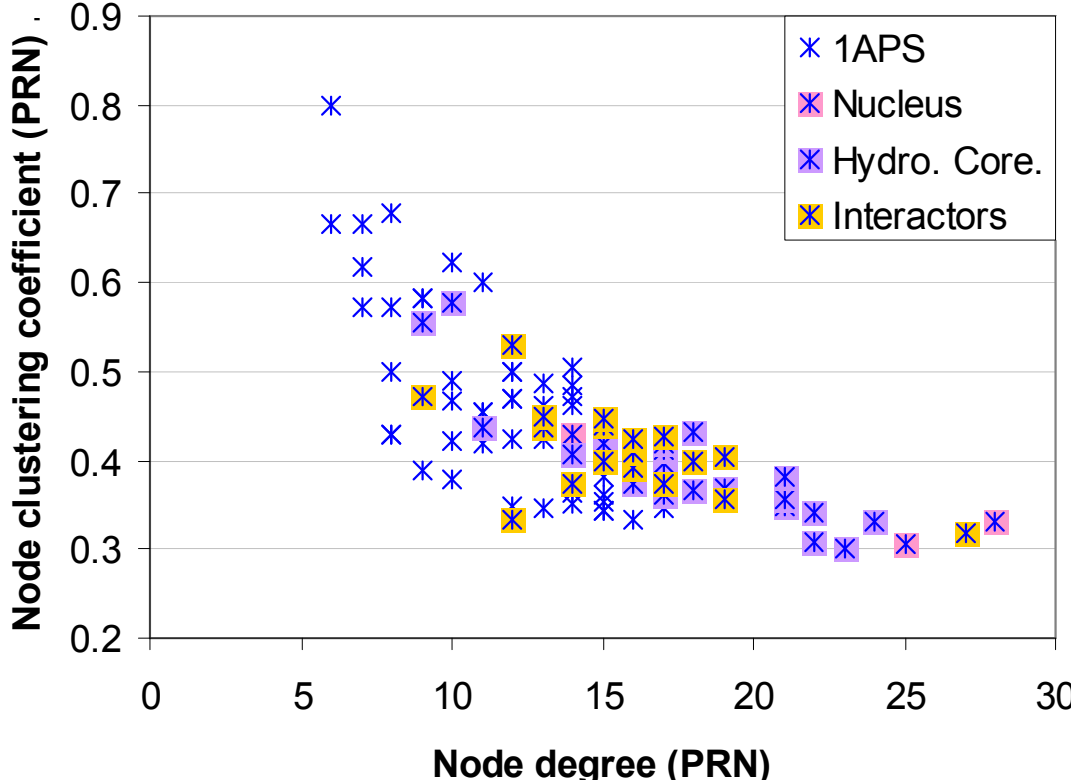

**Fig. B12** The three key folding residues (Y11, P54 and F94) [17] for 1APS are highlighted in pink. Hydrophobic core residues [18] which are not also Nucleus sites are highlighted in purple. Residues making long-range native contacts with the Nucleus sites [17] are highlighted in orange.

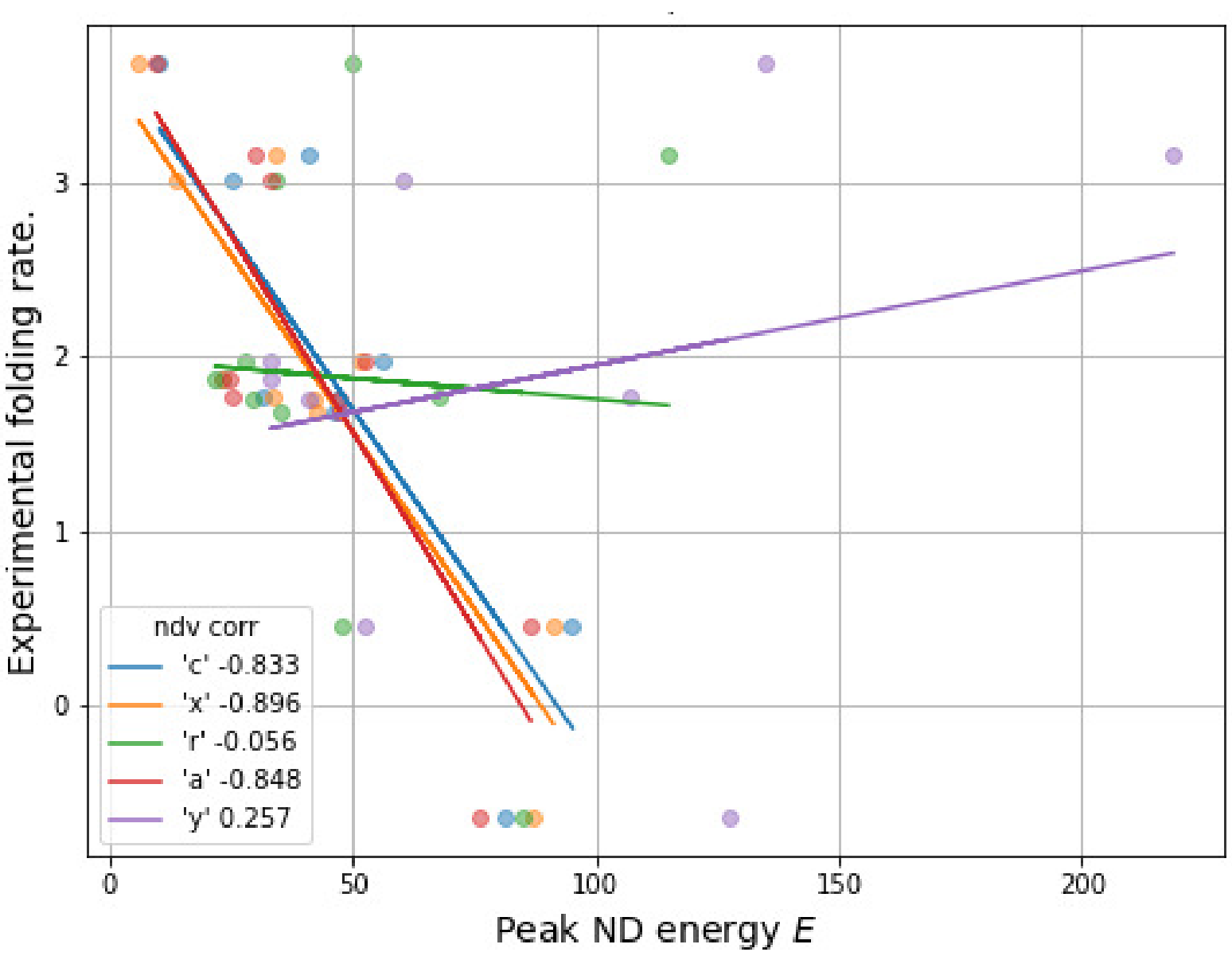

**Fig. C1 Paci dataset.** Pearson correlation between simulated folding rate (peak ND energy) and experimental folding rate. Numbers in the legend give the correlation coefficient by ND variant; the correlation is strongest with the 'x' ND variant.

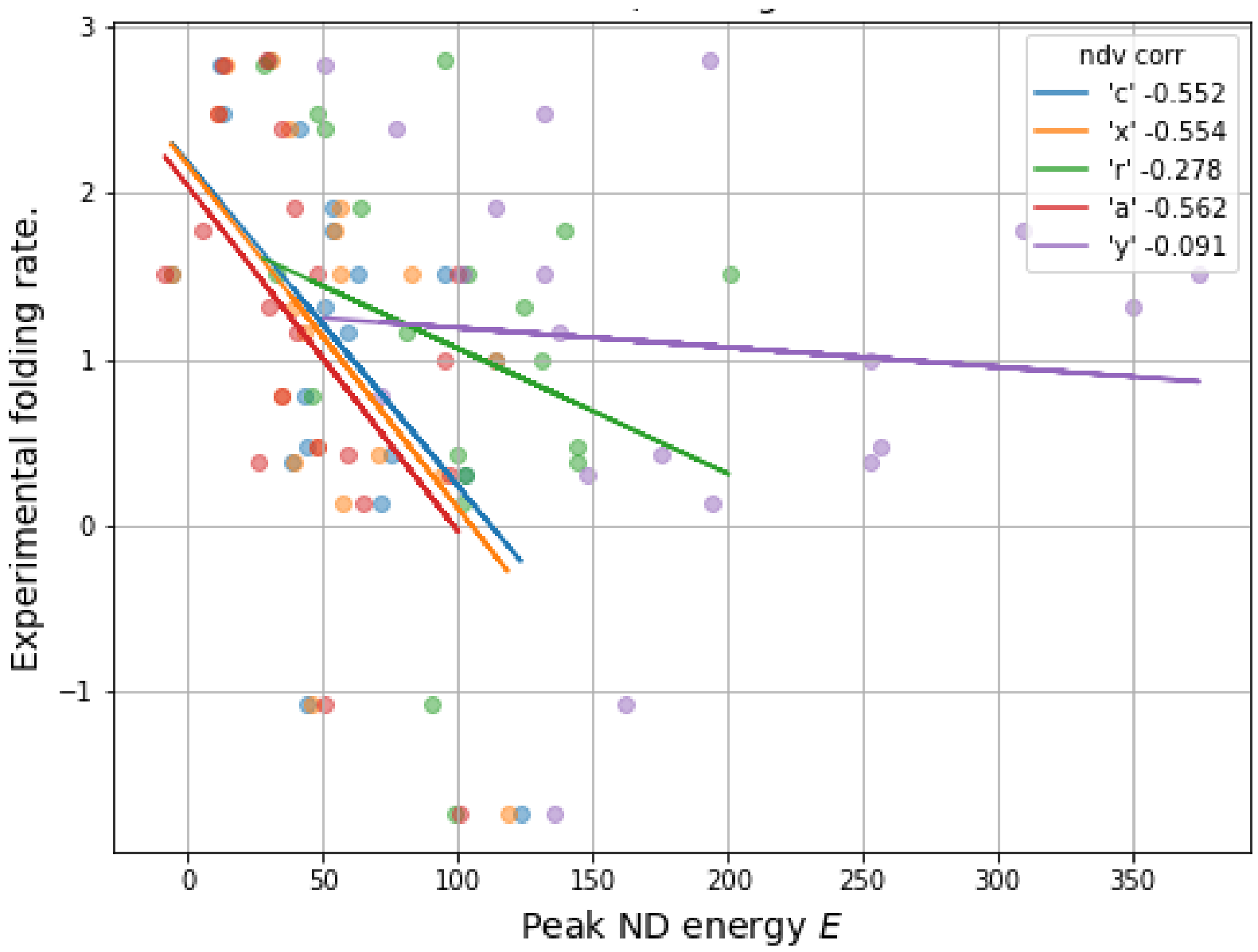

**Fig. C3 Kamagata dataset.** Pearson correlation between simulated folding rate (peak ND energy) and experimental folding rate. Numbers in the legend give the correlation coefficient by ND variant; the correlation is strongest with the 'a' ND variant.

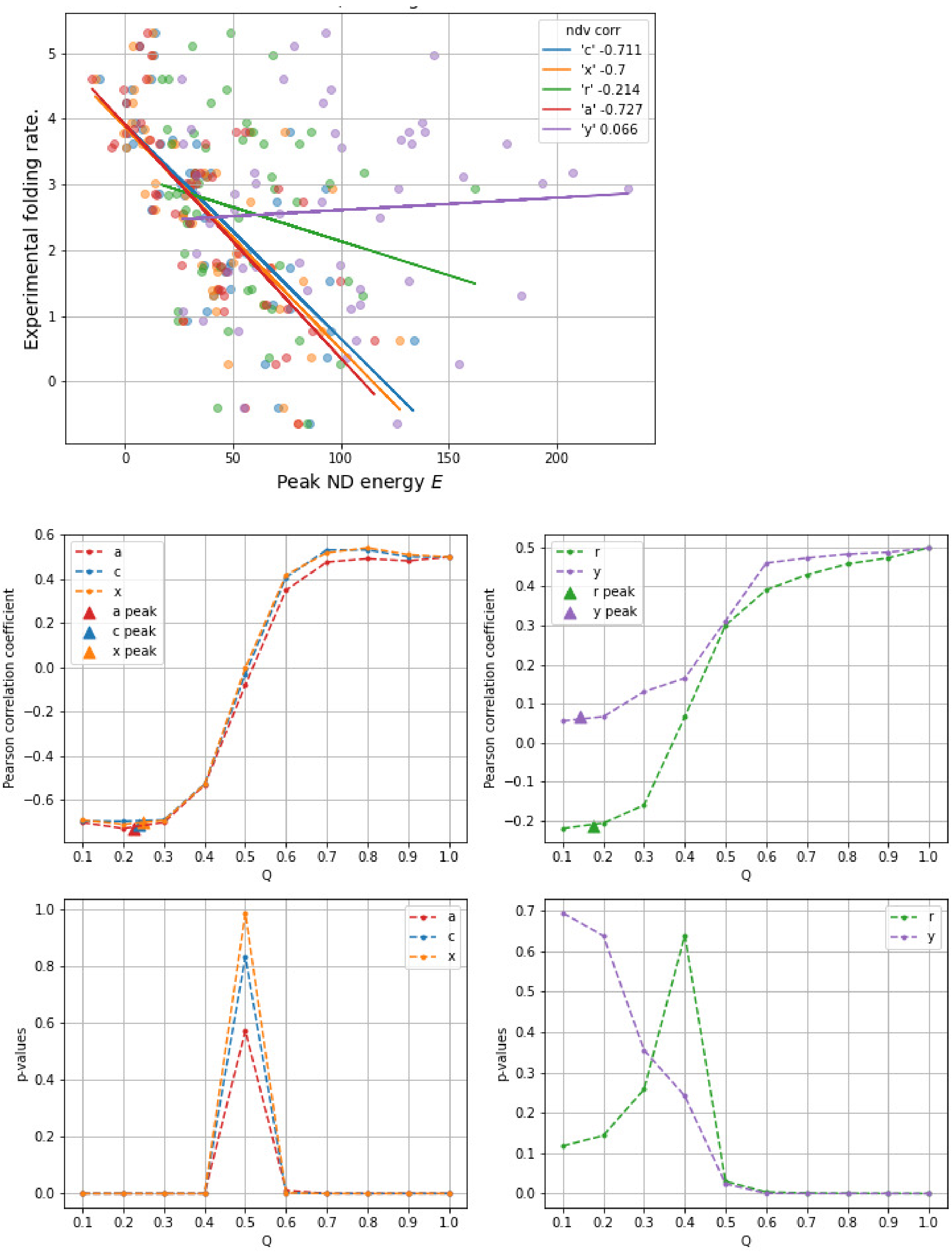

**Fig. C2 Uzunoglu dataset. Top:** Pearson correlation between simulated folding rate (peak ND energy) and experimental folding rate. Numbers in the legend give the correlation coefficient by ND variant; the correlation is strongest with the 'a' ND variant. **Bottom:** Pearson correlation coefficient and its p-value at each $Q$. The correlation is negative and significant for the biased ND models ('a', 'c' and 'x'), within the ND-TS region ($Q < 0.5$). For all ND variants, the smallest correlation coefficient is found around $Q$ where peak $\boldsymbol{E}$s are located.

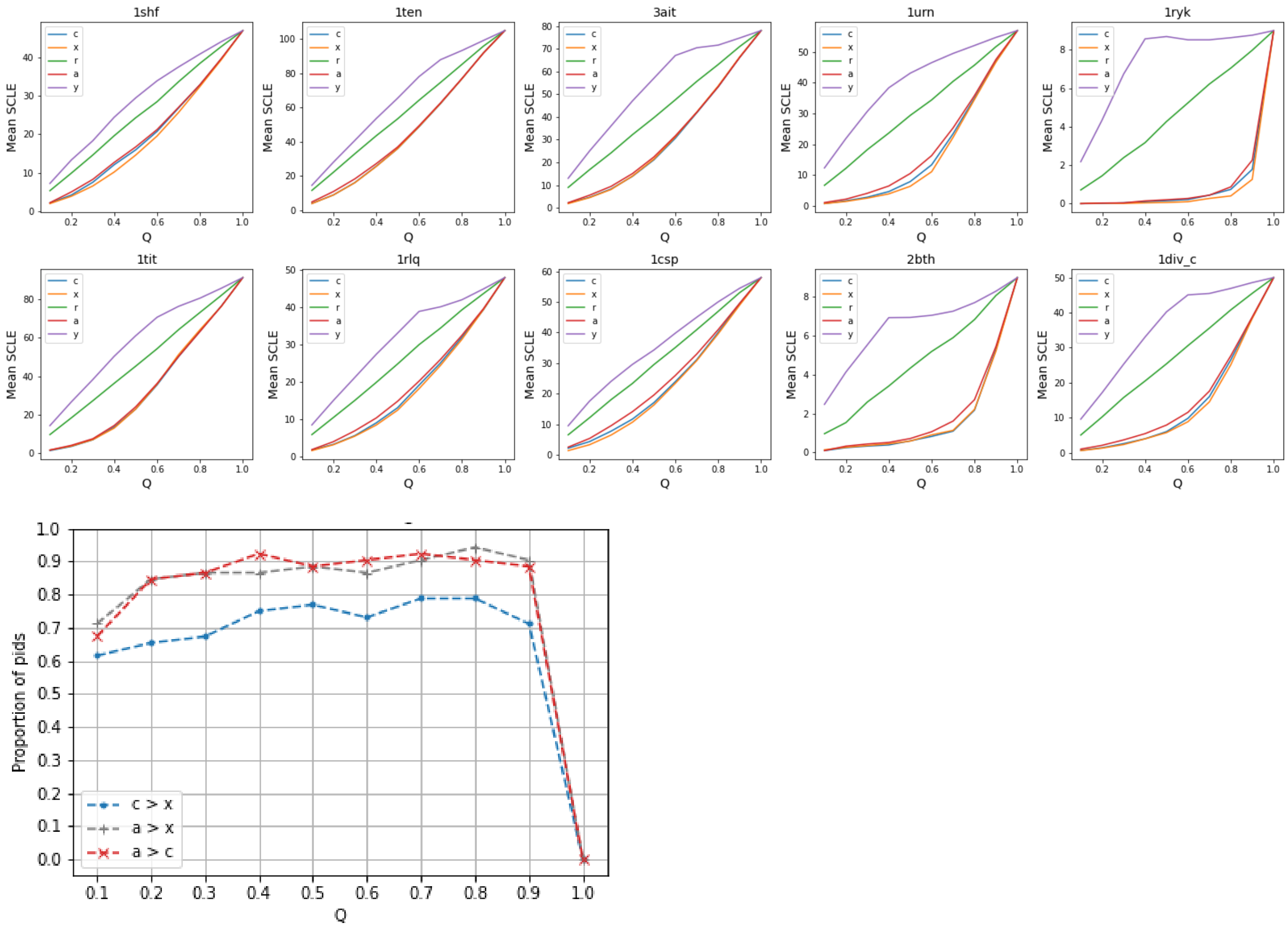

**Fig. C4 Uzunoglu dataset.** These two plots justify the ordering of the ND variants by early to late appearance of long-range native shortcuts (SCLE) as: 'y', 'r', 'a', 'c', 'x'.

**Top:** A random sample of 10 protein chains from the dataset; each plot depicts the average number of long-range native shortcuts (mean SCLE) formed by the five ND variants as a function of $Q$. It is clear from these plots that SCLE form much earlier with the 'y' and 'r' ND variants, than with the biased ND variants.

**Bottom:** Comparison of number of SCLE generated by biased ND variants as a function of $Q$. Each point is the proportion of protein chains (pids) in the dataset whose mean SCLE satisfies an inequality in the legend, at $Q$.
The inequality c > x expresses the condition that mean SCLE generated by the 'c' ND variant is greater than mean SCLE generated by the 'x' ND variant; the other two inequalities are interpreted similarly. For example, at $Q$=0.4, slightly more than 90% of Uzunoglu pids satisfy the condition that more SCLE are detected in 'a' PRNs than in 'c' PRNs.
For all three inequalities, the proportion of satisfying pids is quite consistent within $0.2 \le Q \le 0.9$ and stays above 60% until $Q = 1.0$ when all ND variants are expected to have generated all native shortcuts.

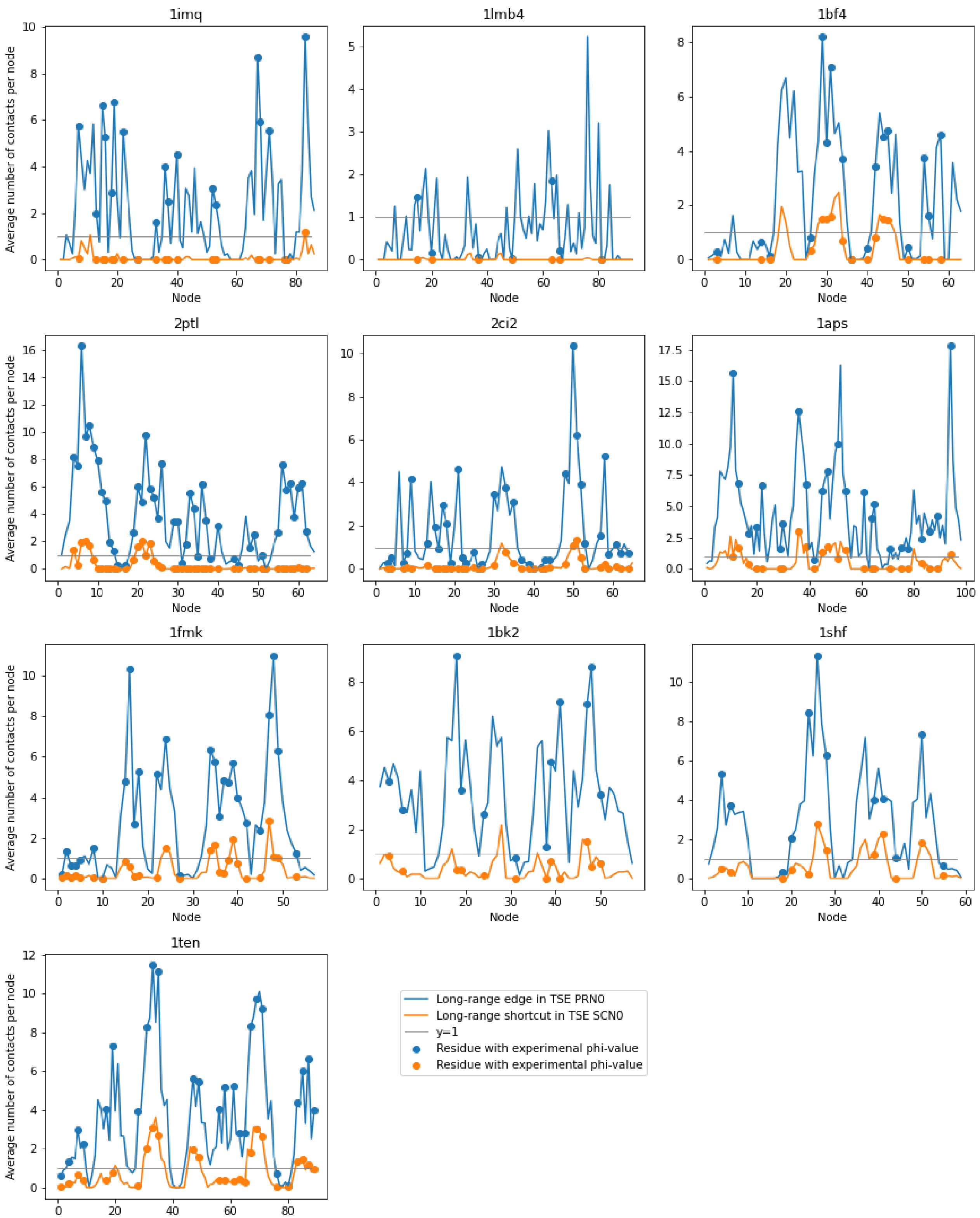

**Fig. C5 Paci dataset.** Average number of long-range contacts (blue line) and of native shortcuts (orange line) by node/residue in TSE PRNs. For each TSE, at least one residue with experimental phi-value has at least one long-range contact and one long-range native shortcut, on average: there is at least one dot (of each colour) above the y=1 line in each plot. Further, residues with experimental phi-value tend to also be nodes with long-range contacts in TSE PRNs: the peaks of the blue lines are frequently crested by blue dots, except for 1lmb4.

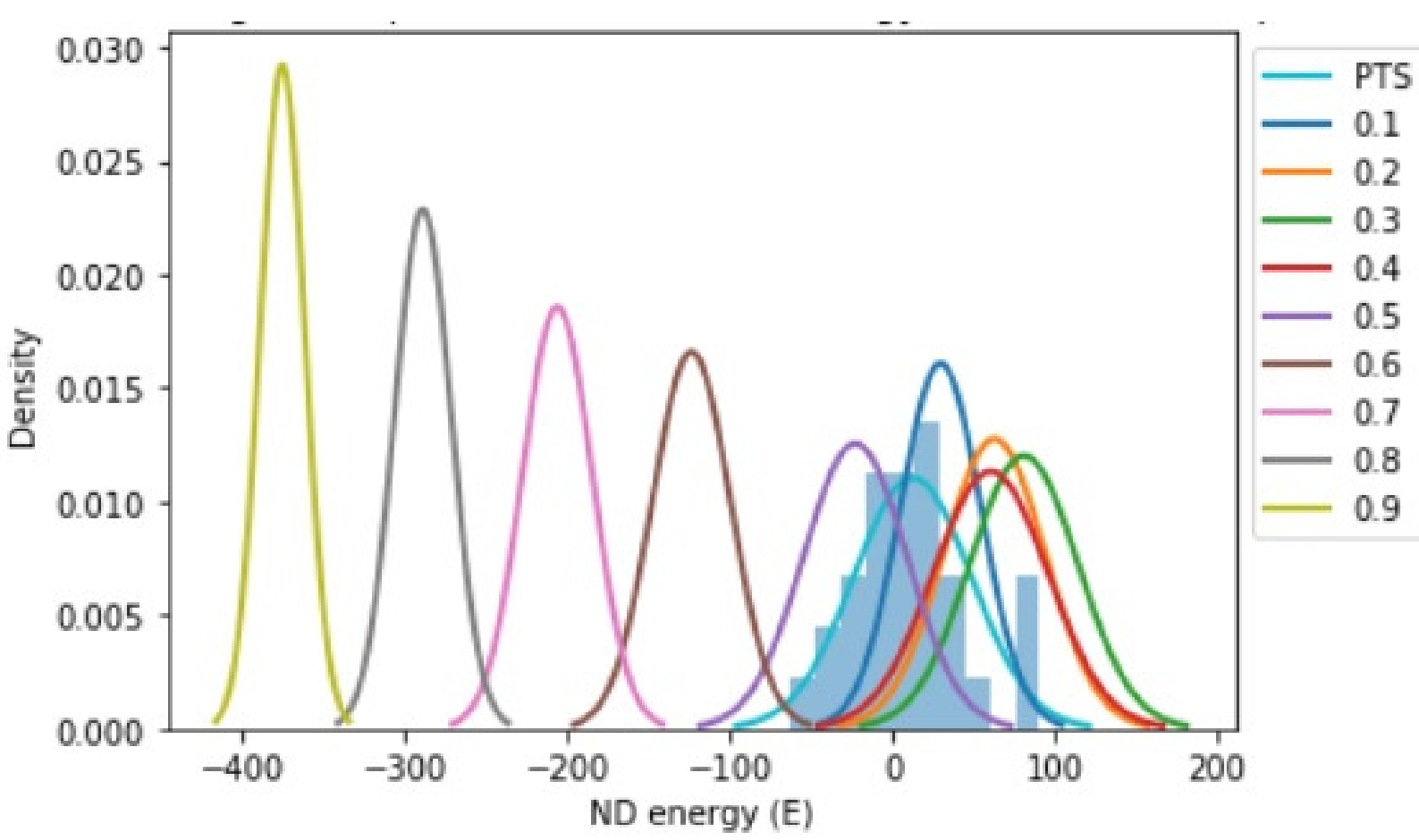

**Fig. D1** Locating the 1APS TSE within the 'c' ND energy Gaussians for 1APS. The histogram is of the TSE $E$ values. In this illustration, the MAP (maximum a posterior probability) estimate of $Q$ is 0.1.

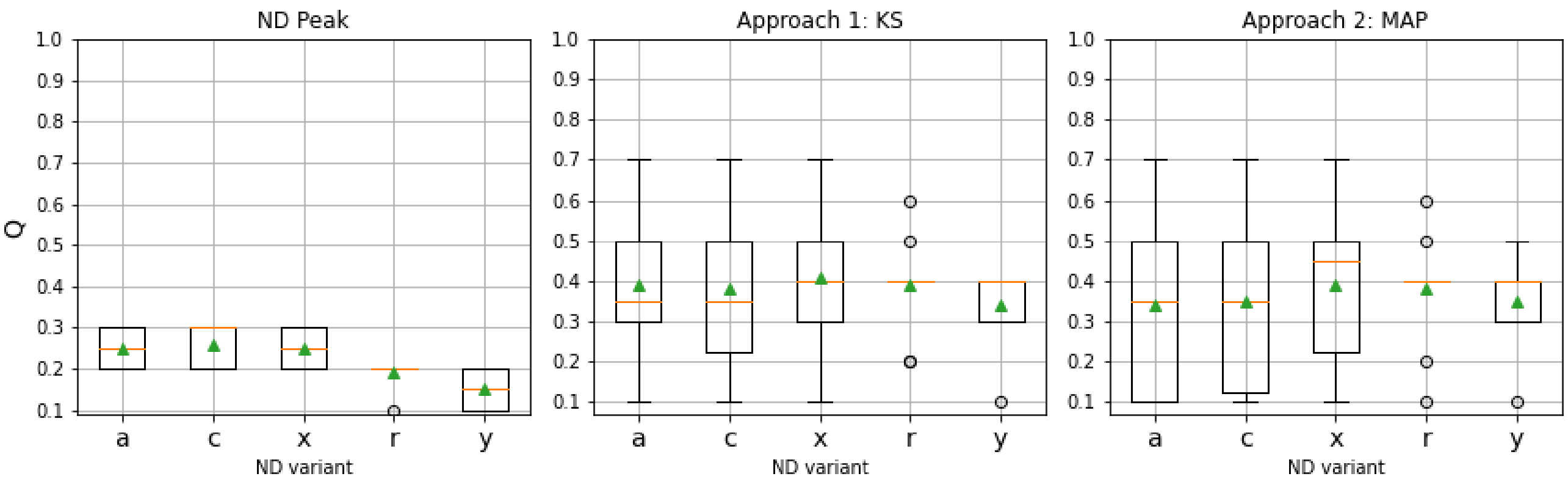

**Fig. D2 Paci dataset.** Locating TSE structures by ND energy. Location of ND peak ***E***s provided for reference only (left). Location of TSEs within ND energy profiles estimated by KS best fit (center) and maximum likelihood (right) approaches. The green triangle marks the mean. The box extends from the lower (Q1) to upper (Q3) quartiles of the data with a line at the median (Q2). The whiskers capture data that lie beyond the box which are within 1.5 times the inter-quartile range (Q3-Q1) in either direction. Points beyond the whiskers are deemed outliers.

**E. Selection of initial fold step** [19]

Residues in a protein sequence are classified as either Helix, Strand or Turn (neither helix nor strand). A secondary structure element (SSE) is a contiguous segment of identically classified residues. An initial fold substructure for a protein sequence is a pair of non-Turn SSEs adjacently located on the protein sequence, and includes any intervening Turn SSE.

A *candidate* initial fold substructure is selected as the initial fold substructure, if it belongs to a combination with the largest $C_{SCN0}$ (network clustering coefficient computed on native shortcuts only). Fig. E1 illustrates. The *selected* initial fold substructure makes the initial fold step. The clustering coefficient $C$ of a graph $G$ is the average clustering coefficient over all its nodes $N$, i.e. $C_G = \frac{1}{N}\sum_{i}^{N} C(i)$. The clustering coefficient of a node is $C(i) = \frac{2e_i}{k_i(k_i - 1)}$ where $e$ is the number of links between node $i$'s $k$ direct neighbors [20].

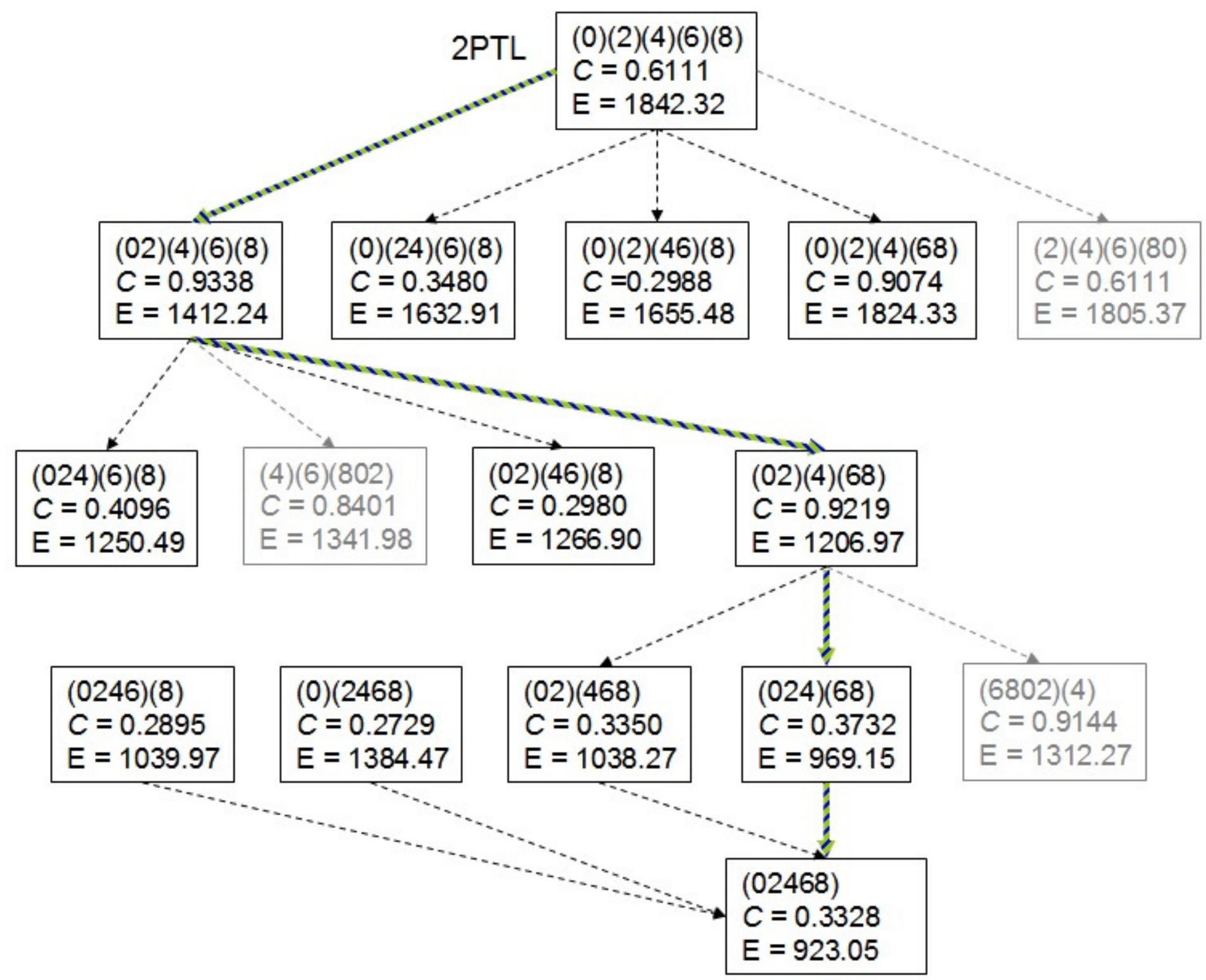

**Fig. E1** The $C_{SCN0}$ folding pathway for 2PTL (marked by the coloured arcs) is $(((0\ 2)_1\ 4)_3\ (6\ 8)_2\ )_4$ or equivalently $(((\beta 1\ \beta 2)_1\ \alpha 1)_3\ (\beta 3\ \beta 4)_2\ )_4$.

Each (non-grayed) box in Fig. E1 represents a feasible combination, and directed arcs represent accessibility between combinations (these design choices were explained previously in [19]). The $C$ and E values in each box respectively denote $C_{SCN0}$ and energy (computed with DeepView [21]) of a combination. The integers **0**, **2**, **4**, **6** and **8** represent the non-Turn SSEs β1, β2, α1, β3 and β4, respectively. These integers skip 1 to signal that there are Turn SSEs between them.

The $C_{SCN0}$ of the initial combination (top node) is due to $C_{SCN0}$ of α1; by themselves Strand SSEs have 0.0 $C_{SCN0}$. The four candidate initial fold substructures: (0 2), (2 4), (4 6) and (6 8) are enumerated in the first row below the top node, within their respective combinations (the rest of the SSEs). (0 2) is selected as the initial fold substructure because it is in the combination with the largest $C_{SCN0}$. The $C_{SCN0}$ of the initial fold substructure (β1 β2) is 0.9338 - 0.6111 = 0.3227. See [19] for calculating the $C_{SCN0}$ of a combination beyond the initial fold step.

**A note on inter-nodal distances in EDS**
The contact formulation for PRN0s considers Euclidean distances between atoms of both the backbone and side-chain parts of amino acids. In this sense, PRN0s can be considered an all-atom topological representation of native protein structures. However, the EDS algorithm only considers Cα-Cα Euclidean distances between nodes. Accounting for distances between side-chain atoms in EDS could yield more pronouncedly different ND folding routes for the topologically similar and highly symmetric proteins G and L [24].

To address this issue, we explored five EDS variants: 'a', 'b', 's', 'l', and 'v'. These EDS variants differ only in terms of how distances between nodes (amino acid residues) in a PRN are calculated. EDS is not used to construct PRNs, but SCNs which form the basis of our proposed $C_{SCN0}$ folding pathways (Fig. E1, Table E1). The EDS algorithm constructs paths on PRNs, and in the process identifies edges in PRNs that act as shortcuts because they make backtracks unnecessary.

The 'a' EDS variant considers Cα-Cα Euclidean distance between node-pairs. The 'b' EDS variant considers Cβ-Cβ Euclidean distance between node-pairs (Cα is used for GLY). The 's' and 'l' EDS variants respectively consider the shorter and the longer of Cα-Cα and Cβ-Cβ Euclidean distances between node-pairs. Finally, the 'v' EDS variant takes the average of Cα-Cα and Cβ-Cβ Euclidean distances between node-pairs.

We find that the number and identity of shortcuts in a PRN0 varies with EDS variant, and this variation is substantial enough to affect the selected initial fold substructure of several single-domain proteins. Importantly, the set of selected initial fold substructures generated by the 'a' EDS variant from native-state PRNs for the set of 14 canonical proteins which were studied in ref. [19], is *not* replicable by any of the other four EDS variants. EDS variant 'l' has the most number of matches with EDS variant 'a', but selects the C-terminus β-hairpin for 2PTL. The number of matches are 8, 9, 12 and 9 for EDS variants 'b', 's', 'l', and 'v' respectively (https://pfwscn.github.io/EDS_variants_native_structures.html). Based on this finding, the ND folding model uses the 'a' EDS variant by default.

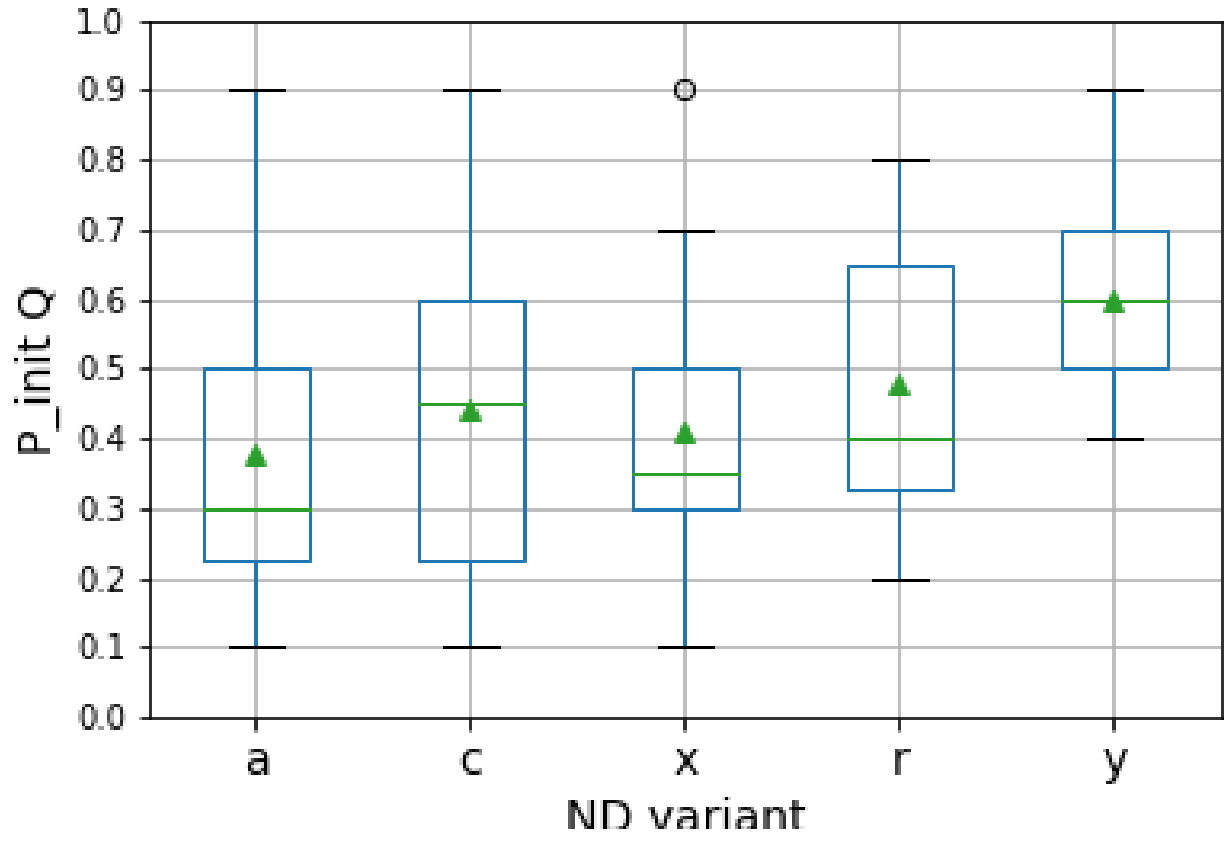

**Fig. E2 Khor dataset.** Distribution of $Q$ values where $P_{init}$ first approaches 50%, for the set of 14 small single-domain proteins, by ND variant. Interpretation of a box-plot is the same as Fig. D2.

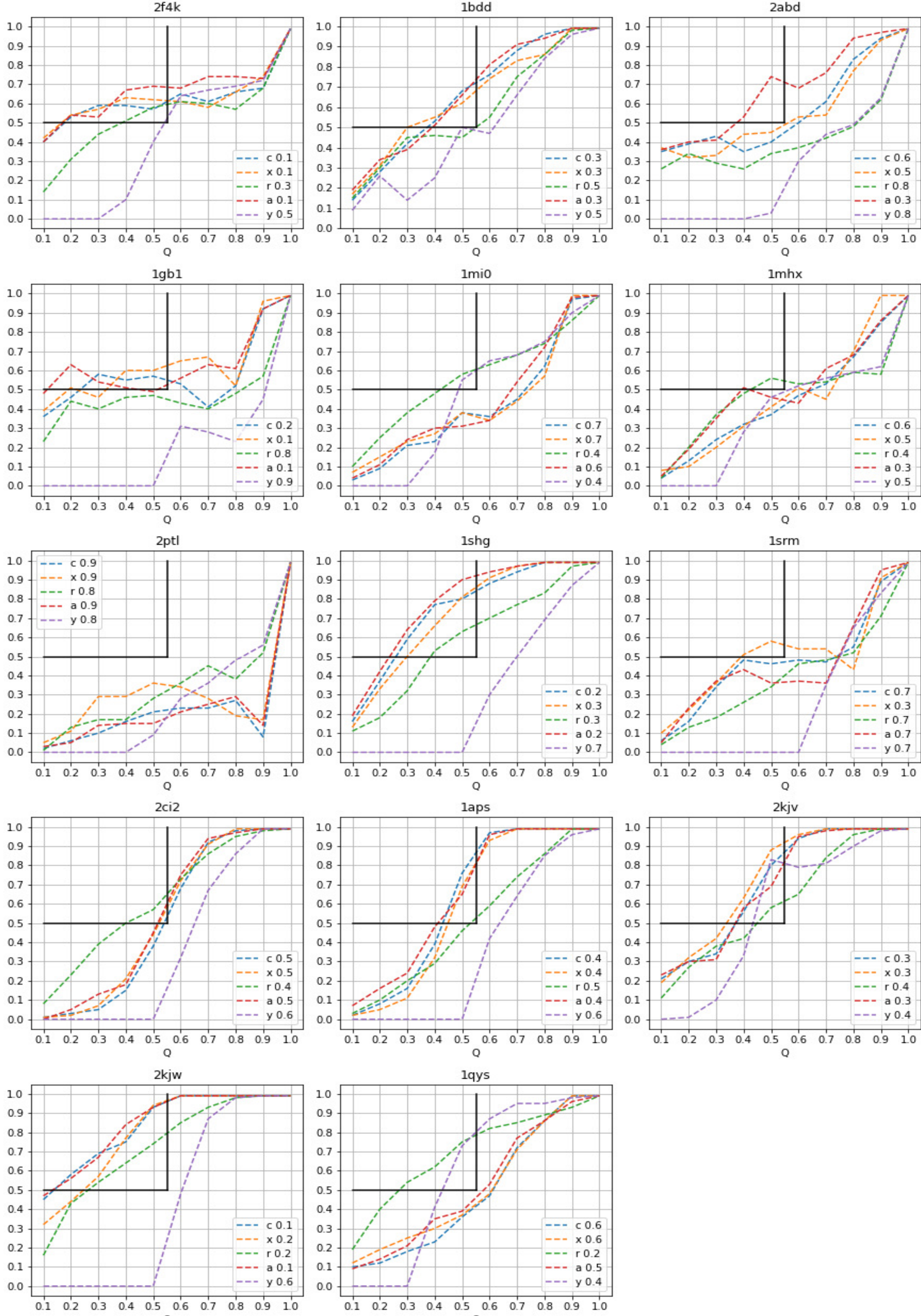

**Fig. E3 Khor dataset.** Proportion of ND runs whose initial fold step agrees with the initial fold step identified from native structure ($P_{\text{init}}$), as a function of $Q$. The numbers in the legend are truncated $Q$ values interpolated at $P_{\text{init}} = 0.5$. Ideally, we would like (at least one of) the 'a', 'c', and 'x' plot lines to cross the upper-left quadrant, since this means that the ND variant produced $P_{\text{init}} \geq 0.5$ at $Q \leq 0.5$, i.e. the initial fold step is made in the ND-TS region with at least 50% probability; and neither of the 'r' or 'y' plot lines cross into this quadrant.

**Table E1 Khor dataset.** (EDS path) substructure centrality for *candidate* initial fold substructures. **Bolded** are the *selected* initial fold substructure, and the largest substructure centrality. Shortest (BFS) path substructure centrality is in *italics* for comparison.

| PDB id | Candidate initial fold substructures | | | | | | First two fold steps<br>Subscripts denote order. |
|---|---|---|---|---|---|---|---|
| 2F4K<br>EDS<br>BFS | **(α1 α2)**<br>**14.76**<br>*9.76* | (α2 α3)<br>12.84<br>*6.47* | | | | | $((\alpha 1\ \alpha 2)_1\ \alpha 3)_2$ |
| 1BDD | (α1 α2)<br>**30.68**<br>*16.57* | **(α2 α3)**<br>29.37<br>*16.28* | | | | | $(\alpha 1\ (\alpha 2\ \alpha 3)_1)_2$ |
| 2ABD | (α1 α2)<br>56.15<br>*27.62* | (α2 α3)<br>**72.69**<br>*35.76* | **(α3 α4)**<br>32.39<br>*17.30* | | | | $(\alpha 2\ (\alpha 3\ \alpha 4)_1)_2$ |
| 1GB1 | (β1 β2)<br>34.89<br>*13.68* | (β2 α1)<br>20.30<br>*10.87* | (α1 β3)<br>26.29<br>*13.87* | **(β3 β4)**<br>**35**<br>*16* | | | $(\beta 1\ \beta 2)_2\ (\beta 3\ \beta 4)_1$ |
| 1MI0 | **(β1 β2)**<br>**53**<br>*23.1* | (β2 α1)<br>28.38<br>*15.92* | (α1 β3)<br>31<br>*17.52* | (β3 β4)<br>48.93<br>*24.5* | | | $(\beta 1\ \beta 2)_1\ (\beta 3\ \beta 4)_2$ |
| 1MHX | **(β1 β2)**<br>**56.4**<br>*26.15* | (β2 α1)<br>30.04<br>*13.88* | (α1 β3)<br>28.2<br>*15.04* | (β3 β4)<br>50.14<br>*25* | | | $(\beta 1\ \beta 2)_1\ (\beta 3\ \beta 4)_2$ |
| 2PTL | **(β1 β2)**<br>40<br>*16.23* | (β2 α1)<br>20.6<br>*11.04* | (α1 β3)<br>26.96<br>*13.03* | (β3 β4)<br>**44.94**<br>*19.61* | | | $(\beta 1\ \beta 2)_1\ (\beta 3\ \beta 4)_2$ |
| 1SHG | (β1 β2)<br>29.53<br>*13.11* | (β2 β3)<br>45.41<br>*22.17* | **(β3 β4)**<br>**50.28**<br>*21.57* | (β4 β5)<br>32<br>*12.85* | | | $(\beta 2\ (\beta 3\ \beta 4)_1)_2$ |
| 1SRM | (β1 β2)<br>14.25<br>*6.04* | **(β2 β3)**<br>**30.07**<br>*13.43* | (β3 β4)<br>28.75<br>*15* | (β4 β5)<br>24.5<br>*9.6* | | | $((\beta 2\ \beta 3)_1\ \beta 4)_2$ |
| 2CI2 | (β1 α1)<br>26.5<br>*14.68* | (α1 β2)<br>29.35<br>*13.56* | **(β2 β3)**<br>59.64<br>*28.88* | (β3 β4)<br>**61.1**<br>*27.25* | | | $((\beta 2\ \beta 3)_1\ \beta 4)_2$ |
| 1APS | (β1 α1)<br>47.48<br>*22.70* | (α1 β2)<br>49.52<br>*20.57* | **(β2 β3)**<br>**81.22**<br>*33.83* | (β3 α2)<br>38.8<br>*19.15* | (α2 β4)<br>48.90<br>*28.13* | (β4 β5)<br>31<br>*16.52* | $(\beta 2\ \beta 3)_1\ (\beta 4\ \beta 5)_2$ |
| 2KJV | (β1 α1)<br>59<br>*24.93* | (α1 β2)<br>41.26<br>*18.33* | **(β2 β3)**<br>**64.67**<br>*33.08* | (β3 α2)<br>47.90<br>*24.57* | (α2 β4)<br>54.14<br>*26.14* | | $(\beta 1\ \alpha 1)_2\ (\beta 2\ \beta 3)_1$ |
| 2KJW | (β3 α2)<br>33.24<br>*16.14* | (α2 β4)<br>55.87<br>*26.71* | **(β4 β1)**<br>**66.1**<br>*27* | (β1 α1)<br>52.27<br>*27.9* | (α1 β2)<br>32.48<br>*16.96* | | $(\beta 3\ \alpha 2)_2\ (\beta 4\ \beta 1)_1$ |
| 1QYS | **(β1 β2)**<br>**86.1**<br>*39.5* | (β2 α1)<br>39.78<br>*21.64* | (α1 β3)<br>40.27<br>*19.92* | (β3 α2)<br>49.30<br>*24.81* | (α2 β4)<br>39.89<br>*22.61* | (β4 β5)<br>70.12<br>*30.5* | $(\beta 1\ \beta 2)_1\ (\beta 4\ \beta 5)_2$ |

In general, shortest (BFS) path substructure centrality follows the same trend as EDS path substructure centrality; the substructure with the largest or second largest EDS path substructure centrality, also has the largest or the second largest shortest path substructure centrality. Shortest path substructure centrality values are about half that of EDS path substructure centrality values because it is necessary to obtain shortest paths in one direction ($i < j$) only since a shortest path from $i$ to $j$ is also a shortest path from $j$ to $i$. This symmetry is not necessarily true for EDS paths; they need to be obtained in both directions for all node pairs. These results are sensitive to the residue range of the protein chain and SSE assignment (context of a residue provided by a protein chain - a context tested by evolution).

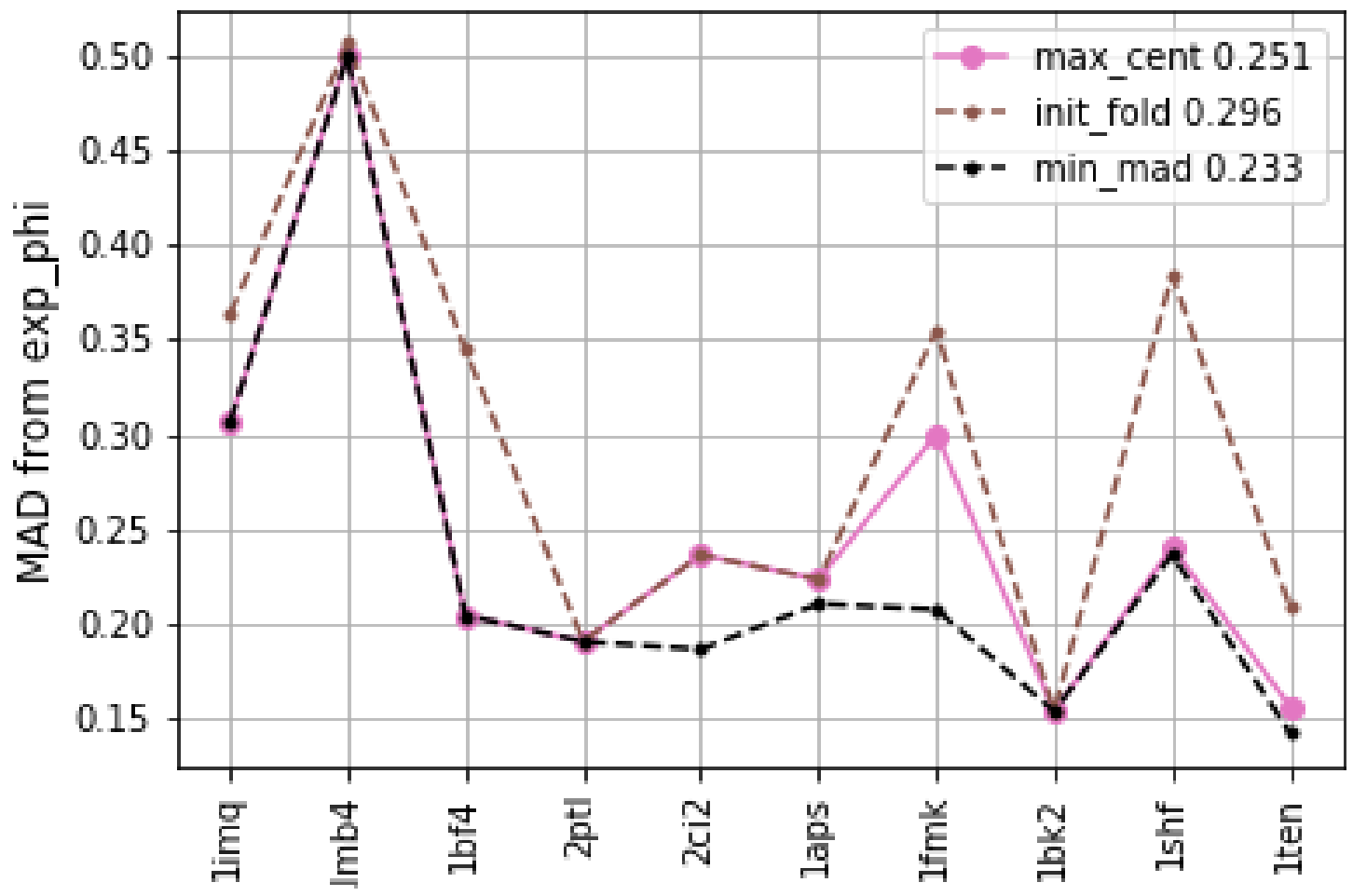

**Fig. F1 Paci dataset.** Mean Absolute Difference (MAD) from experimental phi-values for phi-values calculated from native-state PRNs with normalized local node centrality (*locent*). *locent* is computed from a subset of EDS paths which contributes to substructure centrality (Table E1). The plot compares two different ways for selecting such a subset, with the aim of showing that the subset used in the paper (max_cent) is satisfying: close to optimal and with a clear selection criterion.

With max_cent, *locent* is calculated from EDS paths that contribute to the substructure centrality of the candidate initial fold substructure with the largest substructure centrality. With init_fold, *locent* is calculated from EDS paths that contribute to the substructure centrality of the selected initial fold substructure. 2CI2 in Table E1 is an example where the selected initial fold substructure (β2 β3) does not enjoy largest substructure centrality; this designation belongs to candidate initial fold substructure (β3 β4).

min_mad is the minimum MAD attainable with a subset of EDS paths that contribute to the substructure centrality of a candidate initial fold substructure. All candidate initial fold substructures are evaluated to find the smallest MAD between experimental phi-values and *locent*. There is no *apriori* criterion to select a subset of EDS paths; there is also no observed consistency: for 6/10 proteins min_mad does a max_cent selection, for others, it does neither a max_cent nor an init_fold selection.

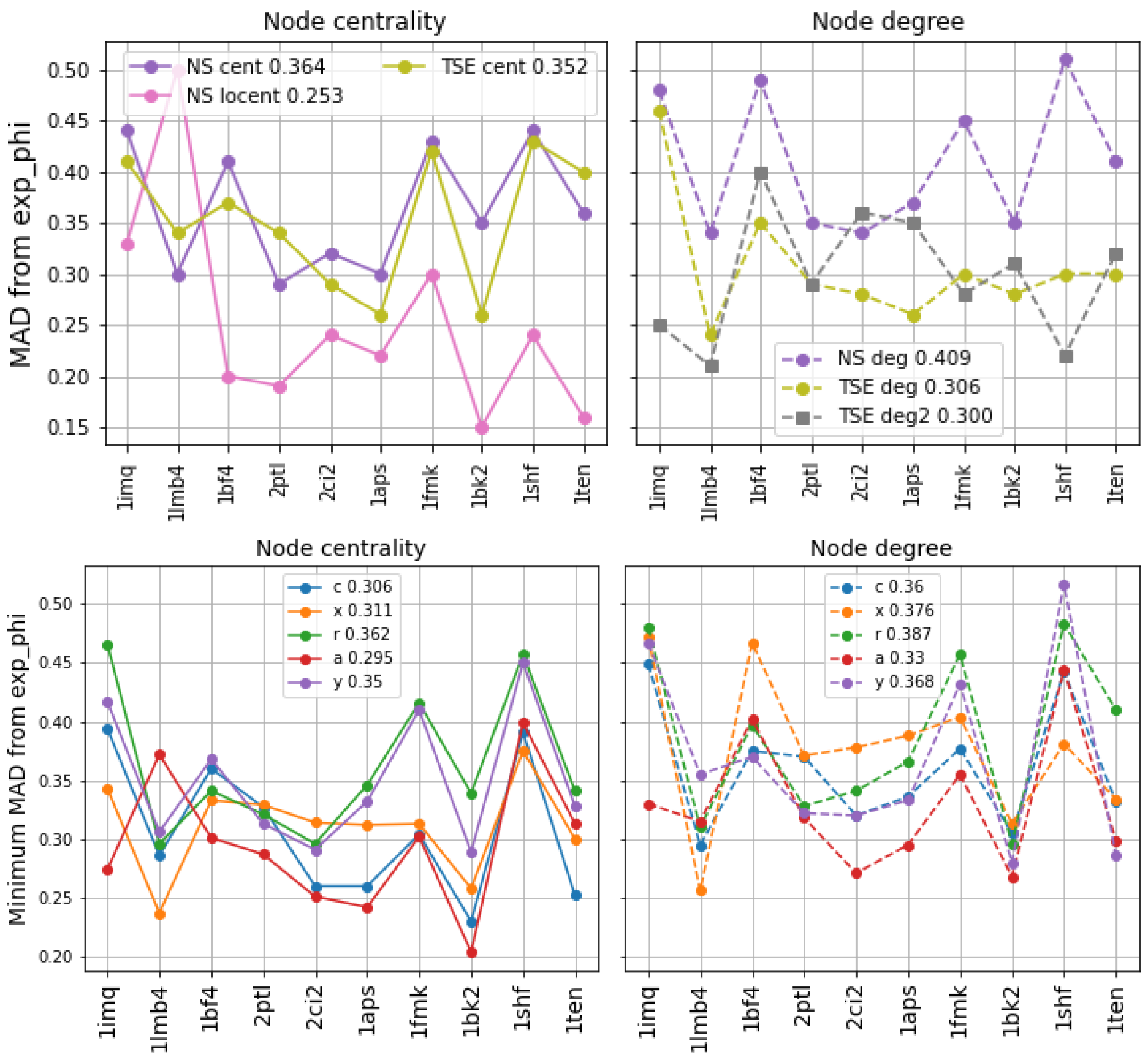

**Fig. F2 Paci dataset.** Summary of results for calculating phi-values as normalized node centrality (*cent*) **(left)** and normalized node degree (*deg*) **(right)** from native-state (NS) and transition-state ensemble (TS) PRNs **(top)**, and from ND generated PRNs **(bottom)**. It is desirable for a method to yield smaller Mean Absolute Difference between calculated and experimental phi-values. The numbers in the legends are the average MAD score for a method.

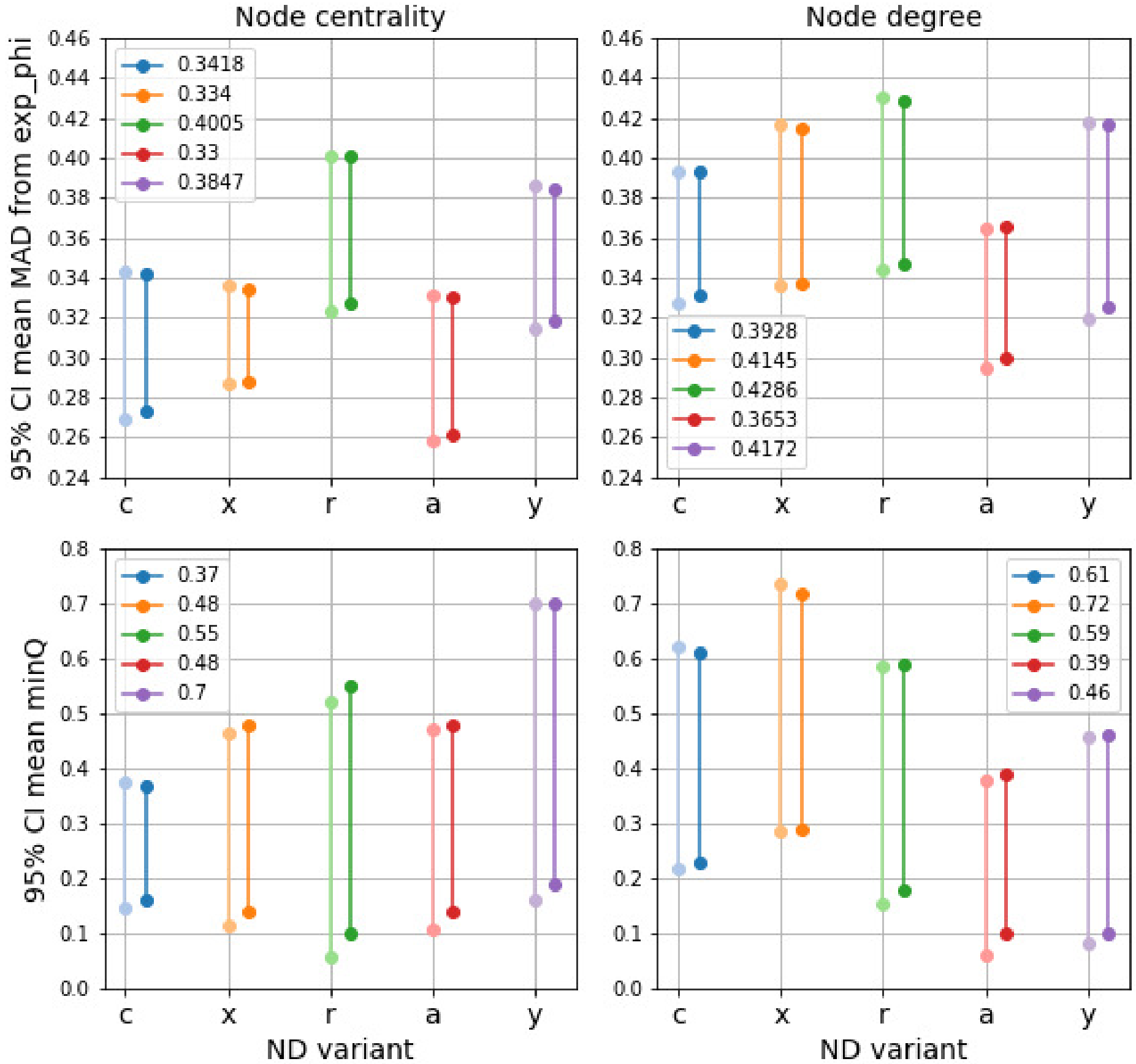

**Fig. F3 Paci dataset.** Summary of results for calculating phi-values as normalized node centrality (*cent*) **(left)** and normalized node degree (*deg*) **(right)** from PRNs generated by the five ND variants investigated. Range estimates (95% CI) are given for the average Mean Absolute Difference (MAD) between experimental and calculated phi-values **(top)**, and for the average min$Q$ ($Q$ where this MAD is minimum) **(bottom)**. The two CIs per method are obtained with CLT and bootstrap resampling methods respectively.

The numbers in the legends give the 97.5 percentile estimate for average MAD, and for average min$Q$. We use these 97.5 percentile estimates to quantify the worst case for an average statistic.

It is desirable for a method (ND variant and node statistic combination) to have smaller average MAD, and average min$Q$ that is clearly within the ND-TS region. For *deg*, 'a' is the only ND variant that satisfies these two criteria, in the worst case. For *cent*, all three biased ND variants satisfy the two criteria, in the worst case.

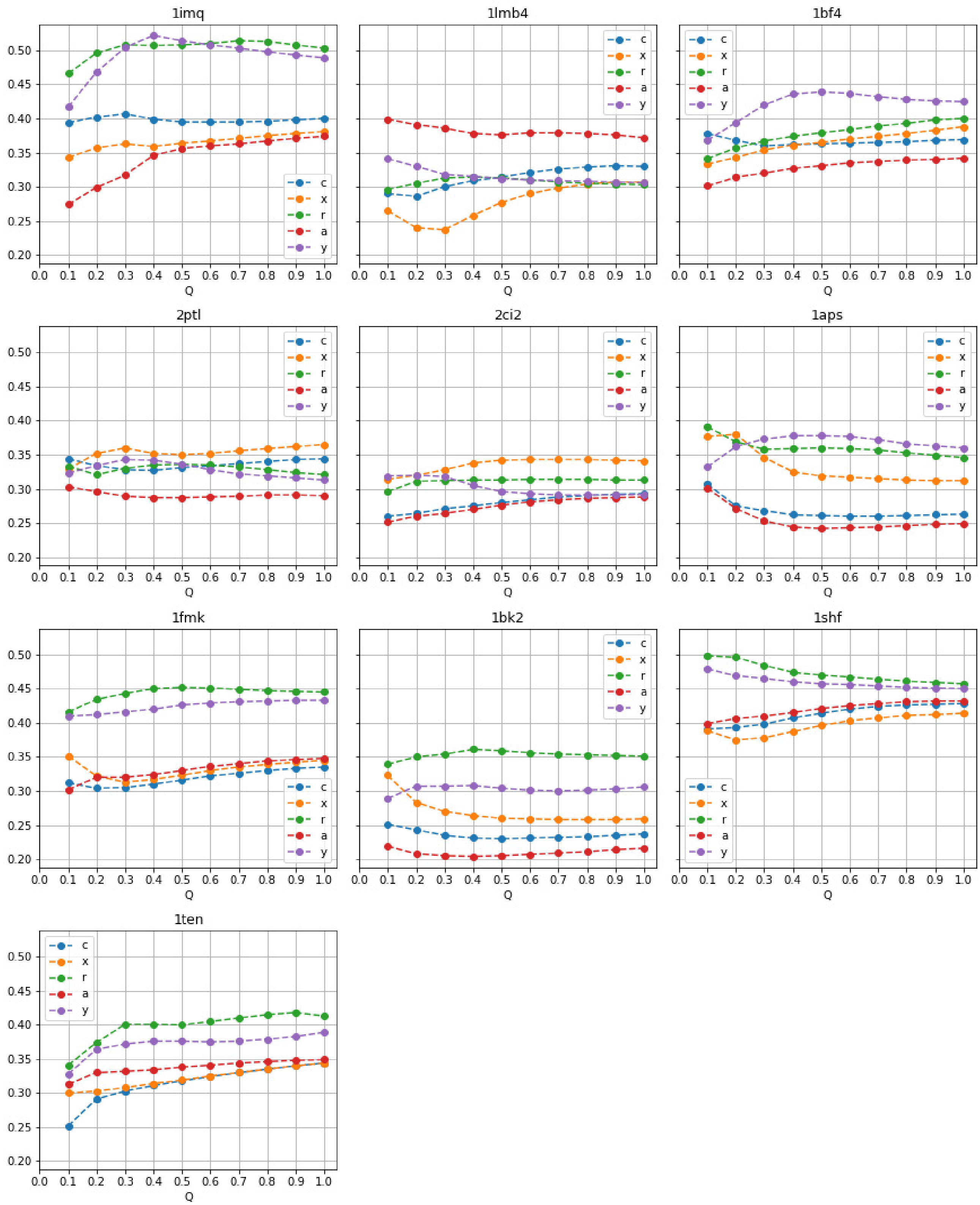

**Fig. F4 Paci dataset.** Mean absolute difference (MAD) between phi-values from experiments, and phi-values calculated as normalized node centrality (*cent*) from ND PRNs, as a function of *Q*. Smaller MAD is desirable; the biased ND variants ('a', 'c', and 'x') tend to produce smaller MAD values. ND node statistics are averaged over the 100 independent ND PRNs generated at each *Q*.

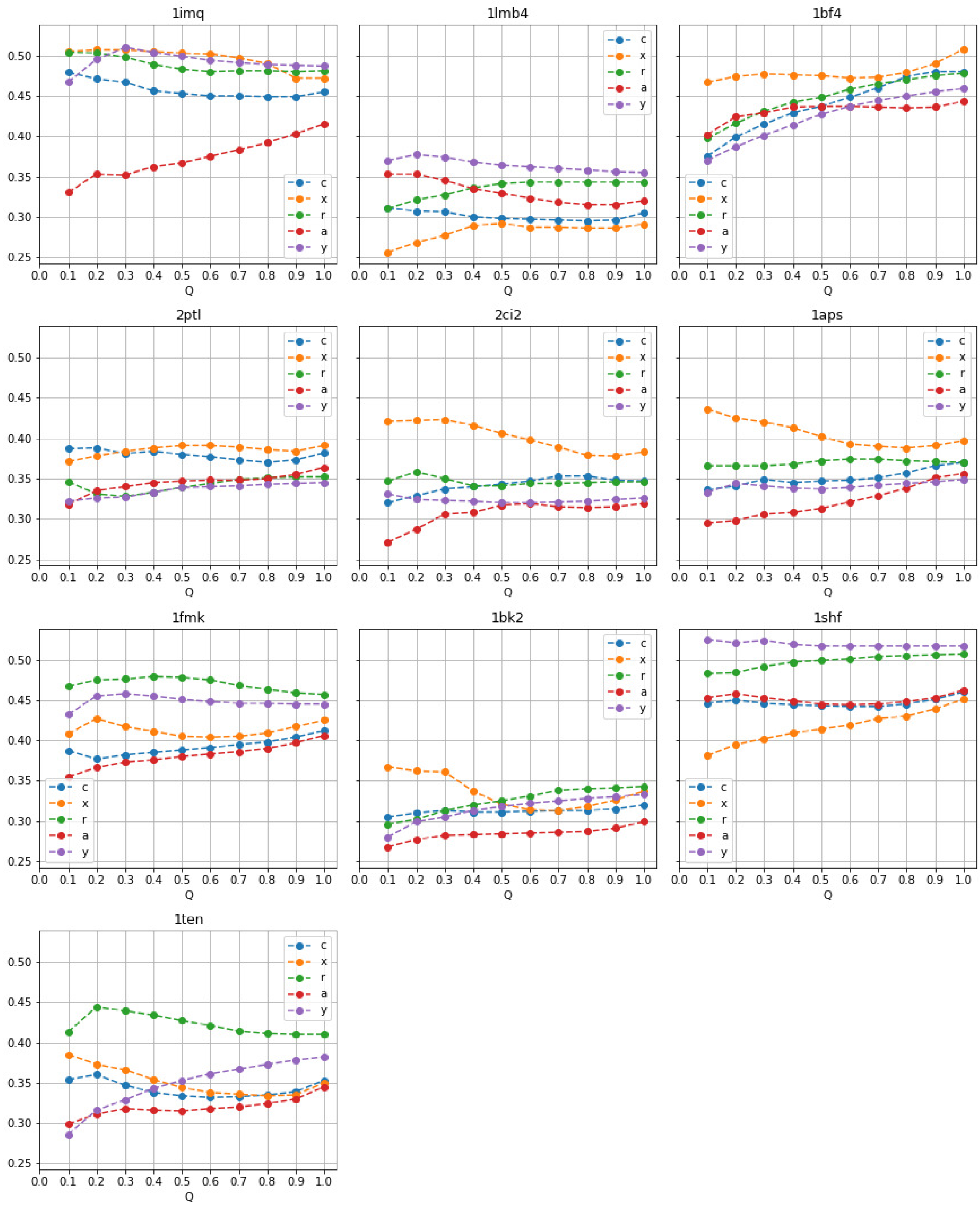

**Fig. F5 Paci dataset.** Mean absolute difference between phi-values from experiments, and phi-values calculated as normalized node degree (*deg*) from ND PRNs, as a function of *Q*. ND node statistics are averaged over the 100 independent ND PRNs generated at each *Q*, before doing the comparison.

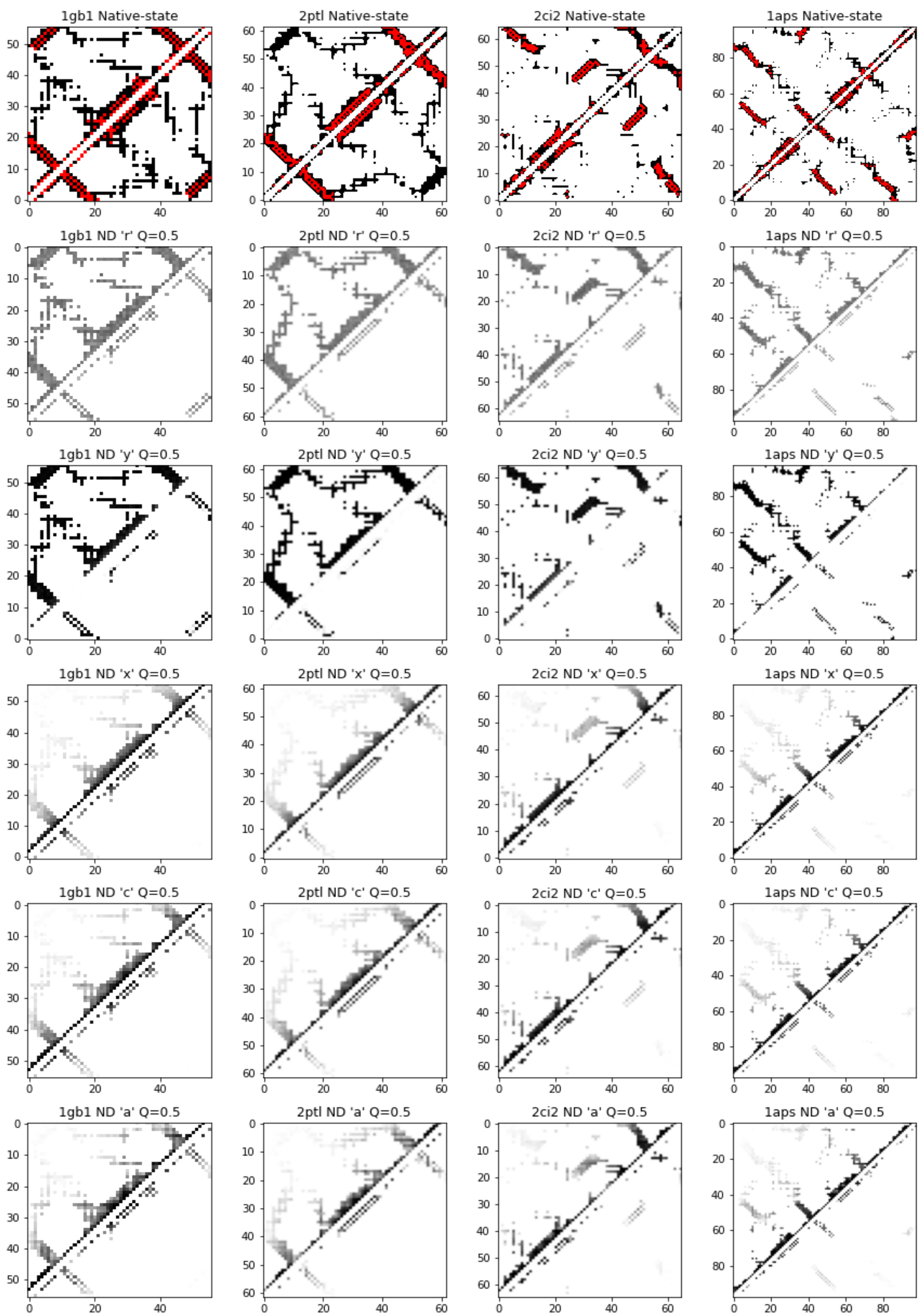

**Fig. G1** ND contact probability maps of PRN edges (upper triangle) and of native shortcuts (lower triangle) at $Q = 0.5$ for four prototypical proteins; higher contact probabilities are associated with darker shades of gray. The topmost row holds native-state contact maps; black and red cells denote PRN0 and SCN0 edges, respectively.

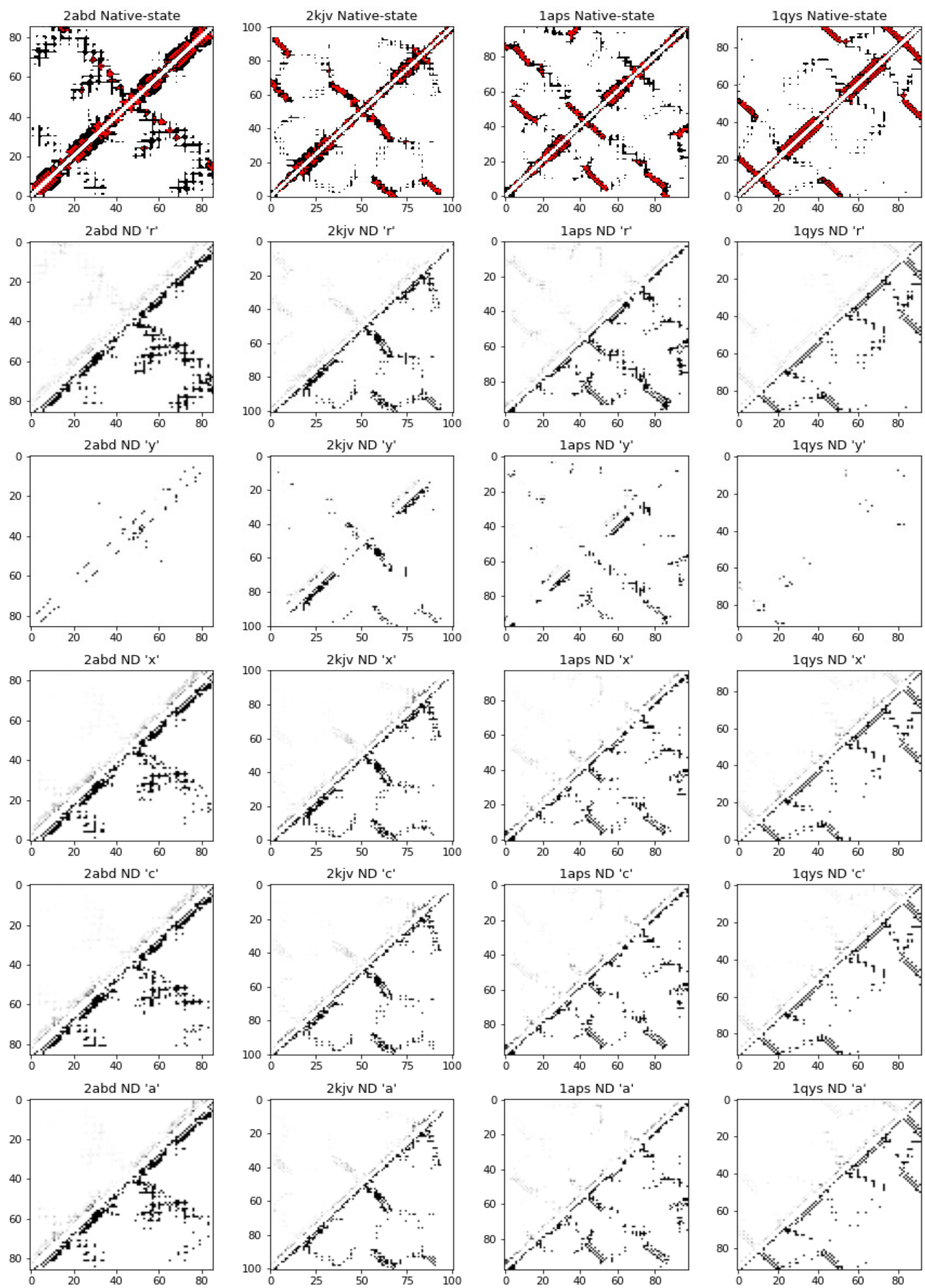

**Fig. G2** ND contact probability maps of non-native shortcuts (upper triangle) at $Q = 0.5$; higher contact probabilities are associated with darker shades of gray. The lower triangles denote by black cells, contacts that appear as non-native shortcuts at least once in 1000 ND networks (100 PRNs at each of the 10 $Q$s). The topmost row holds native-state contact maps; black and red cells denote PRN0 and SCN0 edges, respectively.

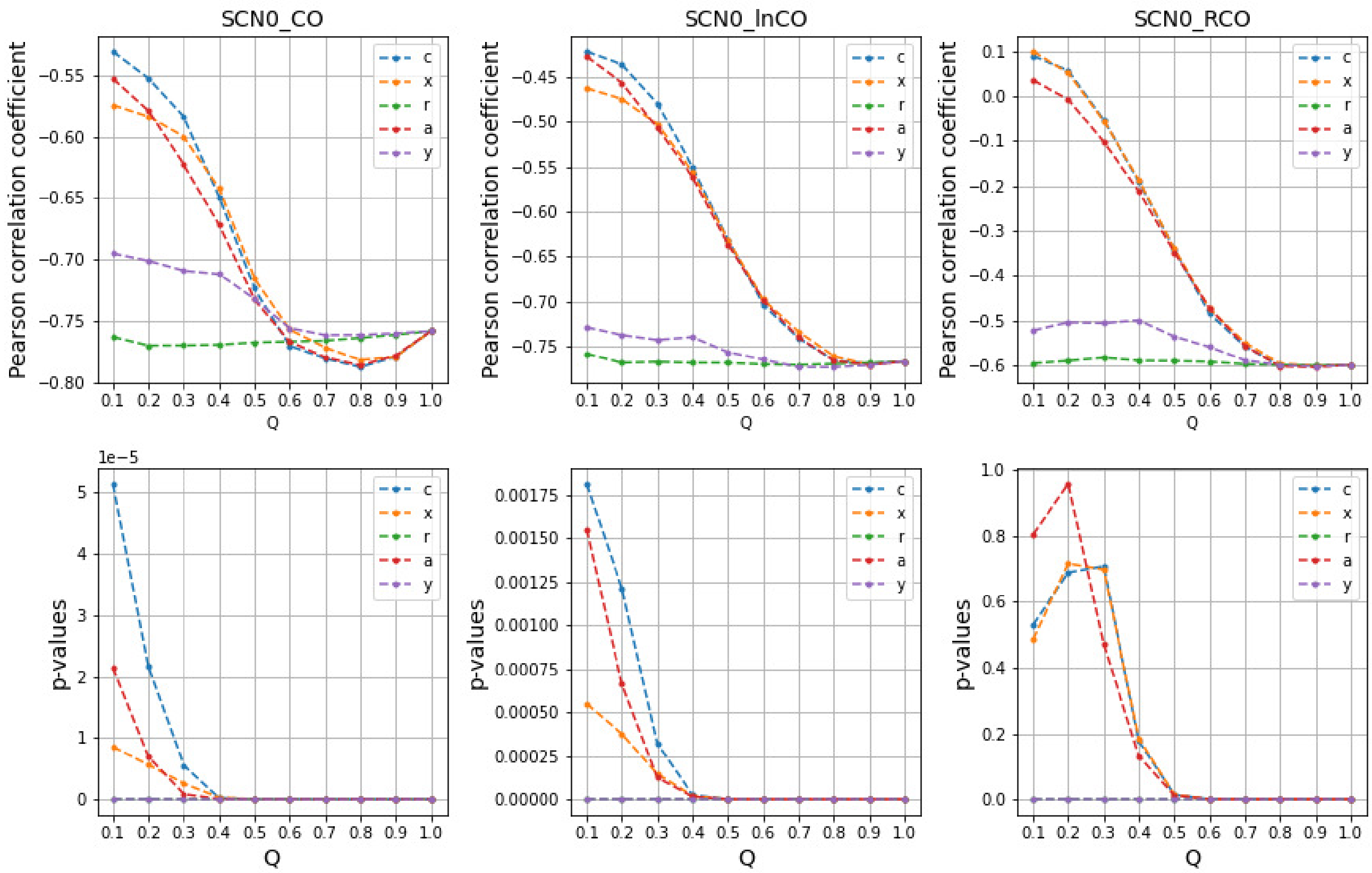

**Fig. G3 Uzunoglu dataset.** Pearson correlation (top: coefficient, bottom: p-value) between experimental folding rate and Contact-Order of native shortcuts, as a function of $Q$. For a set of contacts, CO is the average of their sequence distances, lnCO is the average of the natural logarithm of their sequence distances, and RCO is CO divided by the number of unique nodes in the set of contacts (the denominator varies with $Q$). Note difference in scale on the y-axes.

Compared with the biased ND variants ('a', 'c', 'x'), both the 'r' and 'y' ND variants are more strongly correlated with experimental folding rate within the ND-TS region ($Q < 0.5$). For the biased ND variants, only SCN0_CO and SCN0_lnCO produce a significant correlation within the ND-TS region.

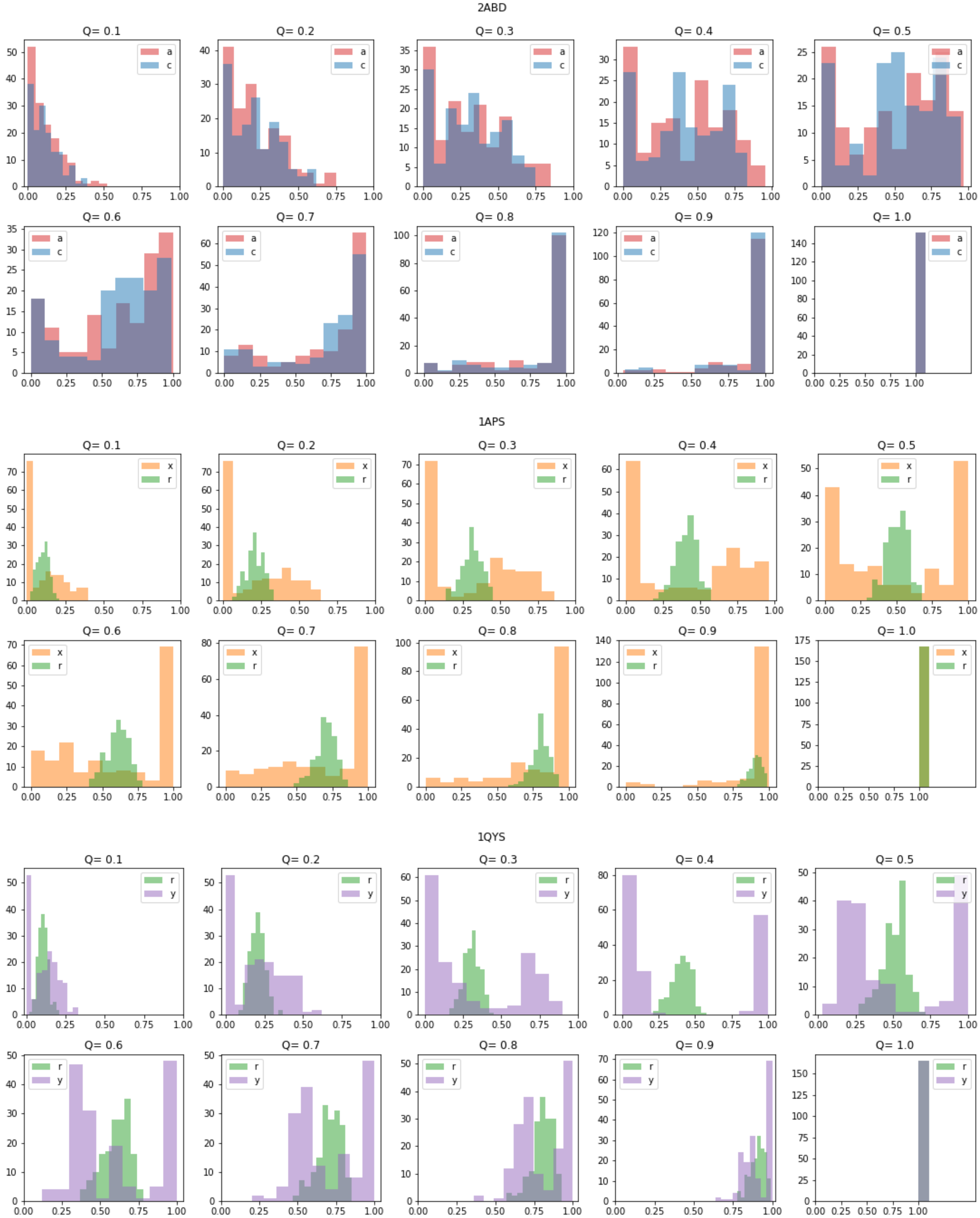

**Fig. G4** Distribution of contact probabilities for native shortcuts as a function of $Q$ for 2ABD (top), 1APS (middle) and for 1QYS(bottom). The 'r' contact probabilities are more homogeneous (smaller variance) than the other four ND variants.

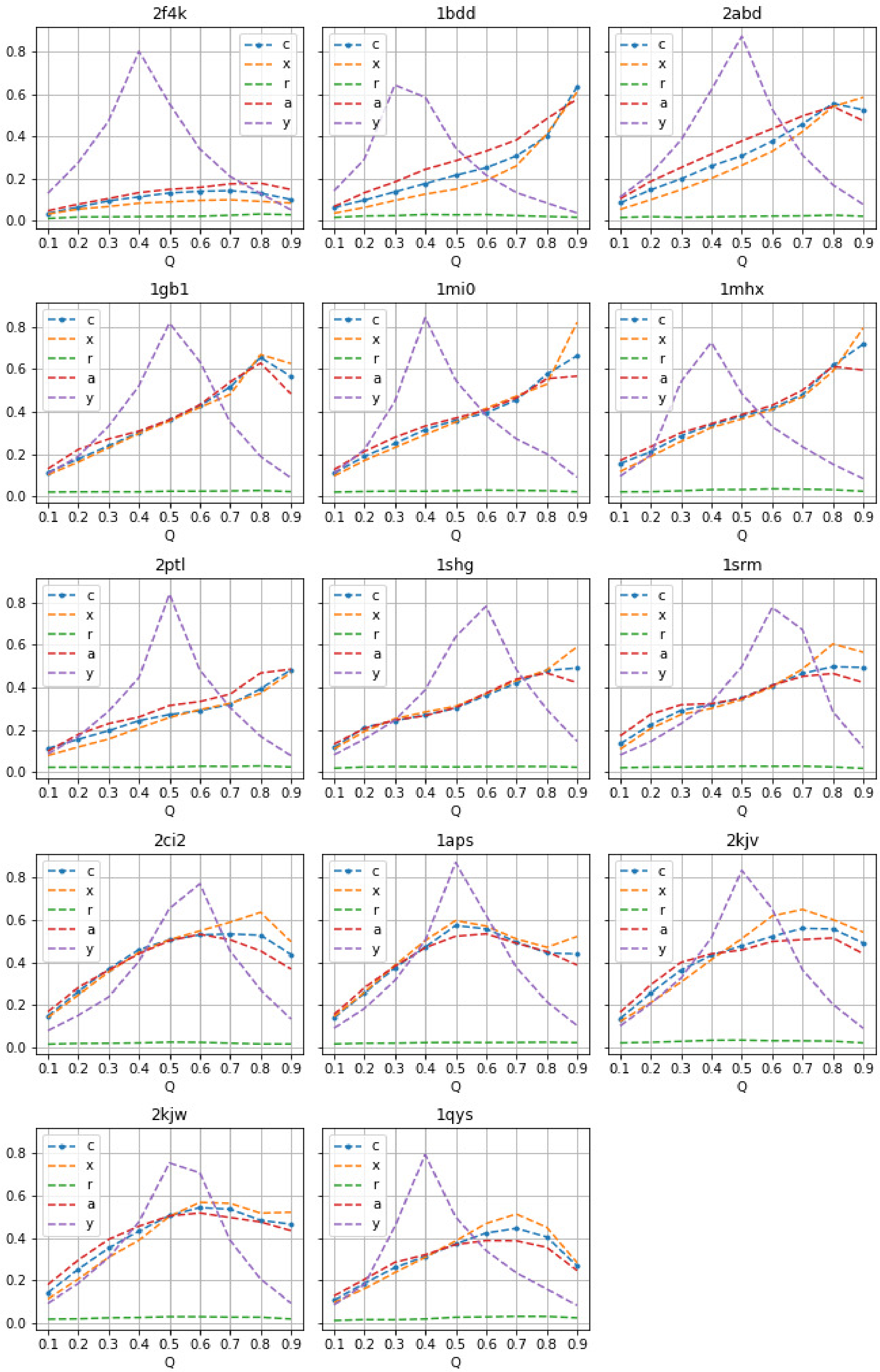

**Fig. G5 Khor dataset**. Route-Order (RO) of native shortcuts generated by the five ND variants. Like the biased ND variants, the (improbable) 'y' ND variant also exhibits route-like behavior, but with RO that peaks about 0.8 in the middle of the reaction coordinate. RO closer to 1.0 indicates fewer folding pathways.

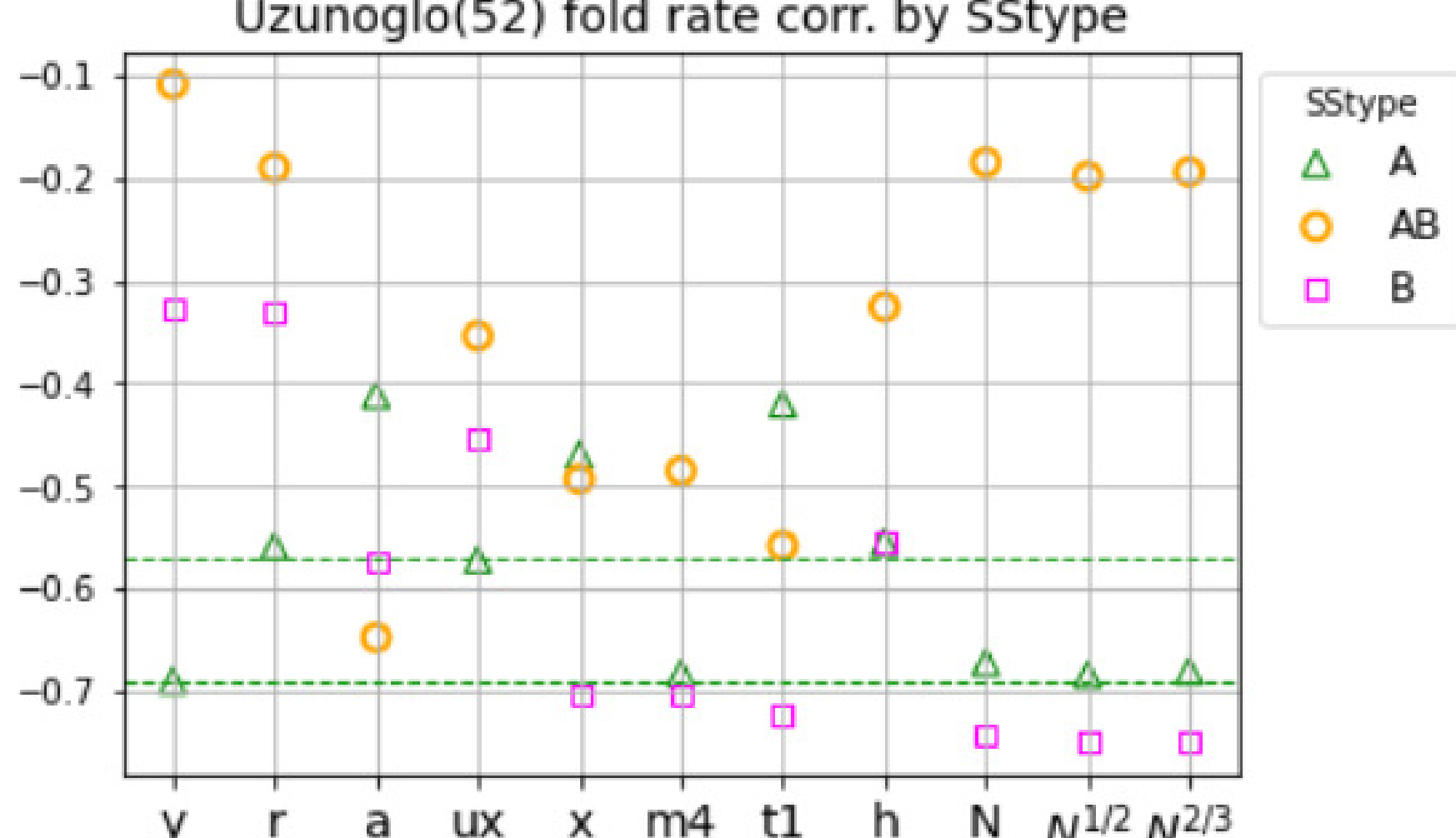

**Fig. H1 Uzunoglu dataset.** Pearson's Correlation coefficient (y-axis) between experimental folding rate with simulated folding rate (peak ***E***) generated by ND variants ('y', 'r', 'a', 'ux', 'x', 'm4', 't1', 'h'), and with scaled chain length *N*. Smaller coefficient is preferable. Notably, the improbable ND variant 'y' produced the strongest correlation for chains with α fold type (SStype = A); this correlation is comparable with scaled *N*. However, using a folding rate determinant like *N* does not provide insight into edge ordering.

'ux' is ND variant 'x' with unscaled $p_e$; its correlation for α fold type chains is competitive with 'r'. 'm4' and 't1' are mixed ND variants. 'm4' correlation for α fold type chains is competitive with 'y' and can be combined with other ND variants to boost overall correlation.

With the 'h' ND variant, PRN0 edges are ordered by their sequence distance in monotonically increasing order and applied in batches by their sequence distance. 'h' is a simplest edge ordering; its overall correlation is -0.542 (p-value < 0.005). For the cost involved in computing peak *E*s for an ND variant, its results need to at least outperform 'h' and preferably also *N* (but *N* gives no information about edge ordering). The overall correlation with $N^{0.5}$ is -0.457 (p-value < 0.005). So both ND variant 'a', and the combination ND variant m4-a-m4 pass this minimum requirement test.

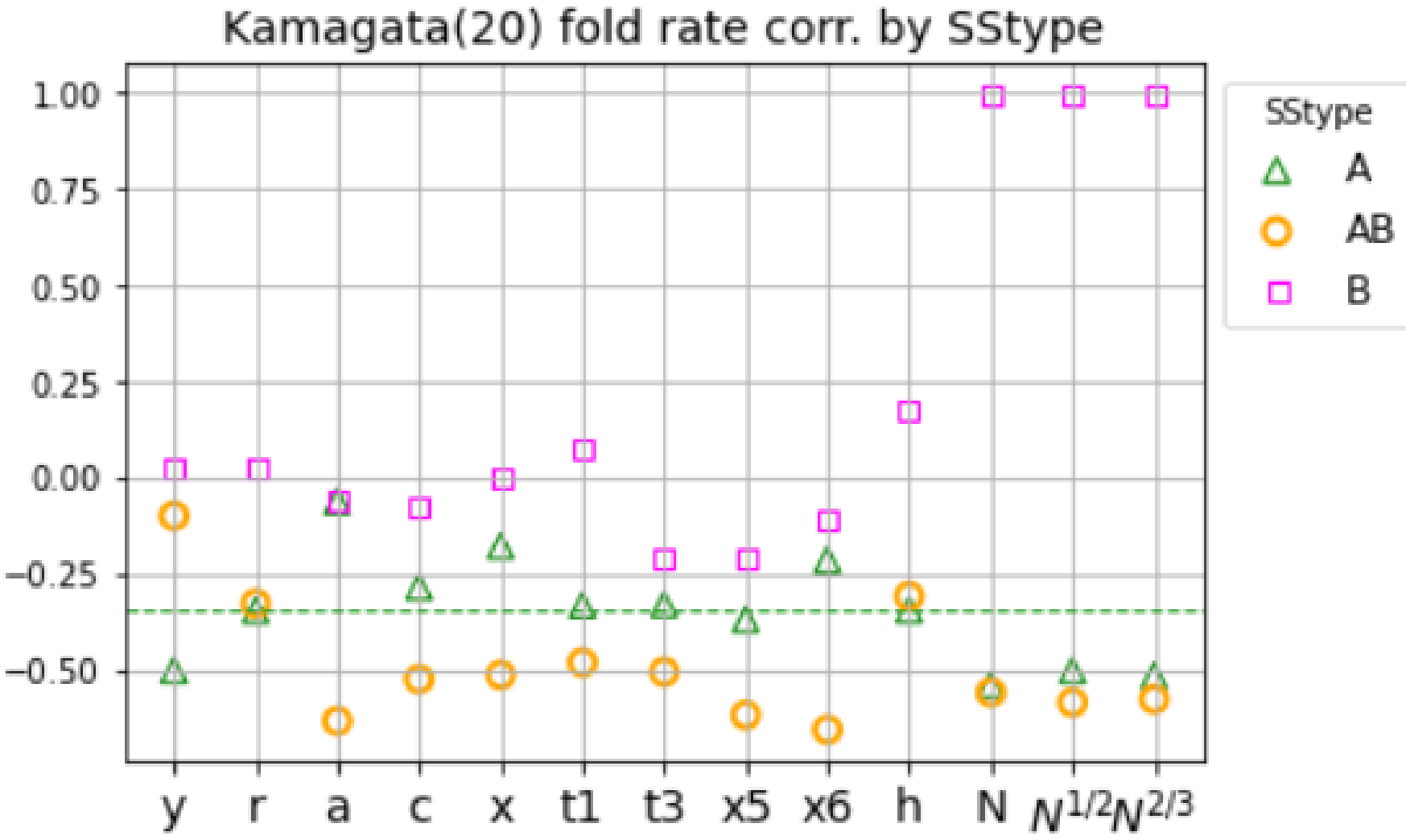

**Fig. H2 Kamagata dataset.** Pearson's Correlation coefficient (y-axis) between experimental folding rate with simulated folding rate (peak ***E***) generated by ND variants ('y', 'r', 'a', 'c', 'x', 't1', 't3', 'x5', 'x6', 'h'), and with scaled chain length *N*. Smaller coefficient is preferable.

The 'h' ND variant (described above) yields an overall correlation of -0.351 (p-value=0.130), while the overall correlation with $N^{0.5}$ is -0.572 (p-value=0.0084). So only the combination ND variant t1-a-t3 passes the minimum requirement test.